\documentclass[aps,pre,reprint,groupedaddress,showpacs,showkeys]{revtex4-1}
\usepackage{amsmath, amssymb}
\usepackage{graphicx}
\usepackage{dcolumn}
\usepackage{enumerate}
\usepackage{bm}
\usepackage{natbib}

\begin{document}

\title{Dynamics of homogeneous shear turbulence -- a key role of the nonlinear transverse
cascade in the bypass concept}

\author{G. Mamatsashvili$^{1,2,3}$}
\email{g.mamatsashvili@hzdr.de}
\author{G. Khujadze$^{4}$}
\author{G. Chagelishvili$^{3,5}$}
\author{S. Dong$^{6}$}
\author{J. Jim\'enez$^{6}$}
\author{H. Foysi$^{4}$}
\affiliation{$^1$Helmholtz-Zentrum Dresden-Rossendorf, P.O. Box 510119, D-01314 Dresden, Germany\\
$^2$Department of Physics, Faculty of Exact and Natural Sciences, Tbilisi State University, Tbilisi 0179, Georgia\\
$^3$Abastumani Astrophysical Observatory, Ilia State University, Tbilisi 0162, Georgia\\
$^4$Chair of Fluid Mechanics, Universit$\ddot{a}$t Siegen, Siegen 57068, Germany\\
$^5$Institute of Geophysics, Tbilisi State University, Tbilisi 0128, Georgia\\
$^6$School of Aeronautics, Universidad Polit\'ecnica de Madrid,
Madrid 28040, Spain}

\date{\today}

\begin{abstract}
To understand the mechanism of the self-sustenance of subcritical
turbulence in spectrally stable (constant) shear flows, we performed
direct numerical simulations of homogeneous shear turbulence for
different aspect ratios of the flow domain with subsequent analysis
of the dynamical processes in spectral, or Fourier space. There are
no exponentially growing modes in such flows and the turbulence is
energetically supported only by the linear growth of Fourier
harmonics of perturbations due to the shear flow nonnormality. This
nonnormality-induced growth, also known as nonmodal growth, is
anisotropic in spectral space, which, in turn, leads to anisotropy
of nonlinear processes in this space. As a result, a transverse
(angular) redistribution of harmonics in Fourier space is the main
nonlinear process in these flows, rather than direct or inverse
cascades. We refer to this new type of nonlinear redistribution as
\emph{the nonlinear transverse cascade}. It is demonstrated that the
turbulence is sustained by a subtle interplay between the linear
nonmodal growth and the nonlinear transverse cascade. This course of
events reliably exemplifies a well-known bypass scenario of
subcritical turbulence in spectrally stable shear flows. These two
basic processes mainly operate at large length scales, comparable to
the domain size. Therefore, this central, small wavenumber area of
Fourier space is crucial in the self-sustenance; we defined its size
and labeled it as \emph{the vital area of turbulence}. Outside the
vital area, the nonmodal growth and the transverse cascade are of
secondary importance - Fourier harmonics are transferred to
dissipative scales by the nonlinear direct cascade. Although the
cascades and the self-sustaining process of turbulence are
qualitatively the same at different aspect ratios, the number of
harmonics actively participating in this process (i.e., the
harmonics whose energies grow more than 10\% of the maximum spectral
energy at least once during evolution) varies, but always remains
quite large (equal to 36, 86 and 209) in the considered here three
aspect ratios. This implies that the self-sustenance of subcritical
turbulence cannot be described by low-order models.
\end{abstract}

\pacs{47.10.-g, 47.20.-k, 47.27.-i, 47.52.+j}

\maketitle

\section{Introduction}

In the 1990s, the nonnormal nature of shear flows and its
consequences became well understood and extensively studied by the
hydrodynamic community (see e.g., Refs.~\cite{Farrell88,
Butler_Farrell92,Reddy_etal93,Trefethen_etal93, Reddy_Henningson93,
Schmid_Henningson01,Criminale_etal03, Schmid07, Zhuravlev14}). As a
result, a new, so-called bypass concept was developed for explaining
the onset and self-sustenance of turbulence in spectrally stable
shear flows
\cite{Farrell_Ioannou94,Gebhardt_Grossmann94,Henningson_Reddy94,Baggett_etal95,Grossmann00,
Chapman02,Eckhardt_etal07,Farrell_Ioannou12,Brandt14}. It is based
on the linear nonnormality-induced, or nonmodal growth of vortical
perturbations due to the nonnormality (shear), which is the only
source of energy for turbulence in such flows
\cite{Henningson_Reddy94}. In this case, the role of nonlinear
processes becomes crucial -- they lie at the heart of
self-sustenance of the turbulence. In the linear regime, without
nonlinear feedback, the nonmodal growth has a transient character
and thus itself alone is incapable of self-sustenance.

Studies of turbulence in constant shear flows, or homogeneous shear
turbulence usually concentrate on dynamics and statistics in
physical space. Coherent vortical structures responsible for the
self-sustenance of the turbulence were identified and described in
\cite{Tavoularis_Corrsin81,Rogers_Moin87,Lee_etal90,
Kida_Tanaka94,Pumir_Shraiman95,Waleffe95,Waleffe97}. The dependence
of the dynamics of these structures, and hence the characteristics
of turbulence, on the aspect ratio of flow domain was extensively
analyzed in \cite{Pumir96,Sekimoto_etal15}. Compared with these
studies, which mostly perform analysis in physical space, the main
idea of our investigation is the analysis of
turbulence dynamics in three-dimensional (3D) spectral, or Fourier (${\bf k}$-) space that allows us:\\
-- to identify harmonics/modes that play a key role in the
self-sustaining process of turbulence;\\
-- to define a wavenumber area in Fourier space that is vital in the
self-sustenance of turbulence;\\
-- to define the range of aspect ratios for which the dynamically
important modes are fully taken into account;\\
-- to reveal a new type of nonlinear cascade process that
principally differs from the canonical ones, i.e., from direct and
inverse cascades;\\
-- to show that the turbulence is sustained by a subtle interplay
between the linear nonmodal growth and the new type of the nonlinear
cascade.

Generally, in spectrally stable shear flows, the classical nonlinear
direct and inverse cascades alone are in fact not able to ensure the
self-sustenance of transiently growing perturbations. In the case of
a specific spectrally stable shear flow, it was demonstrated that
the turbulence can be self-organized and self-sustained by a subtle
interplay of the linear nonmodal and nonlinear processes
\cite{Gebhardt_Grossmann94,Baggett_etal95,Waleffe95,
Waleffe97,Grossmann00,Chapman02, Eckhardt_etal07,
Farrell_Ioannou12}. In this situation, a continuous energy supply
from shear flow into the turbulence is provided by the linear
nonmodal growth mechanism owing to an essential positive feedback
provided by the nonlinear processes. One can say that subcritical
turbulence in shear flows falls outside the validity of the
classical Kolmogorov phenomenology (i.e., direct and inverse
cascades). The deviation from the latter occurs especially in the
region of small and intermediate wavenumbers of perturbations, where
the linear nonnormality-induced energy-exchange and (nonclassical)
nonlinear processes work in cooperation to ensure the sustenance of
shear turbulence.

Actually, the shear-induced linear nonmodal growth of a perturbation
harmonic mainly depends on its wavevector orientation and, to a
lesser degree, on the value (e.g., see Refs.
\cite{Farrell87,Farrell88,Butler_Farrell92,Farrell_Ioannou93,
Schmid_Henningson01,Jimenez13}): the spatial Fourier harmonics that
have a certain orientation of the wavevector with respect to shear
flow draw energy from it and grow, whereas harmonics with other
orientation of the wavevector give energy back to the flow and
decay. This anisotropy of the linear energy-exchange processes with
respect to wavevector orientation (angle), in turn, leads to
anisotropy of nonlinear processes in {\bf k}-space. Specifically, as
revealed in \cite{Chagelishvili_etal02,
Horton_etal10,Mamatsashvili_etal14,Salhi_etal14}, in HD/MHD smooth
shear flows, the main nonlinear process is not a direct or inverse,
but rather a nonlinear transverse cascade, that is
angular/transverse redistribution of perturbation harmonics in {\bf
k}-space. The nonlinear transverse cascade, or transverse cascade
for short, represents an alternative to the classical direct and
inverse cascades in the presence of flow shear.

In the paper by Horton {\it et al.} \cite{Horton_etal10}, the
nonlinear dynamics of coherent and stochastic vortical perturbations
in two-dimensional (2D) plane constant shear flow was studied
numerically and the existence and importance of the transverse
cascade for the self-sustaining dynamics of perturbations was
demonstrated. Specifically, it was shown that its interplay with
linear nonnormality-induced phenomena becomes intricate: it can
realize either positive or negative feedback. In the case of the
former, the transverse cascade repopulates those quadrants in {\bf
k}-space where the shear flow causes linear nonmodal growth. In the
paper by Mamatsashvili {\it et al.} \cite{Mamatsashvili_etal14}, the
dynamics of 2D perturbations in incompressible MHD flows with
constant shear of velocity and uniform magnetic field parallel to
the flow was studied. Investigating subcritical turbulence, the
action of the transverse cascade -- a keystone of the turbulence
self-sustaining (non-decaying) dynamics in this simple open MHD flow
system -- was described in detail. In this spirit, Salhi {\it et
al.} \cite{Salhi_etal14} studied 3D homogeneous turbulence in
rotating shear flows, focusing on the anisotropy of the dynamics in
spectral space, especially, on the nonlinear transverse
redistribution. A refined analysis was done, employing the
polarization and helicity scalars. However, due to the rather short
final time, the DNS had no possibility to characterize a saturated
state of the turbulence and hence had no access to the complete set
characteristic quantities. Nevertheless, putting the line of
research of paper \cite{Salhi_etal14} into perspective, it could
yield important results on the spectral anisotropy of shear
turbulence in the long-time DNS for which the saturated state is
already reached.

In the present paper, we consider the dynamics of homogeneous shear
turbulence in spectral space, extending the knowledge and results
gained from the investigation of a 2D constant shear flow to a more
realistic 3D one. An advantage of the homogeneous shear turbulence
is that it includes the basic effects of shear, while avoiding
complications arising in more realistic cases (e.g., effect of
boundaries, etc.). The proposed investigation of this simple shear
flow system allows us to gain insight into the scheme of realization
of the bypass concept in the case of homogeneous shear turbulence.

The paper is organized as follows. The physical model and derivation
of dynamical equations in spectral space is in Sec. II. The direct
numerical simulations (DNS) of the turbulence dynamics at different
aspect ratios of the flow domain as well as the analysis of the
self-sustaining mechanism in Fourier space are presented in Sec.
III. A summary and discussion are given in Sec. IV.

\section{Physical model and equations}

The motion of an incompressible fluid with constant kinematic
viscosity, $\nu$, is governed by the Navier-Stokes equations
\begin{equation}
\frac{\partial {\bf U}}{\partial t}+\left({\bf U}\cdot \nabla
\right){\bf U}=-\frac{\nabla P}{\rho}+\nu\nabla^2 {\bf U},
\end{equation}
\begin{equation}
\nabla\cdot {\bf U}=0,
\end{equation}
where $\rho,~{\bf U},~P$ are the density, velocity and pressure,
respectively.

Equations (1)-(2) have a stationary equilibrium solution with
velocity profile ${\bf U}_0=(Sy,0,0)$ and spatially constant density
and pressure ($\rho_0, P_0=const.$), i.e., a flow in the streamwise
$x$-direction with a linear shear of velocity in the
vertical/shearwise $y$-direction, while the $z$-axis points in the
spanwise direction. Without loss of generality, the constant shear
rate $S$ is chosen to be positive. Such an idealized configuration
of a flow with a linear shear, despite its simplicity, allows us to
grasp key effects of the shear on the perturbation dynamics and
ultimately on a resulting turbulent state.

Consider 3D perturbations of the velocity and pressure, ${\bf u}$
and $p$, about the background flow: ${\bf u}={\bf U}-{\bf U}_0,
p=P-P_0$. Eqs. (1)-(2) then give the following system of nonlinear
equations for the perturbations
\begin{multline}
\left(\frac{\partial}{\partial t} + Sy \frac{\partial}{\partial x}
\right)u_x = - Su_y - \frac{1}{\rho_0}\frac{\partial p}{\partial
x}-\\\frac{\partial}{\partial x}u_x^2-\frac{\partial}{\partial
y}(u_xu_y)-\frac{\partial}{\partial z}(u_xu_z)+\nu\nabla^2u_x,
\end{multline}
\begin{multline}
\left(\frac{\partial}{\partial t}  + Sy \frac{\partial}{\partial x}
\right)u_y =-\frac{1}{\rho_0}\frac{\partial p}{\partial
y}-\\\frac{\partial}{\partial x}(u_xu_y)-\frac{\partial}{\partial
y}u_y^2-\frac{\partial}{\partial z}(u_yu_z)+\nu\nabla^2u_y,
\end{multline}
\begin{multline}
\left(\frac{\partial}{\partial t}  + Sy \frac{\partial}{\partial x}
\right)u_z =-\frac{1}{\rho_0}\frac{\partial p}{\partial
z}-\\\frac{\partial}{\partial x}(u_xu_z)-\frac{\partial}{\partial
y}(u_yu_z)-\frac{\partial}{\partial z}u_z^2+\nu\nabla^2u_z,
\end{multline}
\begin{equation}\label{4}
\frac{\partial u_x}{\partial x}+\frac{\partial u_y}{\partial
y}+\frac{\partial u_z}{\partial z}=0.
\end{equation}

To investigate the energy balance in the self-sustained turbulent
state, from Eqs. (3)-(6) we derive a dynamical equation for the
kinetic energy density of perturbations, $E= {\rho_0{\bf u}^2}/2$:
\begin{multline}
\left(\frac{\partial}{\partial t} + Sy \frac{\partial}{\partial x}
\right)E =-S\rho_0u_xu_y+\nabla\cdot[{\bf u}
(p+E)]+\\\nu\nabla\cdot(\rho_0{\bf u}\otimes\nabla{\bf
u})-\rho_0\nu[\left(\nabla u_x\right)^2+\left(\nabla u_y\right)^2
+\left(\nabla u_z\right)^2],
\end{multline}
which after averaging over an entire flow domain gives
\begin{equation}
\frac{d}{dt}\langle E \rangle=-S\left\langle
\rho_0u_xu_y\right\rangle -\rho_0\nu\langle \left(\nabla
u_x\right)^2+\left(\nabla u_y\right)^2 +\left(\nabla u_z\right)^2
\rangle,
\end{equation}
where the angle brackets denote a volume average. The first term on
the right hand side of Eq. (8) is the flow shear rate, $S$,
multiplied by the volume-averaged Reynolds stress, $\langle
\rho_0u_xu_y \rangle$, and describes exchange of energy between
perturbations and the background flow. Note that this term
originates from the linear term proportional to shear ($-Su_y$) on
the right hand side of Eq. (3). The second -- viscous dissipation --
term is negative by definition. The nonlinear terms, represented by
divergence in Eq. (7), cancel out in the total energy evolution Eq.
(8) after volume averaging, because there is no net flux of energy
into the flow through the boundaries. Thus, only the Reynolds stress
can supply perturbations with energy, extracting it from the
background flow due to shear. In the case of turbulence studied
below, the Reynolds stress ensures energy injection into turbulent
fluctuations, balancing viscous dissipation. The nonlinear processes
do not directly change the total perturbation energy (extracted from
the background flow by the stress), but redistribute it among
Fourier harmonics of perturbations with different wavenumbers. In
this way, nonlinear processes indirectly contribute to the
perturbation energy balance. Below, we focus on the nonlinear
redistribution process in spectral space to grasp the
self-sustenance scheme of the turbulence.

\subsection{Spectral representation of the dynamical equations}

We derive dynamical equations for the quadratic forms of velocity
perturbations, $(u_x^2$, $u_y^2$, $u_z^2)$, and kinetic energy
density, $E$, in Fourier space by decomposing the perturbations into
spatial Fourier harmonics
\[
f({\bf r},t)=\int \bar{f}({\bf k},t)e^{{\rm i}{\bf k}\cdot{\bf
r}}dk_xdk_ydk_z
\]
where $f\equiv ({\bf u},p)$ denotes the perturbations and
$\bar{f}\equiv (\bar{\bf u}, \bar{p})$ -- their corresponding
Fourier transforms. Then, from Eqs. (3)-(6), we get the following
equations governing the dynamics of perturbation harmonics in
spectral space (the constant density is set to unity, $\rho_0=1$)
\begin{equation}
\left(\frac{\partial}{\partial t}-Sk_x\frac{\partial}{\partial
k_y}\right)\bar{u}_x=-S\bar{u}_y-{\rm i}k_x\bar{p}+Q_x-\nu
k^2\bar{u}_x,
\end{equation}
\begin{equation}
\left(\frac{\partial}{\partial t}-Sk_x\frac{\partial}{\partial
k_y}\right)\bar{u}_y=-{\rm i}k_y\bar{p}+Q_y-\nu k^2\bar{u}_y,
\end{equation}
\begin{equation}
\left(\frac{\partial}{\partial t}-Sk_x\frac{\partial}{\partial
k_y}\right)\bar{u}_z=-{\rm i}k_z\bar{p}+Q_z-\nu k^2\bar{u}_z,
\end{equation}
\begin{equation}
k_x\bar{u}_x+k_y\bar{u}_y+k_z\bar{u}_z=0,
\end{equation}
where $k^2=k_x^2+k_y^2+k_z^2$. These spectral equations contain the
linear as well as the nonlinear, ${\bf Q}({\bf k},t)=[Q_x({\bf
k},t), Q_y({\bf k},t), Q_z({\bf k},t)]$, terms that are the Fourier
transforms of corresponding ones in the original Eqs. (3)-(6). The
latter are given by
\[
Q_i({\bf k},t)=\sum_{j}{k_jq_{ij}({\bf k},t)},~~~~i,j=x,y,z
\]
where
\[
q_{ij}({\bf k},t)=-{\rm i}\int\bar{u}_i({\bf
k^{\prime}},t)\bar{u}_j({\bf k}-{\bf k^{\prime}},t)d^3{\bf
k^{\prime}}
\]
and describe nonlinear triad interactions among harmonics. From Eqs.
(9)-(12) one can eliminate the pressure,
\[
\bar{p}=2{\rm i}S\frac{k_x}{k^2}\bar{u}_y -{\rm i} \frac{{\bf
k}\cdot{\bf Q}}{k^2},
\]
and rewrite them in the following form:
\begin{multline}
\left(\frac{\partial}{\partial t}-Sk_x\frac{\partial}{\partial
k_y}\right)\bar{u}_x=S\left(2\frac{k_x^2}{k^2}-1
\right)\bar{u}_y+\\Q_x-\frac{k_x}{k^2}({\bf k}\cdot{\bf Q})-\nu
k^2\bar{u}_x,
\end{multline}
\begin{multline}
\left(\frac{\partial}{\partial t}-Sk_x\frac{\partial}{\partial
k_y}\right)\bar{u}_y=2S\frac{k_xk_y}{k^2}\bar{u}_y+\\Q_y-\frac{k_y}{k^2}({\bf
k}\cdot{\bf Q})-\nu k^2\bar{u}_y,
\end{multline}
\begin{multline}
\left(\frac{\partial}{\partial t}-Sk_x\frac{\partial}{\partial
k_y}\right)\bar{u}_z=2S\frac{k_xk_z}{k^2}\bar{u}_y+\\Q_z-\frac{k_z}{k^2}({\bf
k}\cdot{\bf Q})-\nu k^2\bar{u}_z,
\end{multline}
or, for the quadratic forms of the Fourier transforms of the
velocity
\begin{multline}
\left(\frac{\partial}{\partial t}-Sk_x\frac{\partial}{\partial
k_y}\right)\frac{|\bar{u}_x|^2}{2}=\\S\left(1-2\frac{k_x^2}{k^2}\right){\cal
H}_k+{\cal N}_x-\nu k^2|\bar{u}_x|^2,
\end{multline}
\begin{equation}
\left(\frac{\partial}{\partial t}-Sk_x\frac{\partial}{\partial
k_y}\right)\frac{|\bar{u}_y|^2}{2}=2S\frac{k_xk_y}{k^2}|\bar{u}_y|^2+{\cal
N}_y-\nu k^2|\bar{u}_y|^2,
\end{equation}
\begin{multline}
\left(\frac{\partial}{\partial t}-Sk_x\frac{\partial}{\partial
k_y}\right)\frac{|\bar{u}_z|^2}{2}=\\S\frac{k_xk_z}{k^2}(\bar{u}_z^{\ast}\bar{u}_y+\bar{u}_z\bar{u}_y^{\ast})+{\cal
N}_z-\nu k^2|\bar{u}_z|^2,
\end{multline}
where ${\cal
H}_k=-(\bar{u}_x\bar{u}_y^{\ast}+\bar{u}_x^{\ast}\bar{u}_y)/2$ is
the spectral density of the Reynolds stress (with minus sign) and
${\cal N}_x, {\cal N}_y, {\cal N}_z$ are the modified nonlinear
transfer functions,
\begin{multline*}
{\cal
N}_i=\frac{1}{2}(\bar{u}_i^{\ast}Q_i+\bar{u}_iQ_i^{\ast})-\\\frac{k_i}{2k^2}[\bar{u}_i^{\ast}({\bf
k}\cdot{\bf Q})+\bar{u}_i({\bf k}\cdot{\bf Q}^{\ast})],~~~i=x,y,z.
\end{multline*}
Using Eqs. (16)-(18) together with the incompressibility condition
(12), we obtain the evolution equation for the spectral energy
density ${\cal E}_k=(|\bar{u}_x|^2+|\bar{u}_y|^2+|\bar{u}_z|^2)/2$,
\begin{equation}
\left(\frac{\partial}{\partial t}-Sk_x\frac{\partial}{\partial
k_y}\right){\cal E}_k=S{\cal H}_k+{\cal N}_k-{\cal D}_k,
\end{equation}
which can be viewed as a counterpart of energy density Eq. (7) in
spectral space. Here ${\cal D}_k=2\nu k^2{\cal E}_k$ describes
energy dissipation and the nonlinear transfer term ${\cal N}_k$ for
the spectral energy is the sum of the modified transfer functions
\[
{\cal N}_k = {\cal N}_x+{\cal N}_y+{\cal N}_z = \frac{1}{2}[{\bf
\bar{u}^{\ast}}\cdot{\bf Q}+{\bf \bar{u}}\cdot{\bf Q^{\ast}}].
\]
Eqs. (16)-(19) fully determine the nonlinear dynamics of the
considered system in Fourier space and are the basis for subsequent
analysis. According to them, four basic processes underly the
perturbation dynamics:
\begin{enumerate}
\item
The second terms with $Sk_x\partial/\partial k_y$ in the brackets on
the lhs are the fluxes of the corresponding quantities parallel to
the $k_y$-axis. These terms are of linear origin, coming from the
convective derivative on the left hand sides of the main Eqs.
(3)-(5) and correspond to the advection by the background flow. In
other words, the background shear flow makes the spectral quantities
(Fourier transforms) ``drift'' in {\bf k}-space: harmonics with
$k_x>0$ and $k_x<0$ drift in opposite directions along the
$k_y$-axis at a speed $S|k_x|$, $k_y(t)=k_y(0)-Sk_xt$, whereas the
ones with $k_x=0$ are not advected by the flow. Since $\int d^3{\bf
k}\partial (k_x{\cal E}_k)/\partial k_y = 0$, this drift only
transports harmonics parallel to the $k_y$-axis without changing the
total kinetic energy.
\item
The first terms on the rhs are associated with flow shear, i.e.,
originate from the linear term proportional to the shear rate on the
rhs of Eq. (3). They describe the direct influence of the background
flow on the evolution of perturbation harmonics, while the first
term on the rhs of Eq. (19) describes the exchange of energy between
the background flow and individual harmonics. This term in the
spectral energy equation is related to the volume-averaged Reynolds
stress entering Eq. (8) through
\[
-\langle u_xu_y\rangle=\int {\cal H}_k({\bf k},t)d^3{\bf k},
\]
and hence serves as the only source of energy for harmonics. This
shear-induced nonmodal growth process is linear by nature and has a
transient character when nonlinearity is not included
\cite{Chagelishvili_etal03,Tevzadze_etal03,Umurhan_Regev04,Jimenez13,Zhuravlev14}.
The Reynolds stress spectral density, ${\cal H}_k({\bf k},t)$,
describes injection of kinetic energy into turbulent eddies as a
function of wavenumbers. Areas of positive and large values of
${\cal H}_k({\bf k},t)$ correspond to intensive pumping of the
kinetic energy into turbulent eddies and therefore are dynamically
important. Due to the ``drift'' along the $k_y$-axis, the harmonics
enter and leave the energy injection areas in spectral space. The
first terms on the rhs of Eqs. (16)-(18) describe the same process
for the quadratic forms of perturbation velocity components.
Obviously, nonlinear transfer of kinetic energy to the dynamically
important areas from other areas of {\bf k}-space -- i.e.,
repopulation of the dynamically important areas -- is vital for the
self-sustenance of turbulence.
\item
The second terms on the rhs of these equations, ${\cal N}_x$, ${\cal
N}_y$ and ${\cal N}_z$, describe, nonlinear transfers
(redistributions), respectively, of the streamwise,
$|\bar{u}_x|^2/2$, shearwise, $|\bar{u}_y|^2/2$, and spanwise
$|\bar{u}_z|^2/2$ spectral energies in ${\bf k}$-space. Their net
effect in the spectral energy budget over all wavenumbers is zero in
time:
\begin{equation}
\int [{\cal N}_x({\bf k},t)+{\cal N}_y({\bf k},t)+{\cal N}_z({\bf
k},t)]d^3{\bf k}=0,
\end{equation}
which is a consequence of vanishing of the nonlinear terms in the
total energy evolution equation in physical space (Eq. 8). Although
nonlinearity does not directly change the total (i.e., summed over
all wavenumbers) spectral energy, it plays a central role in shear
flow turbulence. ${\cal N}_k$ determines transfer, or cascade of the
spectral energy in {\bf k}-space. These transfer functions are one
of the main focus of the present study. We explore them by adopting
the approach developed in
\cite{Chagelishvili_etal02,Horton_etal10,Mamatsashvili_etal14},
whose main idea is performing a full 3D Fourier analysis of
individual terms in the dynamical Eqs. (16)-(18), thus allowing for
the flow shear-induced anisotropy of spectra and cascades.
\item
The third terms on the rhs describe the viscous dissipation that
becomes important at large wavenumbers $k \gtrsim
\sqrt{S/\nu}=k_{\cal D}$.
\end{enumerate}

Concluding this section, the only source for the total perturbation
energy is the integral of the stress over an entire spectral space
$\int {\cal H}_kd^3{\bf k}$ -- the flow energy extraction and
perturbation growth mechanisms are essentially linear by nature. The
role of nonlinearity is to continually repopulate those harmonics in
{\bf k}-space that are able to undergo nonmodal growth and in this
way continually feed the nonlinear state over long times. This
scenario of a self-sustained state, based on a subtle interplay
between linear and nonlinear processes, is a keystone of the bypass
concept of subcritical turbulence in spectrally stable shear flows
\cite{Gebhardt_Grossmann94,Baggett_etal95,Grossmann00,Chapman02,Rempfer03,Eckhardt_etal07}.

\begin{table*}[t]
\caption{Properties of the numerical simulations: box aspect ratio,
number of grid points, ratio of the grid cell sizes to the temporal
mean of the dissipation scale, volume- and time-averaged values of
the energy, $\tilde{E}$, and Reynolds stress, $-\widetilde{u_xu_y}$,
in the fully developed turbulence. The Reynolds number, $Re_z=4930$,
is the same in all the runs.}
\begin{ruledtabular}
\begin{tabular}{cccccc}
$(A_{xz},A_{yz})$ & $N_x\times N_y\times N_z$ & $(\Delta
x\times \Delta y\times \Delta z)/\eta$ & $\tilde{E}$ & -$ \widetilde{u_xu_y}$ \\
\hline $(1,1)$ & $128\times 128\times 128$ & $1.66\times 1.66\times
1.66$ & 2.37 & 0.71 \\ $(3,2)$ &
$256\times 128\times 64$ & $2.15\times 2.86\times 2.86$ & 1.36 & 0.38 \\
$(1,2)$ & $128\times 256\times 128$ &
$1.49\times 1.49\times 1.49$ & 1.51 & 0.43 \\
\end{tabular}
\end{ruledtabular}
\end{table*}

\section{Results}

In this section, we present the results of numerical analysis of the
nonlinear evolution of perturbations. Main emphasis will be on the
spectral aspect of the dynamics based on the mathematical formalism
outlined in the previous section. The numerical simulations are
performed for the flow in a rectangular domain with size $0 \leq x
\leq L_x, -L_y/2\leq y \leq L_y/2, 0\leq z \leq L_z$ and $N_x\times
N_y\times N_z$ number of grid points. The corresponding resolutions
in these directions are $\Delta x = L_x/N_x, \Delta y = L_y/N_y,
\Delta z=L_z/N_z$. In this domain, we solve the main Eqs. (3)-(6)
using the code developed by Sekimoto et al. \cite{Sekimoto_etal15}
at the School of Aeronautics, Universidad Polit\'ecnica de Madrid,
which solves the Navier-Stokes equations in the velocity-vorticity
representation as in \cite{Kim_etal87}. The streamwise and spanwise
directions are periodic, so that the spectral method is applied in
these directions with the ``3/2'' dealiasing rule. The vertical
direction is shear-periodic and is numerically realized by adding
``shift'' factors in the compact-finite-difference matrices which
guarantee spectral like resolution \cite{Lele92}. The method does
not require the remeshing procedure and therefore does not lead to
the loss of enstrophy \cite{Rogallo81}. To assess how well our
simulations are resolved, we compare the dissipation length scale
$\eta=(\nu^3/\langle \varepsilon \rangle)^{1/4}$, where
$\langle\varepsilon \rangle=\nu \langle(\nabla \times {\bf
u})^2\rangle$ is the volume-averaged dissipation rate of the
perturbation energy, to the grid cell sizes $(\Delta x, \Delta y,
\Delta z)$ in the computational box (see Table I). Since in the
$y$-direction the boundary conditions are shear-periodic
corresponding to time-dependent, or drifting $k_y$ of each harmonic,
a standard Fast Fourier Transform (FFT) technique generally cannot
be applied for calculating Fourier transforms along this direction
during post-processing. We circumvent this by using the method of
Refs. \cite{Mamatsashvili_Rice09,Heinemann_Papaloizou09} that allows
for the time-variation of the shearwise wavenumber when calculating
Fourier transforms (see Appendix).

All physical variables are normalized by the dynamical, or shear
time $S^{-1}$ and the spanwise size of the box, $L_z$. The latter is
the relevant lengthscale in the homogeneous shear turbulence
dynamics, because it sets the typical length and velocity scales of
turbulent structures \cite{Sekimoto_etal15}. The problem has two
dimensionless free parameters -- aspect ratios of the domain,
$A_{xz}=L_x/L_z$, $A_{yz}=L_y/L_z$. Table I lists the
characteristics of the three simulations with different box aspect
ratios that we performed. The Reynolds number is defined in terms of
the box spanwise size, $Re_z=SL_z^2/\nu$, and is the same in all the
runs, $Re_z=4930$. The ratio of the grid cell sizes to the
time-averaged dissipation scale varies from 1.49 for the box
$(A_{xz},A_{yz})=(1,2)$ up to 2.86 for the box
$(A_{xz},A_{yz})=(3,2)$. This implies that large and intermediate
lengthscales, where, as we will show, the self-sustaining dynamics
is concentrated, are well resolved in our simulations. The
wavenumbers $k_x,k_y,k_z$ are normalized, respectively, by $\Delta
k_x=2\pi/L_x, \Delta k_y=2\pi/L_y$ and $\Delta k_z=2\pi/L_z$, that
is, $(k_x/\Delta k_x, k_y/\Delta k_y, k_z/\Delta k_z)\rightarrow
(k_x,k_y,k_z)$. Since $\Delta k_x, \Delta k_y$ and $\Delta k_z$ are
the grid cell sizes in Fourier space, the normalized streamwise and
spanwise wavenumbers become integers $k_x,k_z=0,\pm 1, \pm 2, ...$,
while $k_y$, although changes with time due to drift, is integer at
discrete moments $t_m=mA_{xz}/(|k_x|A_{yz})$, where $m$ is a
positive integer.

A self-sustained turbulent state is achieved in each simulation,
starting from initial random perturbations. Our main goal is to
investigate underlying self-sustaining process of the turbulence and
how its dynamics depends on the box aspect ratio. This allows us to
reveal general patterns of the self-organization as well as it
specificities for each model parameters. Finally, it should be
stressed that our study is based on the DNS of turbulence and in
this respect is general, not relying on simplifying assumptions used
in low-order models of self-sustaining processes (e.g.,
\cite{Waleffe95,Waleffe97}).
\begin{figure}
\includegraphics[width=\columnwidth]{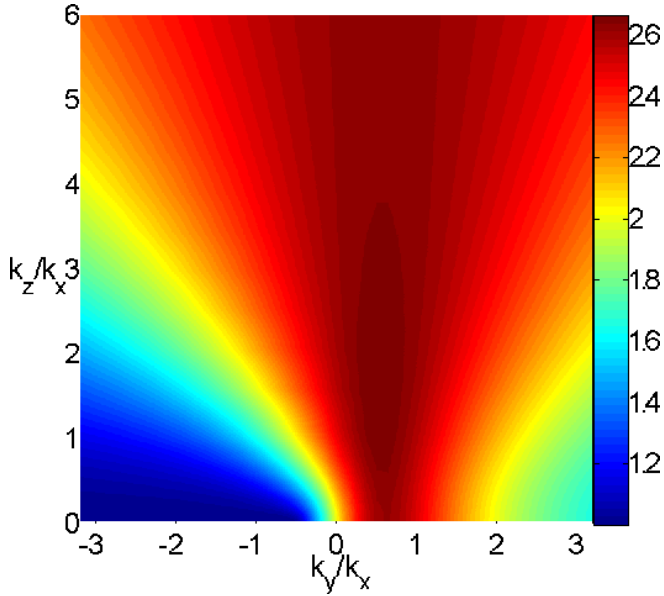}
\caption{(Color online) Transient growth factor of optimal
harmonics' energy, ${\cal E}_k(t_d)/{\cal E}_k(0)$, as a function of
wavenumber ratios. It is calculated during the dynamical time
$t_d=S^{-1}$ with the linearized equations of motion.}
\end{figure}
\begin{figure}
\includegraphics[width=\columnwidth]{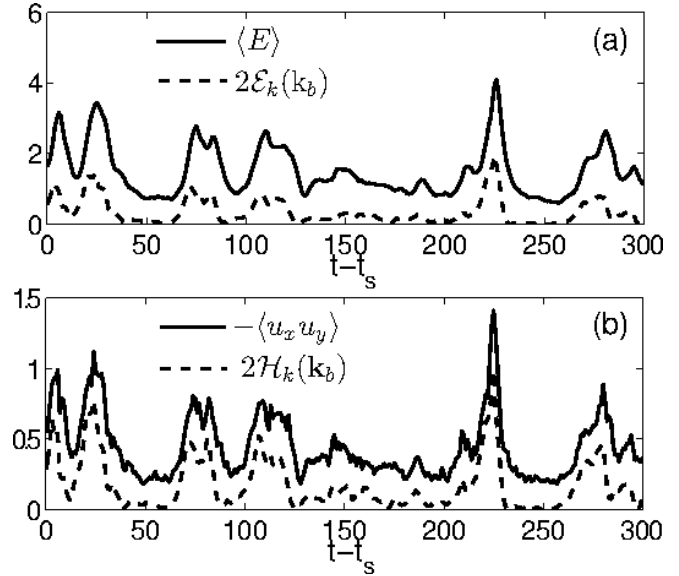}
\caption{Evolution of (a) the volume-averaged total kinetic energy,
$\langle E \rangle$, and the energy of the basic mode with ${\bf
k}_b=(0,0,\pm 1)$, $2{\cal E}_k({\bf k}_b)$, as well as (b) the
volume-averaged Reynolds stress, $-\langle u_xu_y\rangle$ and the
stress corresponding to the basic mode, $2{\cal H}_k({\bf k}_b)$, in
the cubic box.}
\end{figure}
\begin{figure*}
\includegraphics[width=5.6cm]{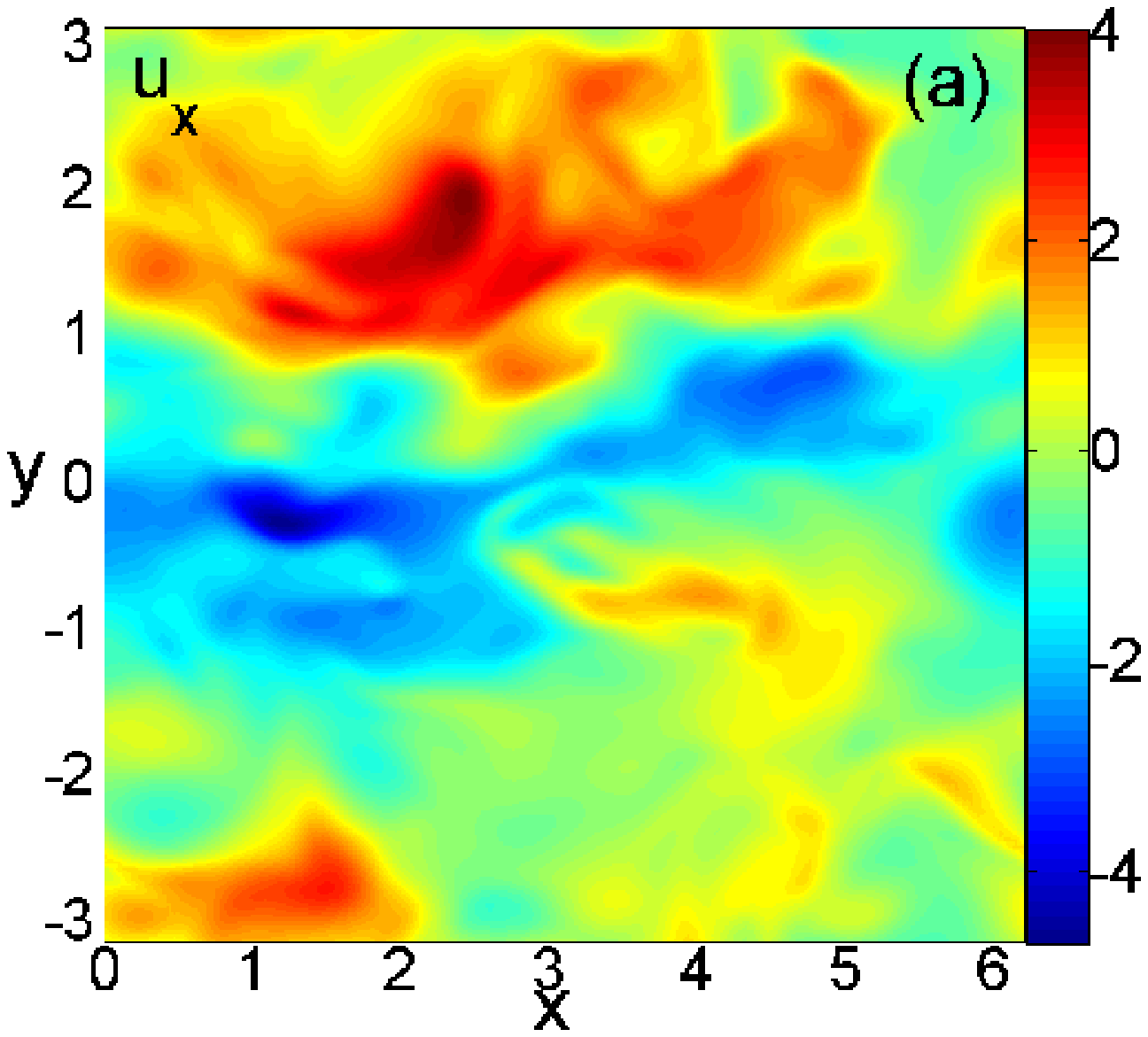}
\includegraphics[width=5.6cm]{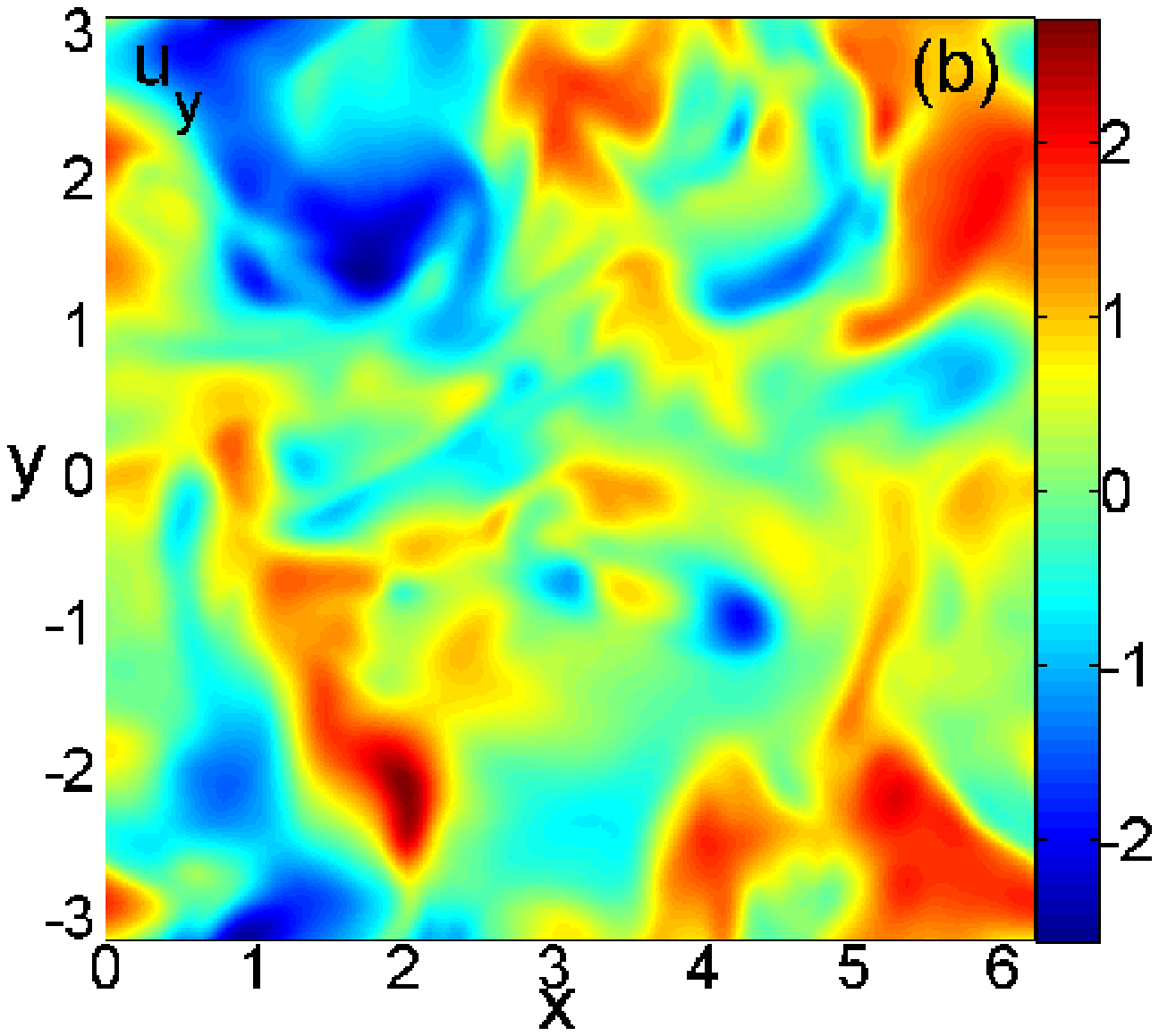}
\includegraphics[width=5.6cm]{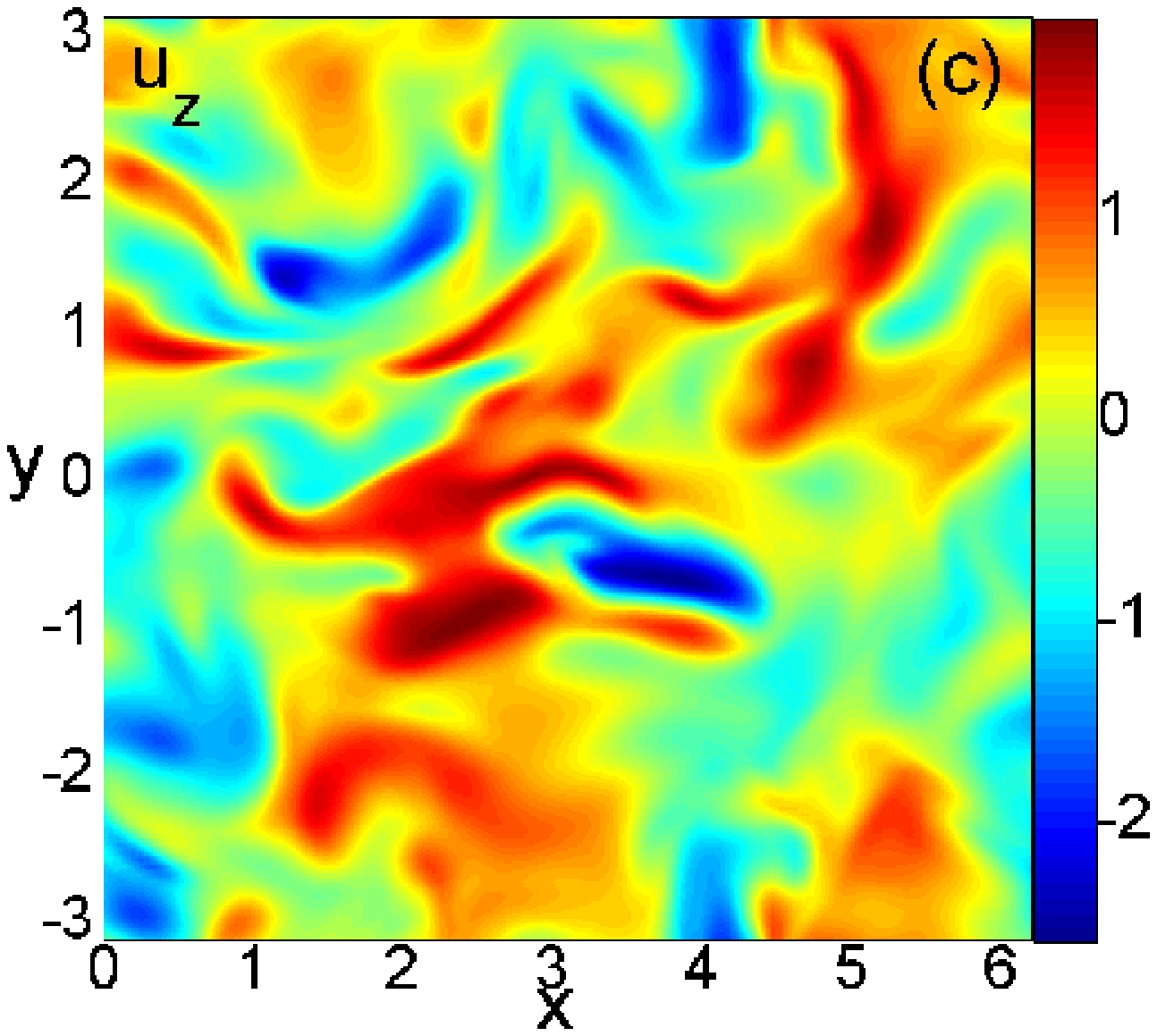}
\includegraphics[width=5.6cm]{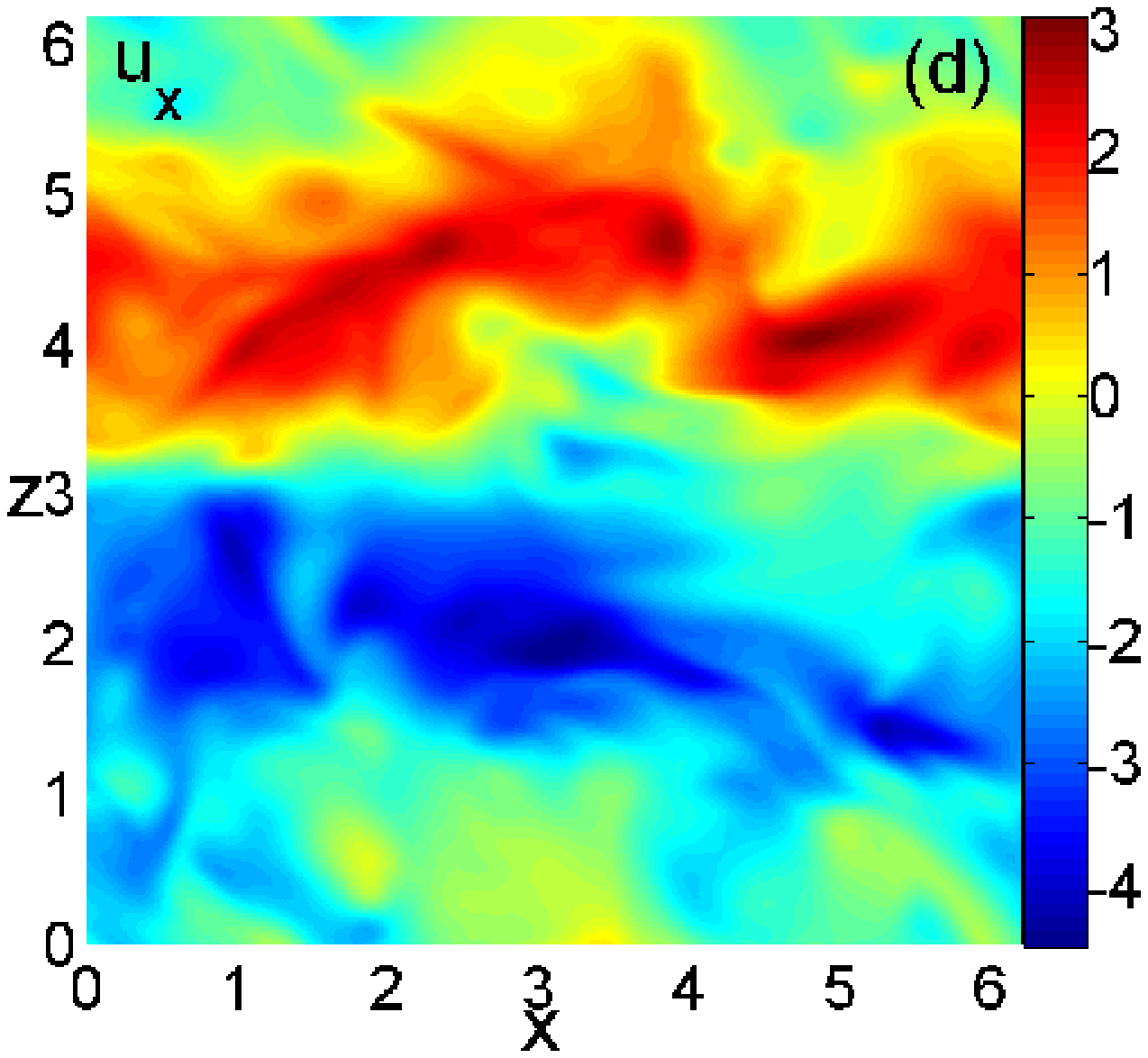}
\includegraphics[width=5.6cm]{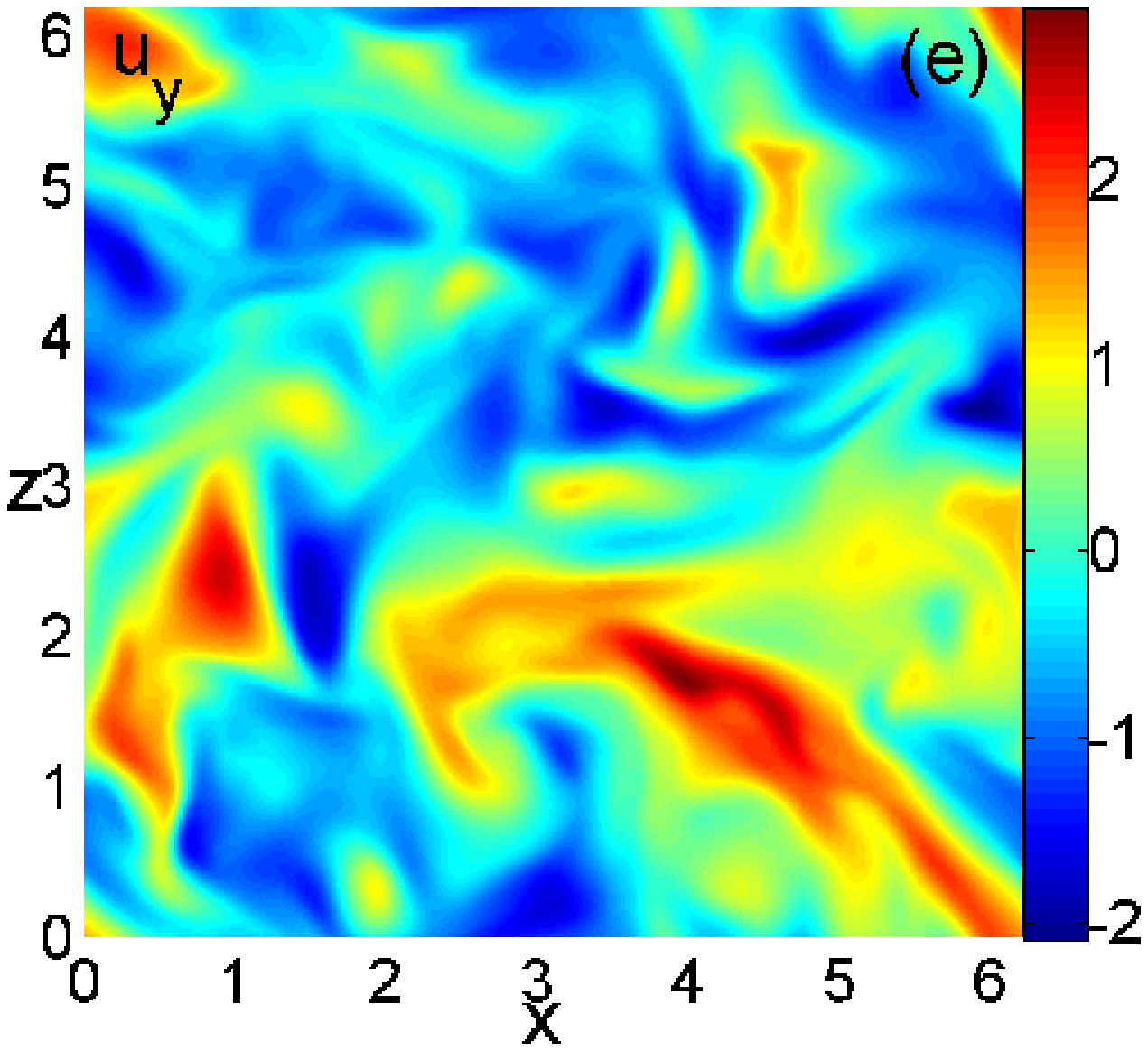}
\includegraphics[width=5.6cm]{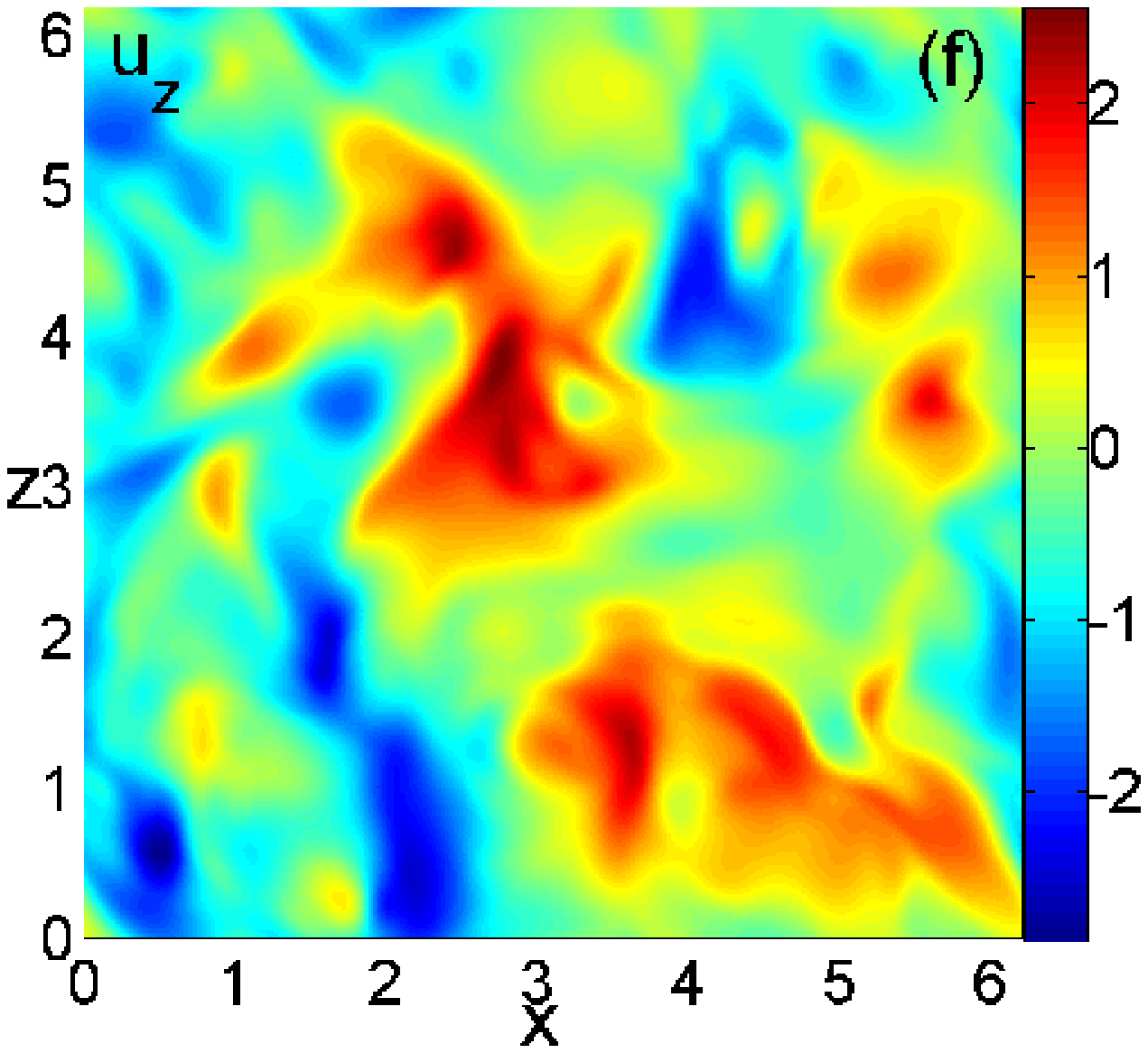}
\caption{(Color online) Distribution of the velocity components in
the fully developed turbulence (at $t-t_s=33$) in the cubic box.
Shown are $(x,y)$- and $(x,z)$-slices at $z=\pi$ and $y=0$,
respectively. The streamwise component $u_x$ is larger than $u_y$
and $u_z$. In the $(x,z)$ slices of $u_y$ and especially of $u_x$, a
signature of the basic harmonic with large spanwise scale can be
discerned.}
\end{figure*}
\begin{figure}
\includegraphics[width=\columnwidth]{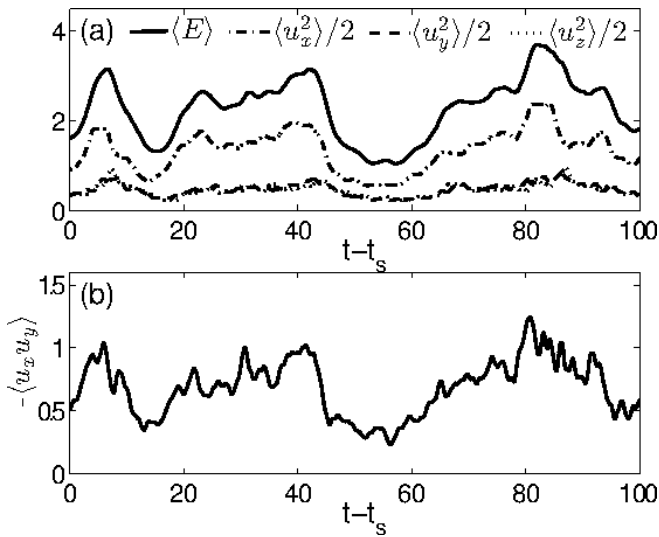}
\caption{Evolution of (a) the volume-averaged perturbation kinetic
energy (solid line) and the quadratic forms of streamwise, $\langle
u_x^2\rangle/2$ (dot-dashed line), shearwise/vertical, $\langle
u_y^2\rangle/2$ (dashed line) and spanwise, $\langle u_z^2\rangle/2$
(doted line) velocities as well as (b) the Reynolds stress in the
cubic box.}
\end{figure}
\begin{figure*}
\includegraphics[width=5.9cm]{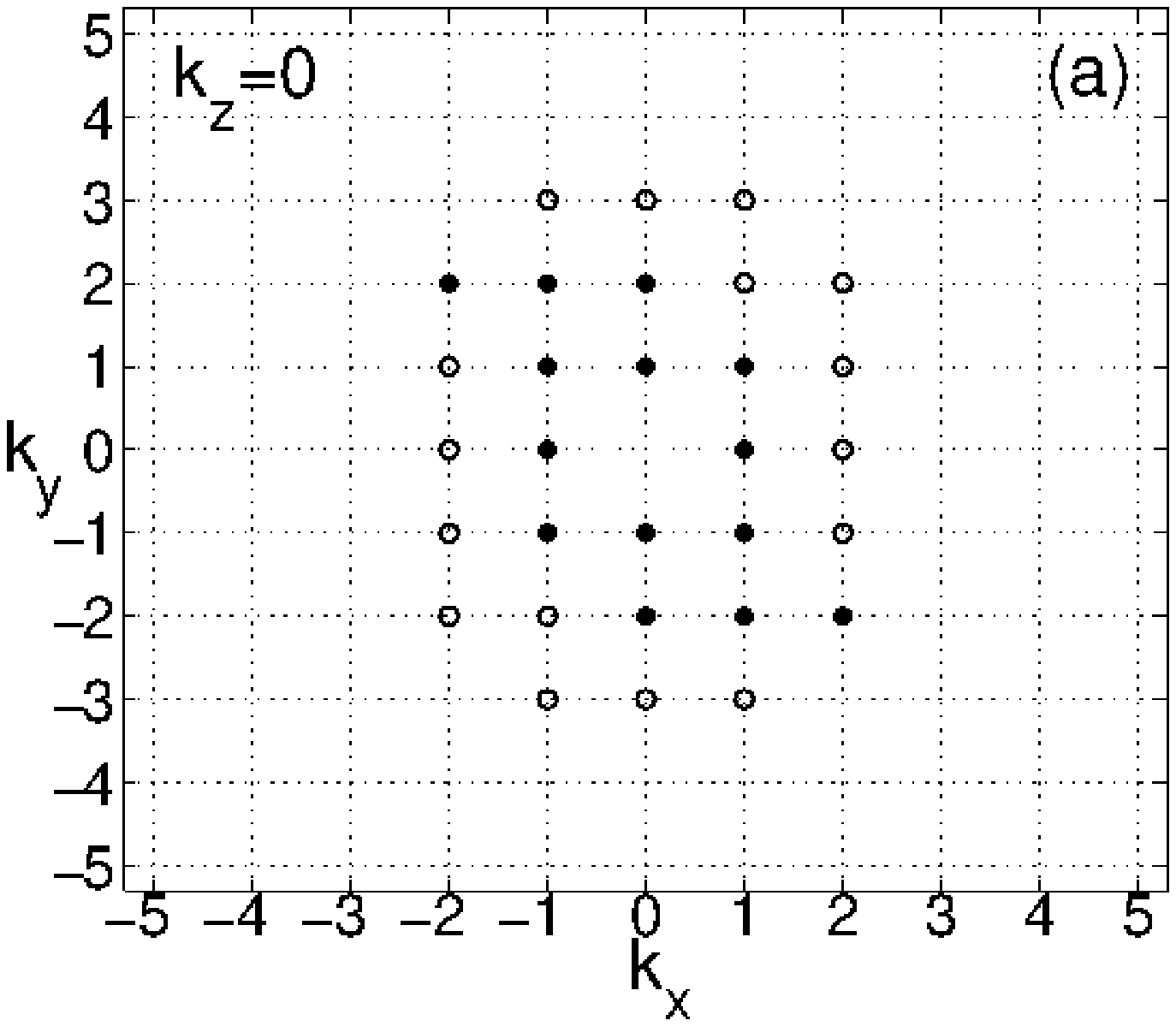}
\includegraphics[width=5.9cm]{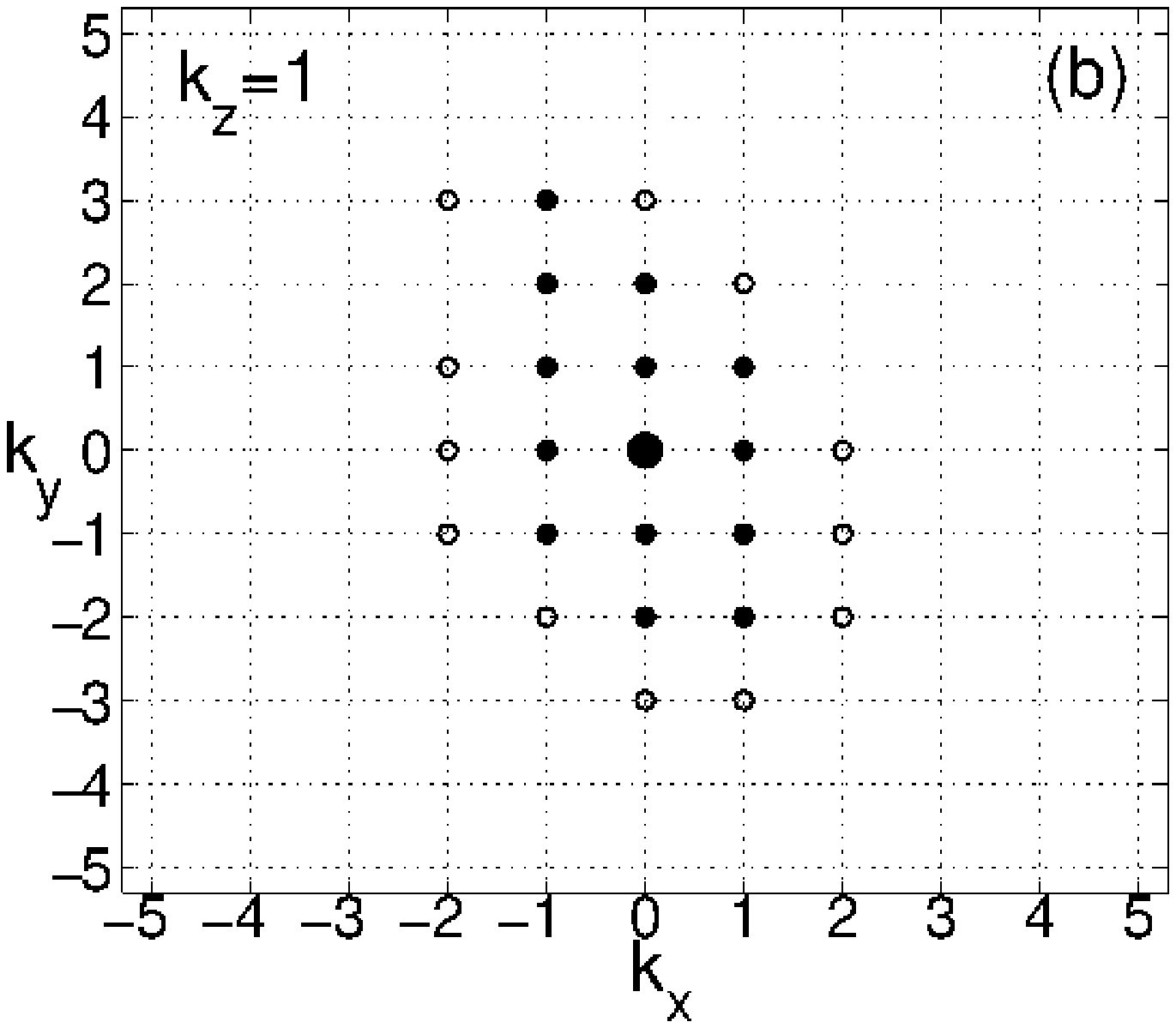}
\includegraphics[width=5.9cm]{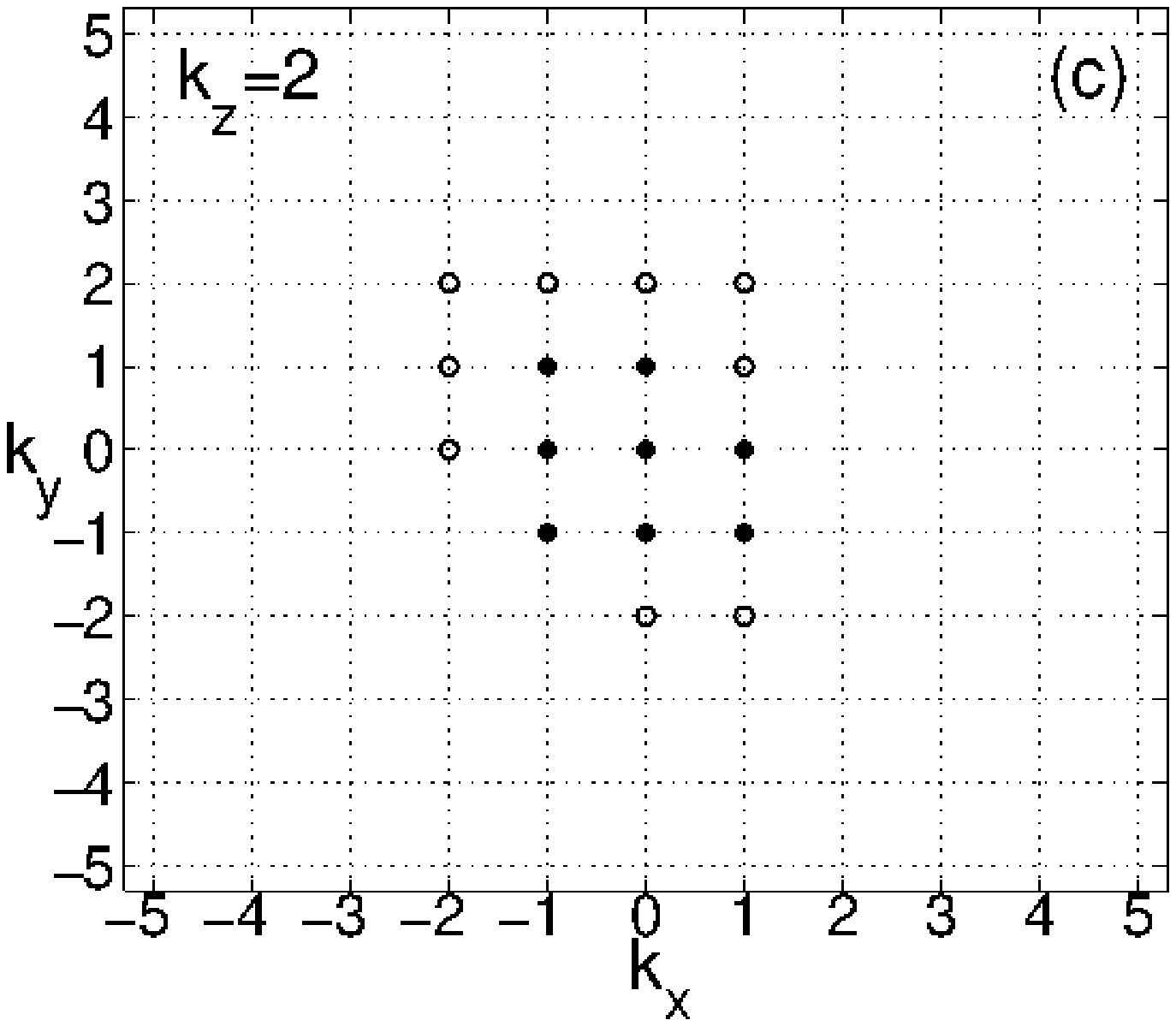}
\caption{Distribution of the modes carrying dominant energy in {\bf
k}-space at (a) $k_z=0$ (b) $k_z=1$ and (c) $k_z=2$ in the cubic
box. Black dots represent the harmonics whose energies grow more
than $10 \%$ of the maximum spectral energy at least once during
evolution. The biggest black dot in the plot (b) corresponds to the
basic mode with the maximum spectral energy at all times. Circles
represent the harmonics whose energies grow from 5\% to 10\% of the
maximum spectral energy.}
\end{figure*}

It is well-known from the linear nonmodal theory of shear flows that
optimal perturbations characterized by a maximum nonmodal growth
during the dynamical or eddy turnover time, which can be assumed of
the order of shear time, $t_d \simeq S^{-1}$ (or, $t_d \simeq 1$ in
normalized units) are responsible for most of the energy extraction
from the background flow
\cite{Schmid_Henningson01,Farrell_Ioannou96}. Figure 1 shows the
growth factor of energy for optimal harmonics calculated from the
linearized inviscid equations of motion, i.e., using only the first
linear terms on the rhs of Eqs. (13)-(15). These terms, and hence
the linear dynamics of harmonics in the inviscid case, depends only
on the ratios of wavenumbers, $k_y/k_x$ and $k_z/k_x$. It is seen
from this figure that the optimal growth factor is largest at $1
\lesssim k_z/k_x < 4$ and $0<k_y/k_x \lesssim 1$. Consequently, one
can encompass as many of these optimal perturbations as possible and
thus better account for their major role in the energy exchange
processes between turbulence and background flow if the
computational box aspect ratios satisfy the conditions: $A_{xz}\leq
A_{yz}$ and $A_{xz}\geq 1$.

In agreement with the previous study of homogeneous shear turbulence
by Pumir \cite{Pumir96}, in our simulations we found that in
addition to the non-symmetric ($k_x\neq 0$) optimal harmonics, a
harmonic with a large scale spanwise variation, which is uniform in
the streamwise and shearwise directions, $k_x=k_y=0, k_z=\pm 1$,
plays a central role in the turbulence dynamics. Due to its
importance, we call this harmonic the basic mode. Other harmonics,
whose energy grows more than 10\% of the maximum value of the
spectral energy at least once during the evolution, are considered
to be actively participating in and forming the self-sustaining
dynamics of turbulence. In particular, we show that one of the
characteristic features of the homogeneous shear turbulence --
quasi-periodic bursts of energy and Reynolds stress \cite{Pumir96}
-- are determined by a competition between the basic mode and other
dynamically ``active'' harmonics, whose number generally depends on
the box aspect ratio and is usually large.

Below we focus on three different boxes with the following aspect
ratios: $(A_{xz}, A_{yz})=(1,1)$ (cubic box),  $(3,2)$, $(1,2)$ and
analyze in detail the specific spectra and self-sustaining dynamics
of turbulence in wavenumber/spectral space by calculating the
Fourier transforms of individual linear and nonlinear terms in
governing equations using the DNS data. In all simulations, the
time-averages of the spectral quantities are calculated over an
entire time of the saturated turbulent state, discarding an initial
transient phase, which typically lasts for about $30S^{-1}$.

\subsection{Aspect ratio $(A_{xz}, A_{yz})=(1,1)$}

At the early stage of evolution, each harmonic contained in the
initial perturbations grows due to the flow nonnormality, leading to
the growth of the total energy and Reynolds stress. Then, after
about $t_s=30S^{-1}$, the amplitudes of the perturbations become
sufficiently large in the nonlinear regime and eventually the flow
settles into a self-sustained turbulence. In this cubic box, the
basic mode/harmonic with wavenumber ${\bf k}_b=(0,0,\pm 1)$,
dominates over other ones, i.e., its spectral energy, ${\cal
E}_k({\bf k}_b)$, and stress, ${\cal H}_k({\bf k}_b)$, are,
respectively, maximum of ${\cal E}_k$ and ${\cal H}_k$ during an
entire course of evolution (see also Ref. \cite{Pumir96}). Figure 2
shows the evolution of the volume-averaged total energy and stress
as well as the energy and stress of the basic mode in the saturated
state, which display similar patterns of quasi-periodic bursts in
time. The volume-averaged Reynolds stress (with negative sign),
$-\langle u_xu_y\rangle$, and the stress of the basic mode, ${\cal
H}_k({\bf k}_b)$ are positive at all times and non-decaying as
required by the self-sustaining process according to Eq. (8). The
bursts in the total energy and stress closely follow amplifications
(peaks) of ${\cal E}_k({\bf k}_b)$ and stress ${\cal H}_k({\bf
k}_b)$ in the basic mode. This indicates that these bursts are
related to the dynamics of the basic mode, which thus appears to be
mainly responsible for the energy pumping from the background flow
into turbulence. During the bursts, the contribution of ${\cal
H}_k({\bf k}_b)$ to the total Reynolds stress is as much as 60\% -
70\%, while other modes become significant, although still remain
smaller than the basic mode, in quiescent intervals between the
bursts, when the total energy and stresses as well as the energy and
stress of the basic mode are low. So, for the aspect ratio $(1,1)$,
most of the turbulent energy is produced through the interaction of
the basic mode with the background shear flow.

\begin{figure*}
\includegraphics[width=5.9cm]{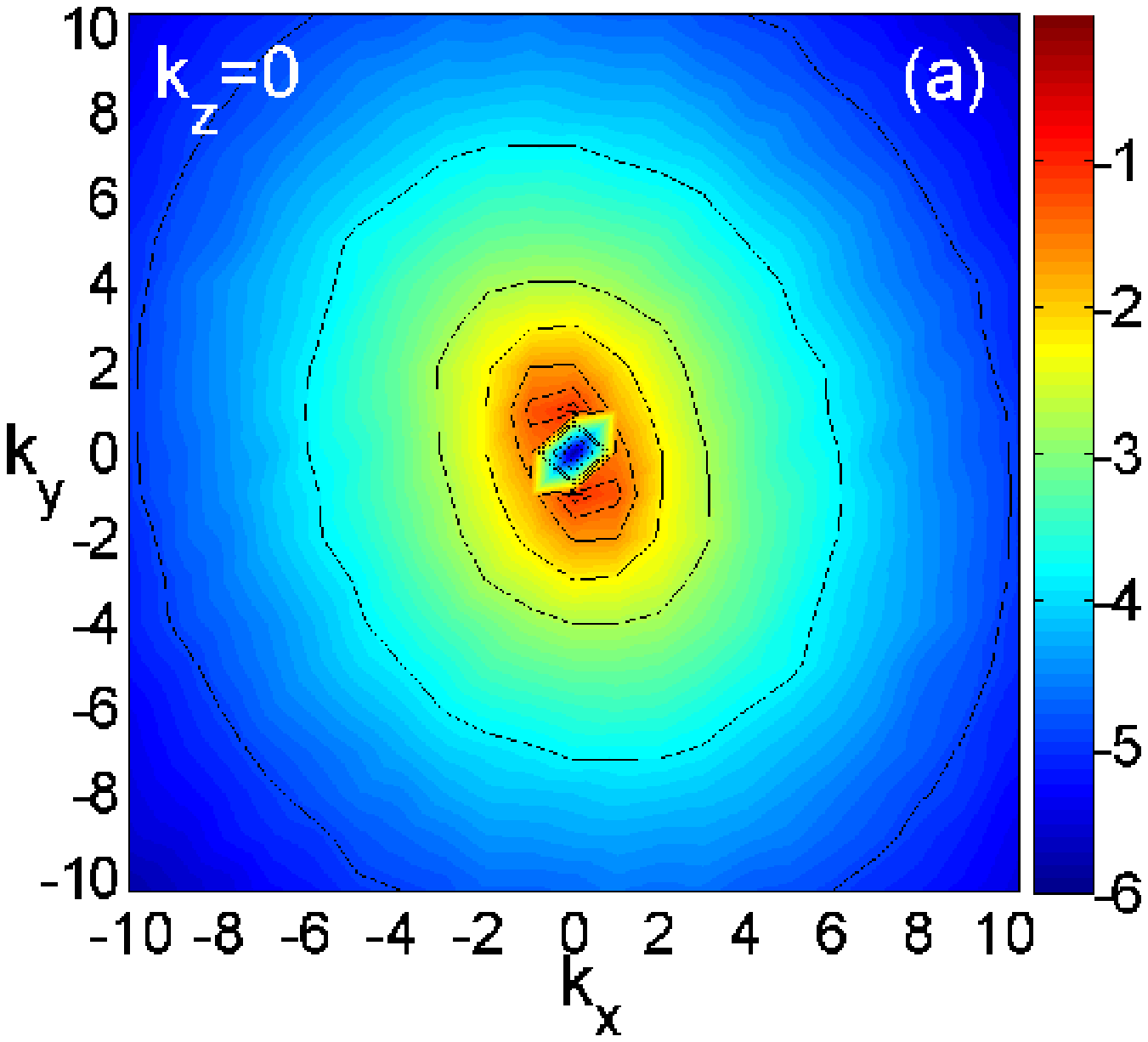}
\includegraphics[width=5.9cm]{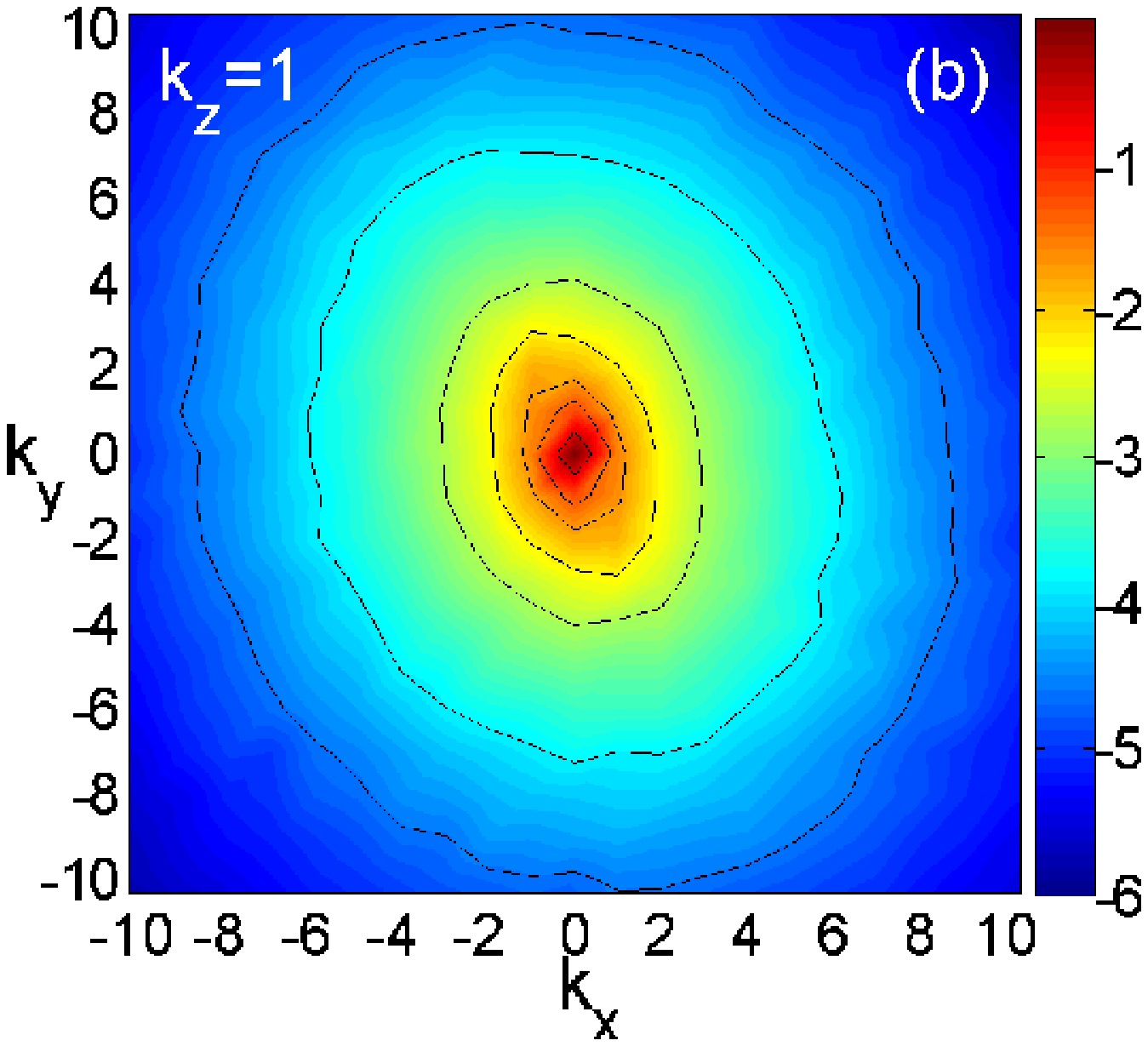}
\includegraphics[width=5.9cm]{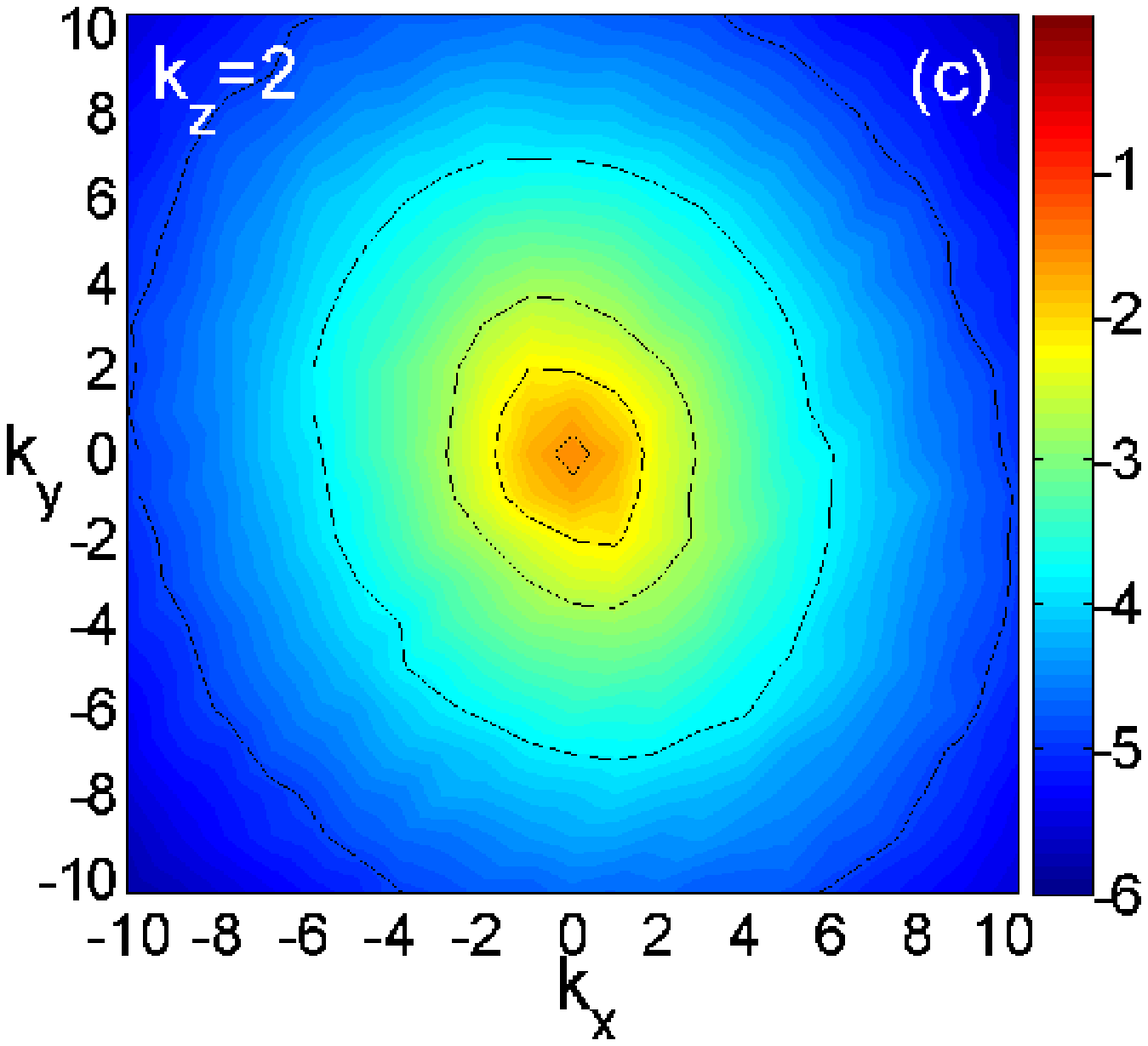}
\caption{(Color online) Logarithm of the time-averaged spectrum of
the kinetic energy, $log_{10}({\cal E}_k)$, in the cubic box. The
$(k_x,k_y)$-slices are at (a) $k_z=0$, (b) $k_z=1$ and (c) $k_z=2$.
The spectrum has an anisotropic character, having larger power in
the $k_y/k_x<0$ part of Fourier space at a given $k_x$. The spectral
energy has maximum at $k_z=1$ and decreases with increasing $k_z$.}
\end{figure*}
\begin{figure}
\includegraphics[width=\columnwidth]{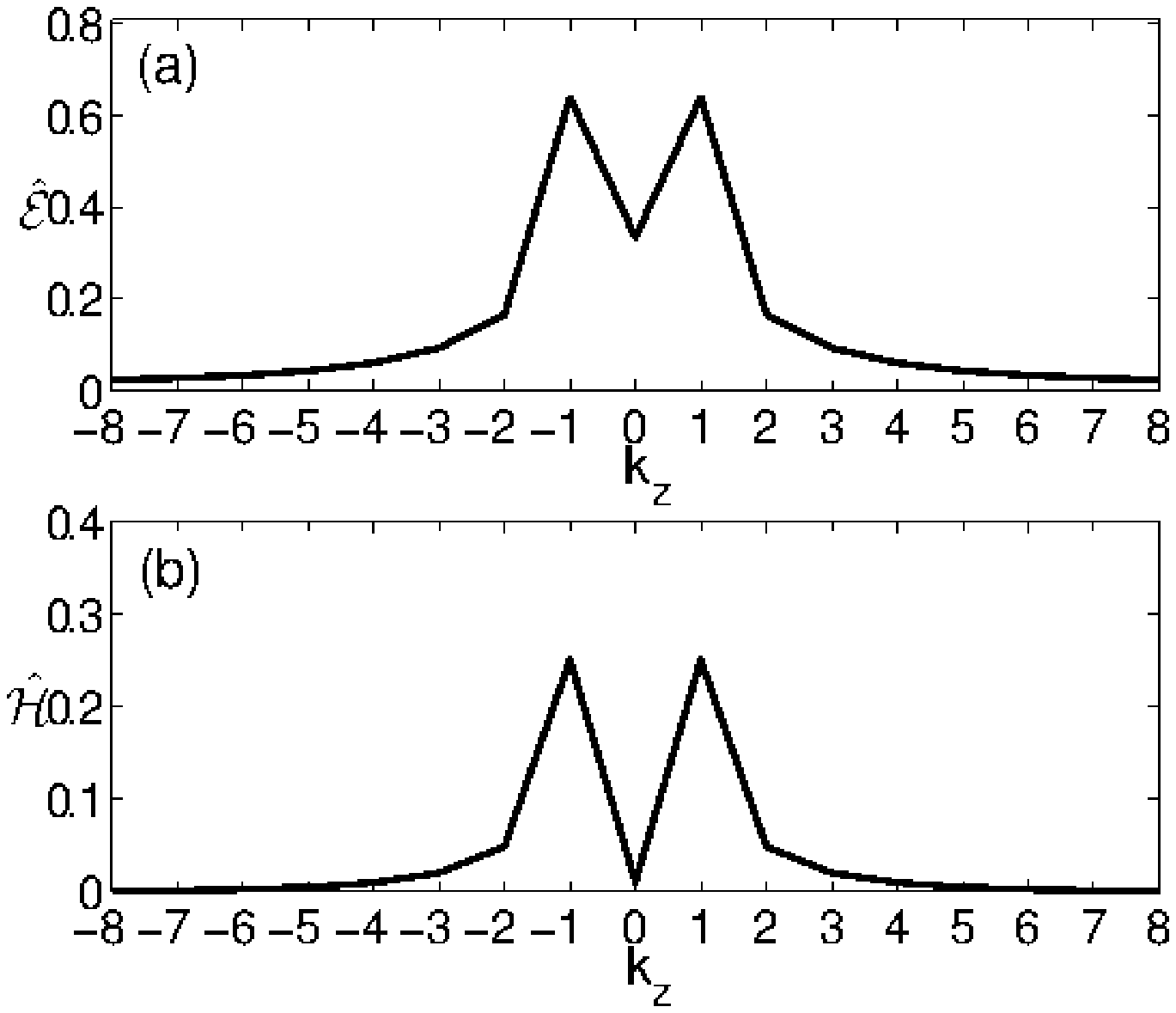}
\caption{Integrated in $(k_x,k_y)$-plane time-averaged (a) kinetic
energy spectrum, $\hat{\cal E}$, and (b) Reynolds stress, $\hat{\cal
H}$ vs. $k_z$.}
\end{figure}

The structure of the velocity components in the fully developed
turbulence in physical space is presented in Fig. 3. The evolution
of the volume-averaged perturbation kinetic energy and quadratic
forms of streamwise, $\langle u_x^2\rangle/2$, shearwise/vertical,
$\langle u_y^2\rangle/2$, and spanwise, $\langle u_z^2\rangle/2$,
velocities as well as the Reynolds stress are shown in Fig. 4. From
these plots it is clear that the streamwise velocity prevails over
the shearwise and spanwise ones. In the $(x,z)$ slices of $u_y$ and,
especially, of $u_x$, we clearly see the signatures of the dominant
basic harmonic with small scale fluctuations due to other harmonics
superimposed on it.

To understand the role of other modes in the turbulence dynamics, it
is important to define the main, energy-containing area in {\bf
k}-space, which we also refer to as \emph{the vital area} of
turbulence. For this purpose we define two sets of harmonics whose
energy grows from 5\% to 10\% and those larger than 10\% of the
maximum spectral energy at least once during the evolution; the
second set is referred to as ``active modes''. These
energy-carrying, or dynamically important harmonics are displayed in
Fig. 5. They have small wavenumbers $|k_x|\leq 2, |k_y|\leq 3,
|k_z|\leq 3$, which basically form the vital area of turbulence (at
$k_z=3$, not shown in this figure, mode energies always remain less
than 10\% of the maximum energy). Harmonics with larger wavenumbers
$|k_x|>2, |k_y|>3, |k_z|>3$ lie outside the vital area and always
have energy and stress less than 5\% of the maximum value,
therefore, not playing a major role in the energy exchange process
between the background flow and turbulence. Essentially, only these
large scale modes in the vital area take part in the self-sustaining
dynamics of turbulence, but the basic mode/harmonic still remains
dominant -- it is the most active among the active harmonics. We
emphasize that the total number of the large scale active modes
(black dots in Fig. 5) is equal to 36, implying that the dynamics of
homogeneous shear turbulence cannot be reduced to low-order models
of the self-sustaining processes.

Figure 6 shows the time-averaged spectrum of the kinetic energy in
$(k_x,k_y)$-plane at different $k_z$. The spectrum is anisotropic
due to shear -- the isolines have elliptical shape inclined opposite
to the $k_y$-axis. This fact indicates that on average modes with
$k_y/k_x<0$ have more energy than those with $k_y/k_x>0$ at fixed
$k_x$. The plots show that the time-averaged spectral kinetic energy
at fixed $k_z$ is larger for small $k_x$ and $k_y$ and decays at
least by two order of magnitude at $\sqrt {k_x^2+k_y^2} \approx 2$.
So, the active modes are located just within this range of small
wavenumbers, as also seen in Fig. 5. As mentioned above, the maximum
of the spectral energy is reached at $k_x=k_y=0, k_z=\pm 1$
corresponding to the basic mode. However, it decays significantly
already for the next spanwise modes with $k_z=2$ and further
decreases rapidly with increasing $k_z$. This behavior is also seen
in Fig. 7, which shows the integrated in $(k_x,k_y)$-plane the
time-averaged spectral energy, $\hat{\cal E}(k_z)=\int {\cal E}_k
dk_xdk_y$, and stress, $\hat{\cal H}(k_z)=\int {\cal H}_k dk_xdk_y$,
vs. $k_z$. Both reach a maximum at $k_z=\pm 1$ and rapidly decrease
with $|k_z|$.

The anisotropic nature of the kinetic energy spectrum, which clearly
distinguishes it from the isotropic spectrum in the classical
(without shear) Kolmogorov phenomenology, arises as a result of the
specific action of linear and nonlinear processes in {\bf k}-space.
As we show below, these processes are anisotropic over wavenumbers
due to shear, resulting in a new phenomenon -- the transverse
redistribution of power in spectral space.

\begin{figure*}
\includegraphics[width=0.32\textwidth]{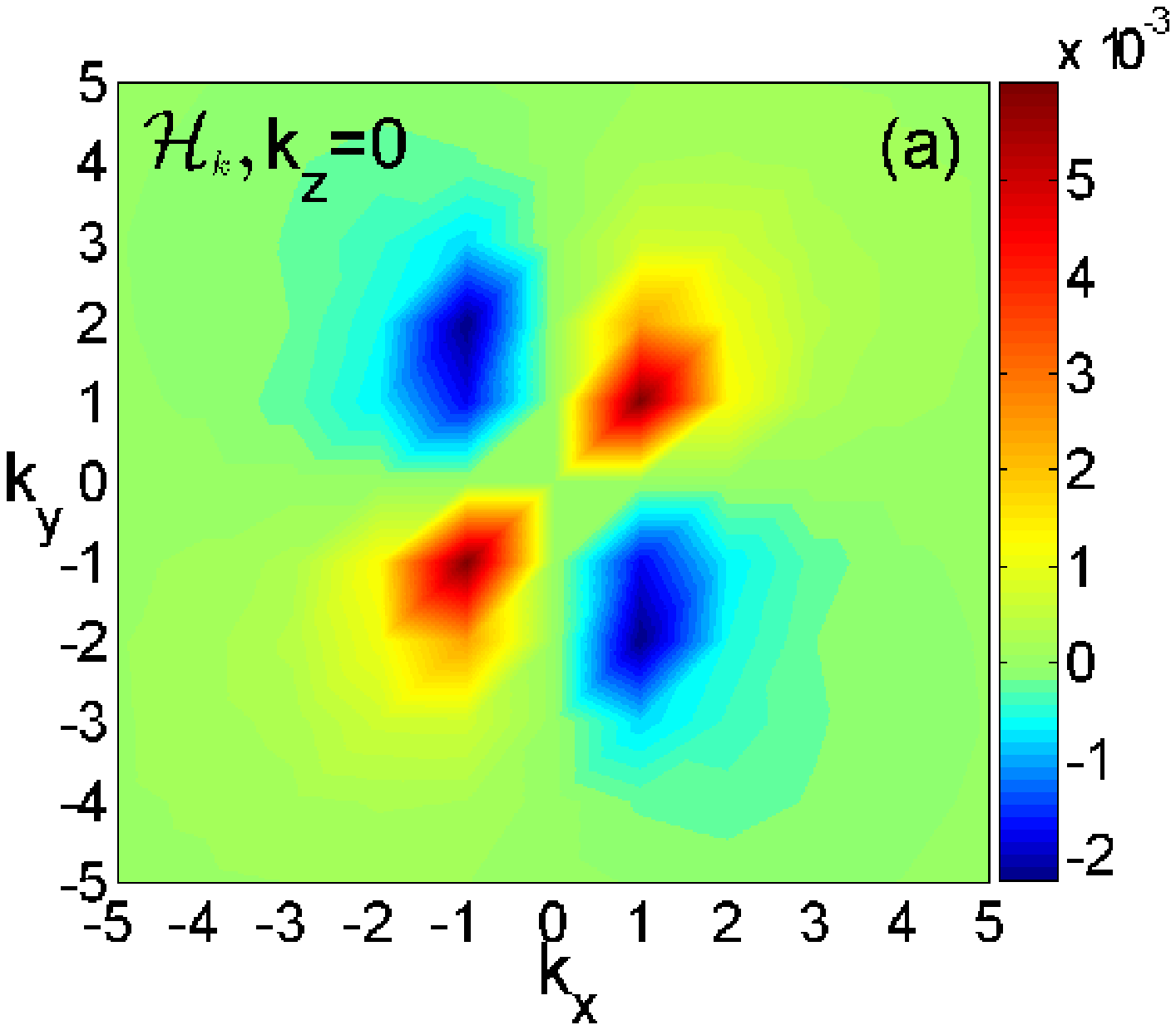}
\includegraphics[width=0.32\textwidth]{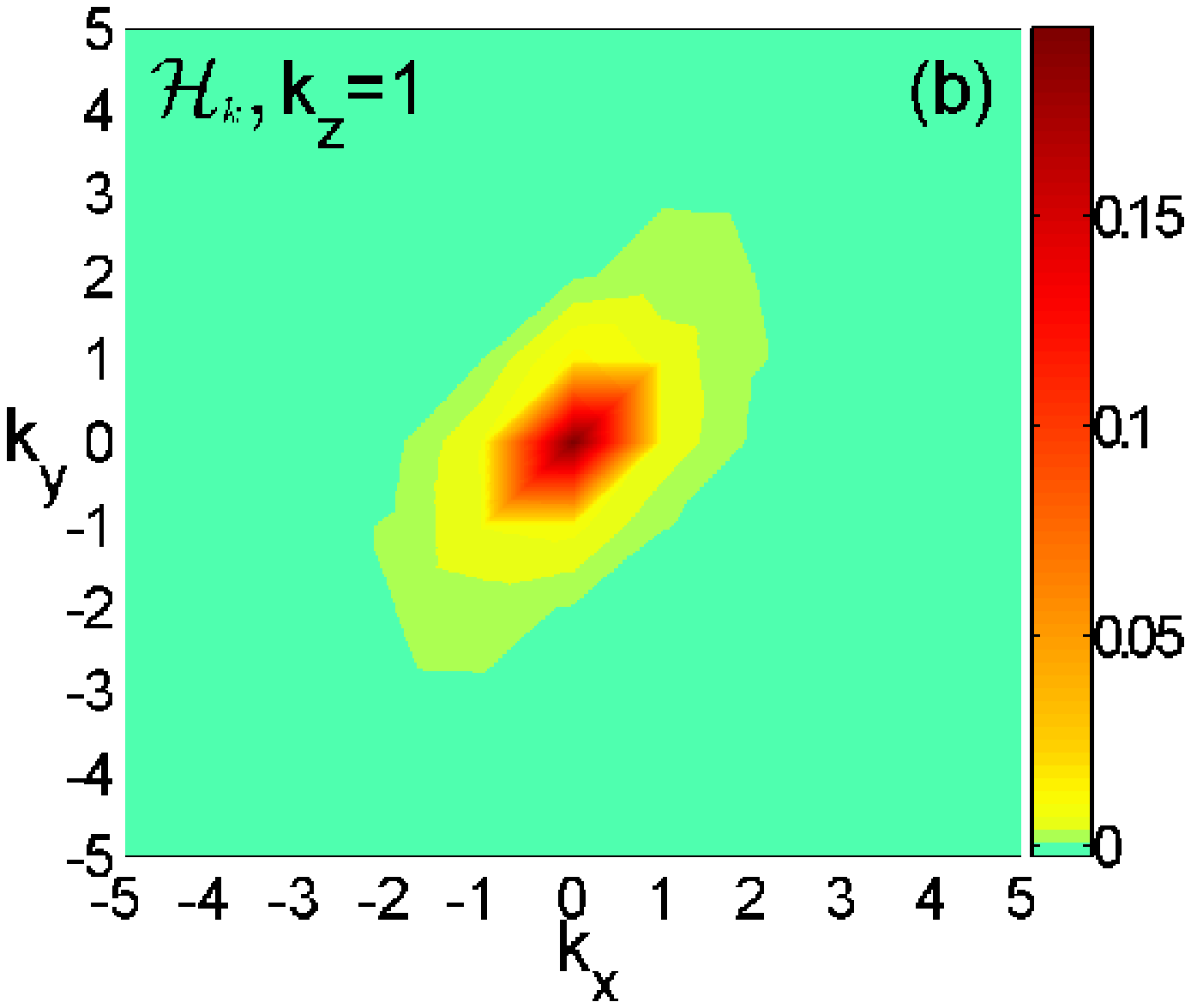}
\includegraphics[width=0.32\textwidth]{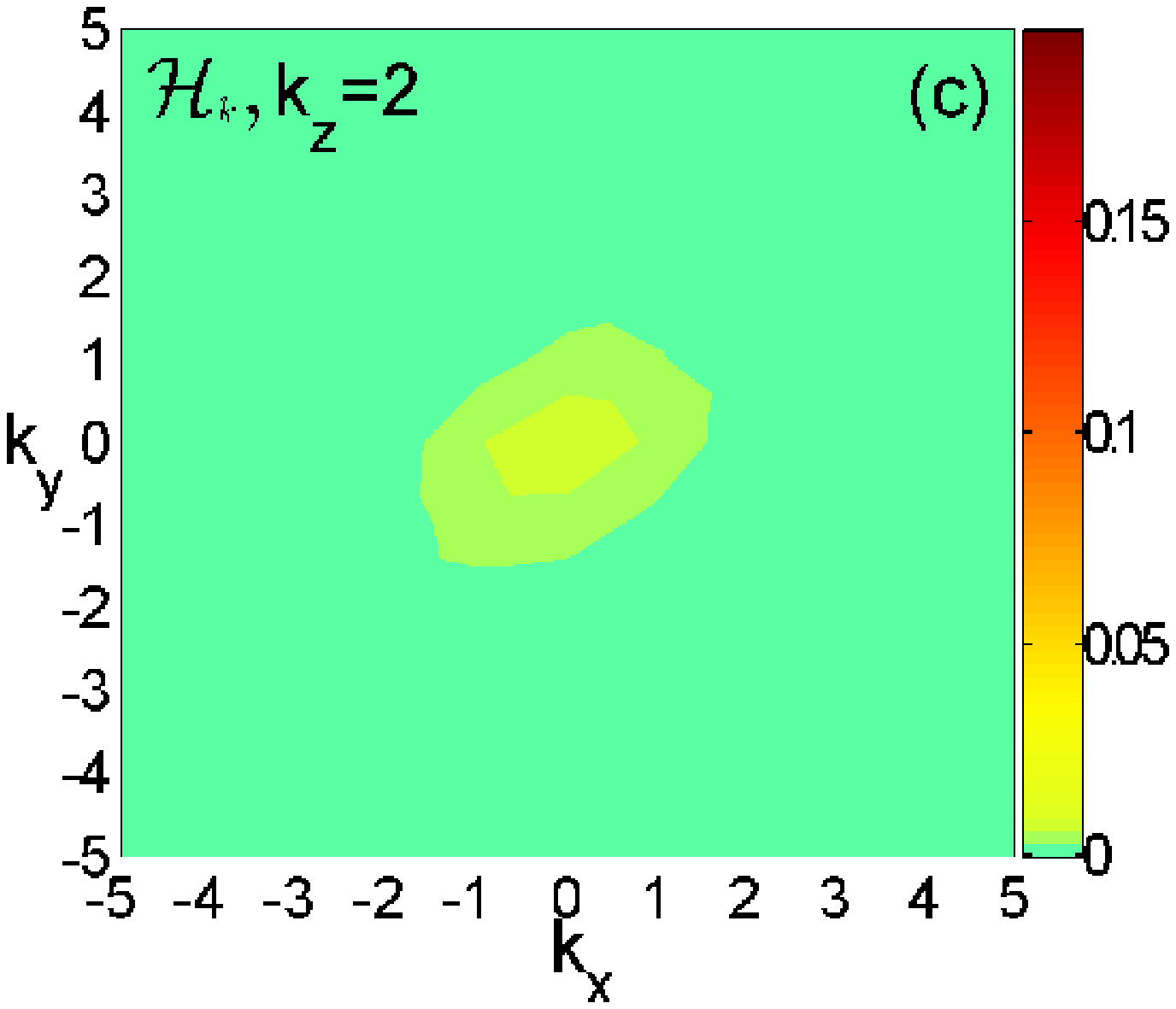}
\includegraphics[width=0.32\textwidth]{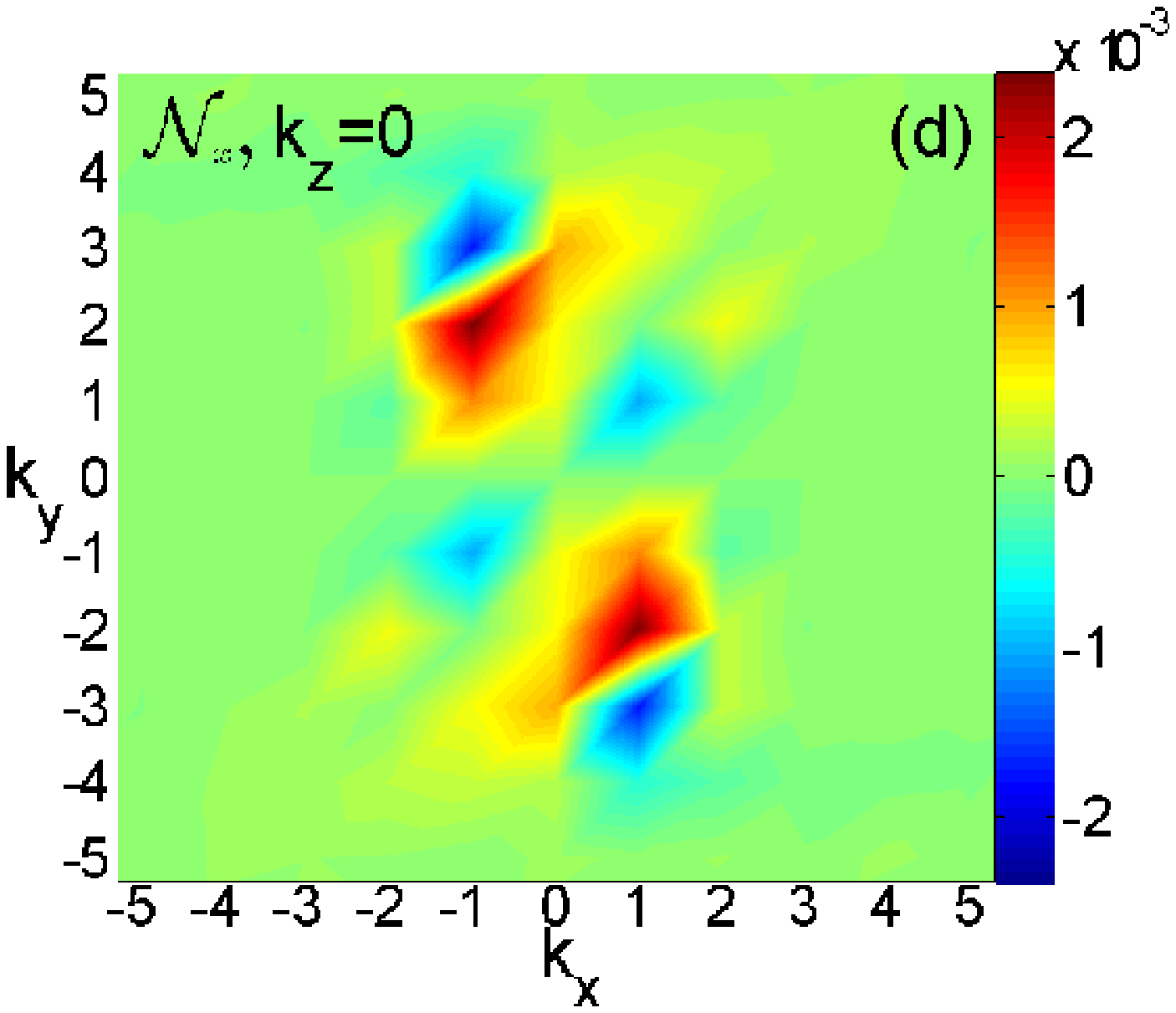}
\includegraphics[width=0.32\textwidth]{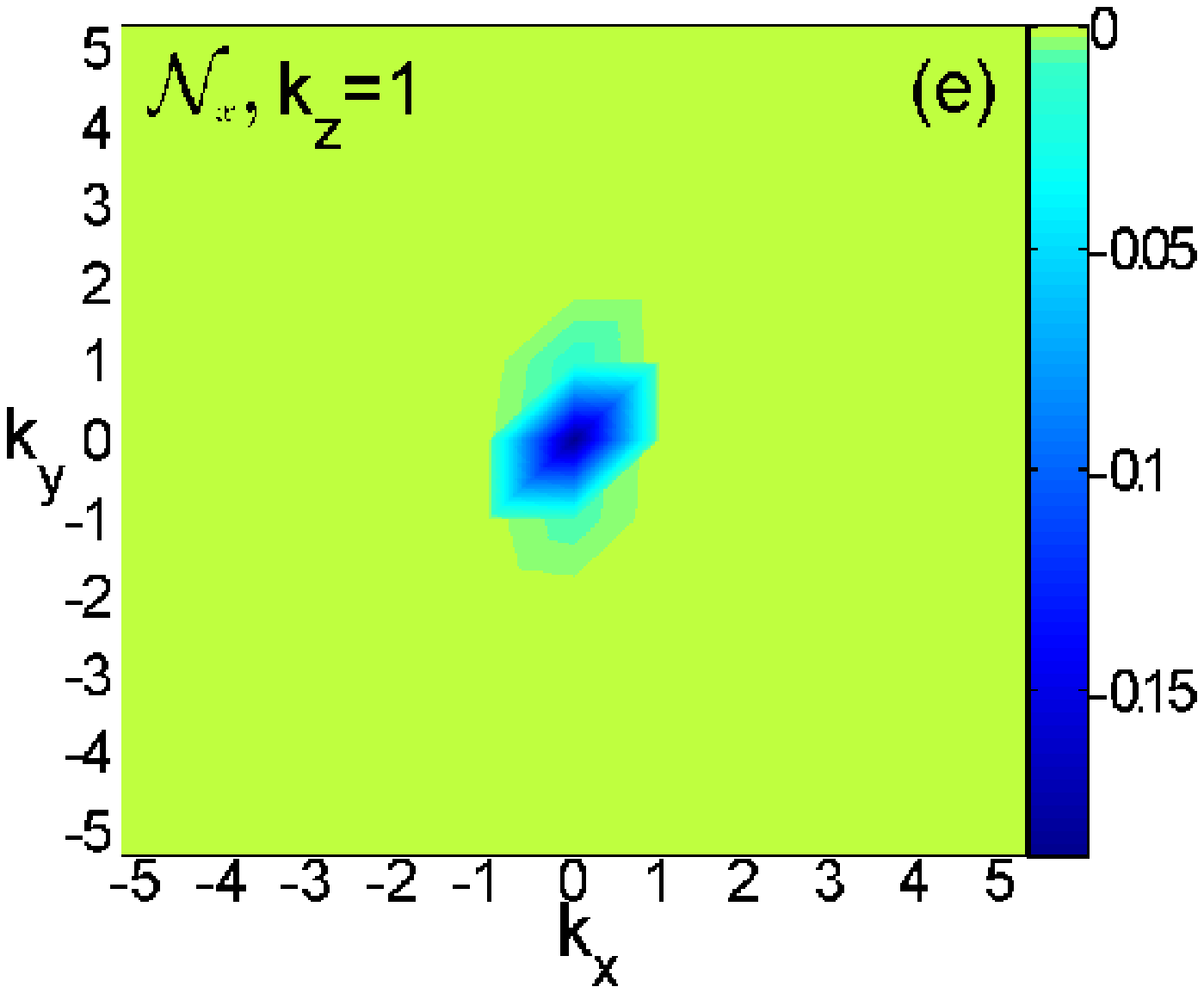}
\includegraphics[width=0.32\textwidth]{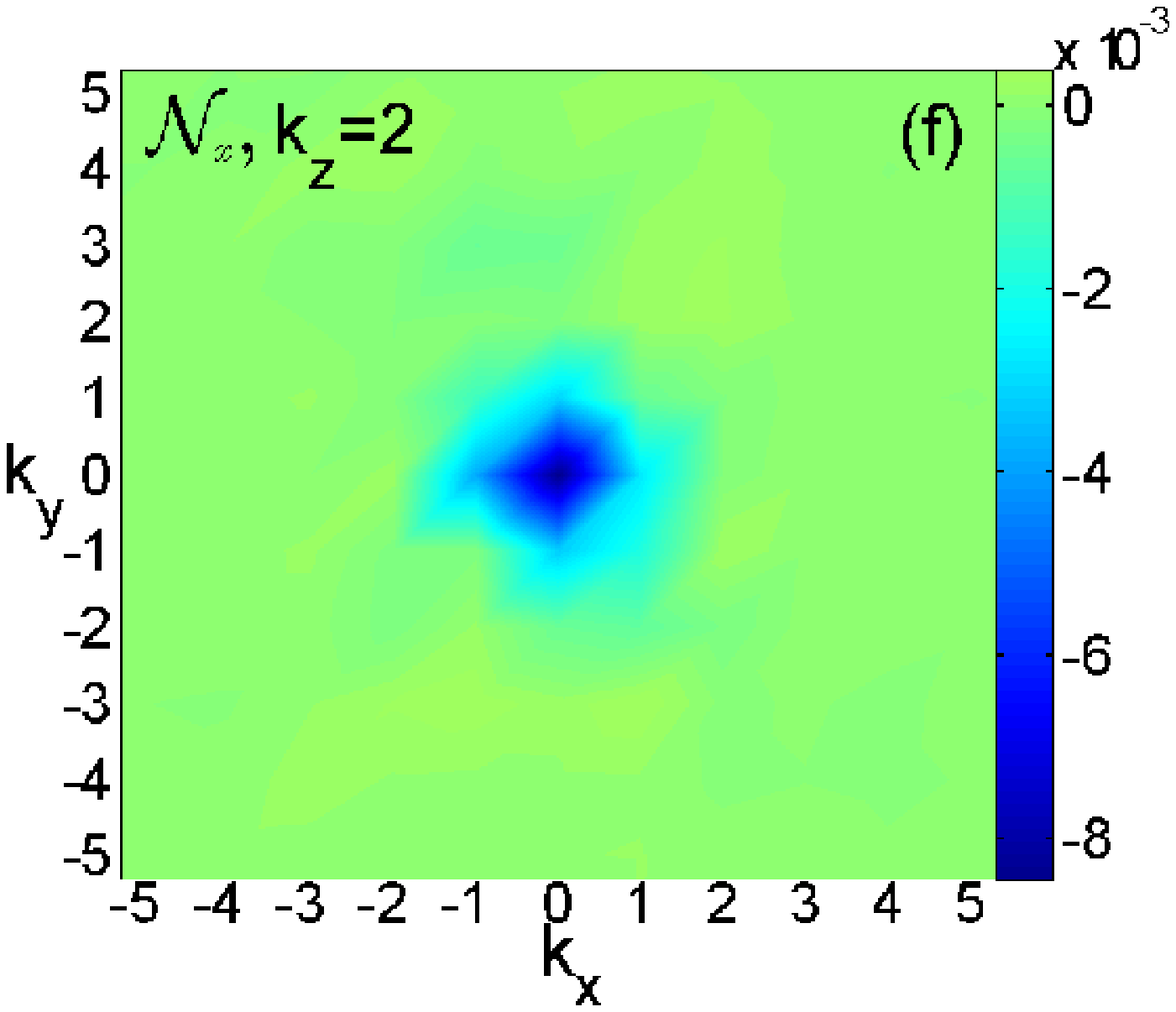}
\includegraphics[width=0.32\textwidth]{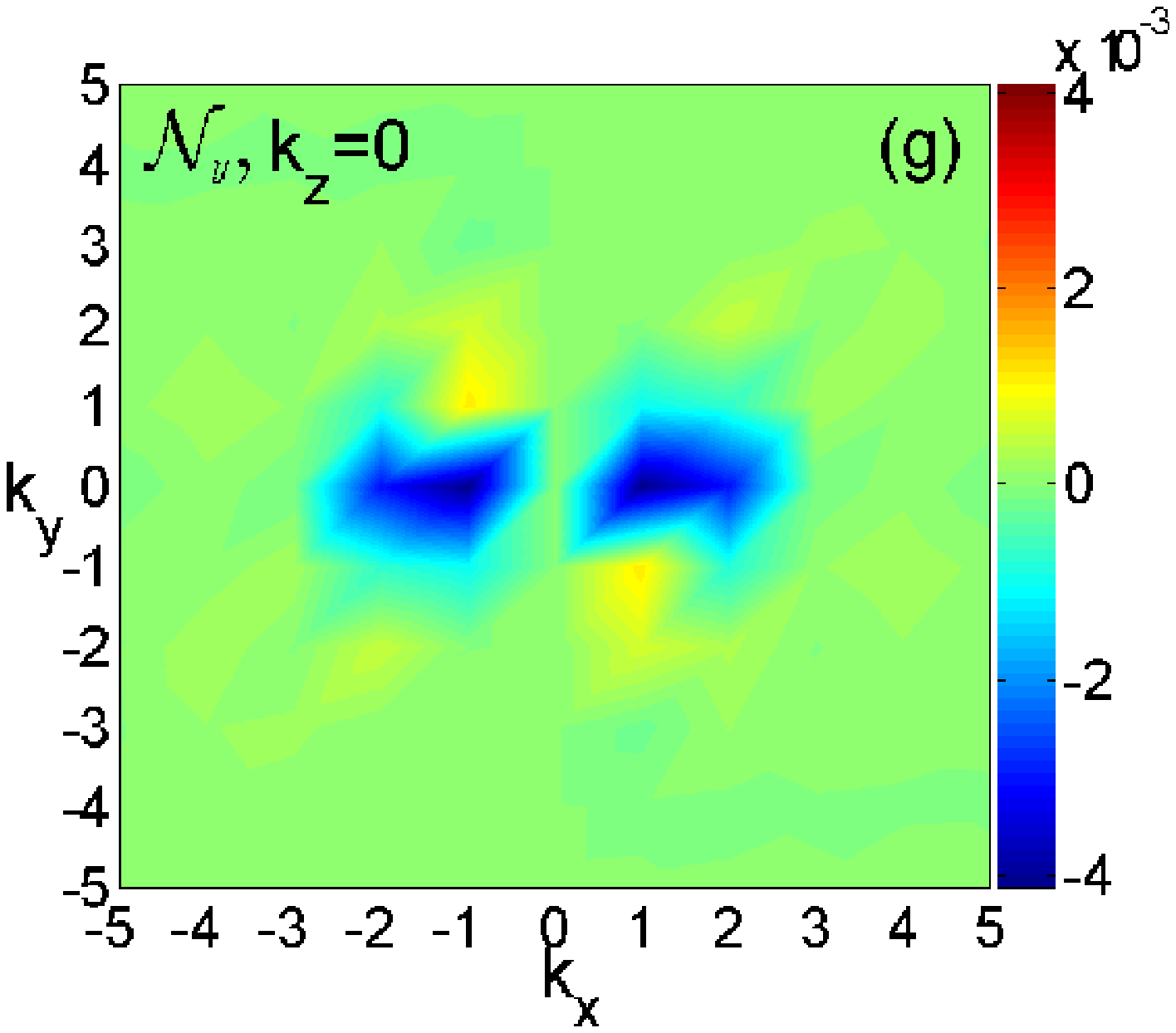}
\includegraphics[width=0.32\textwidth]{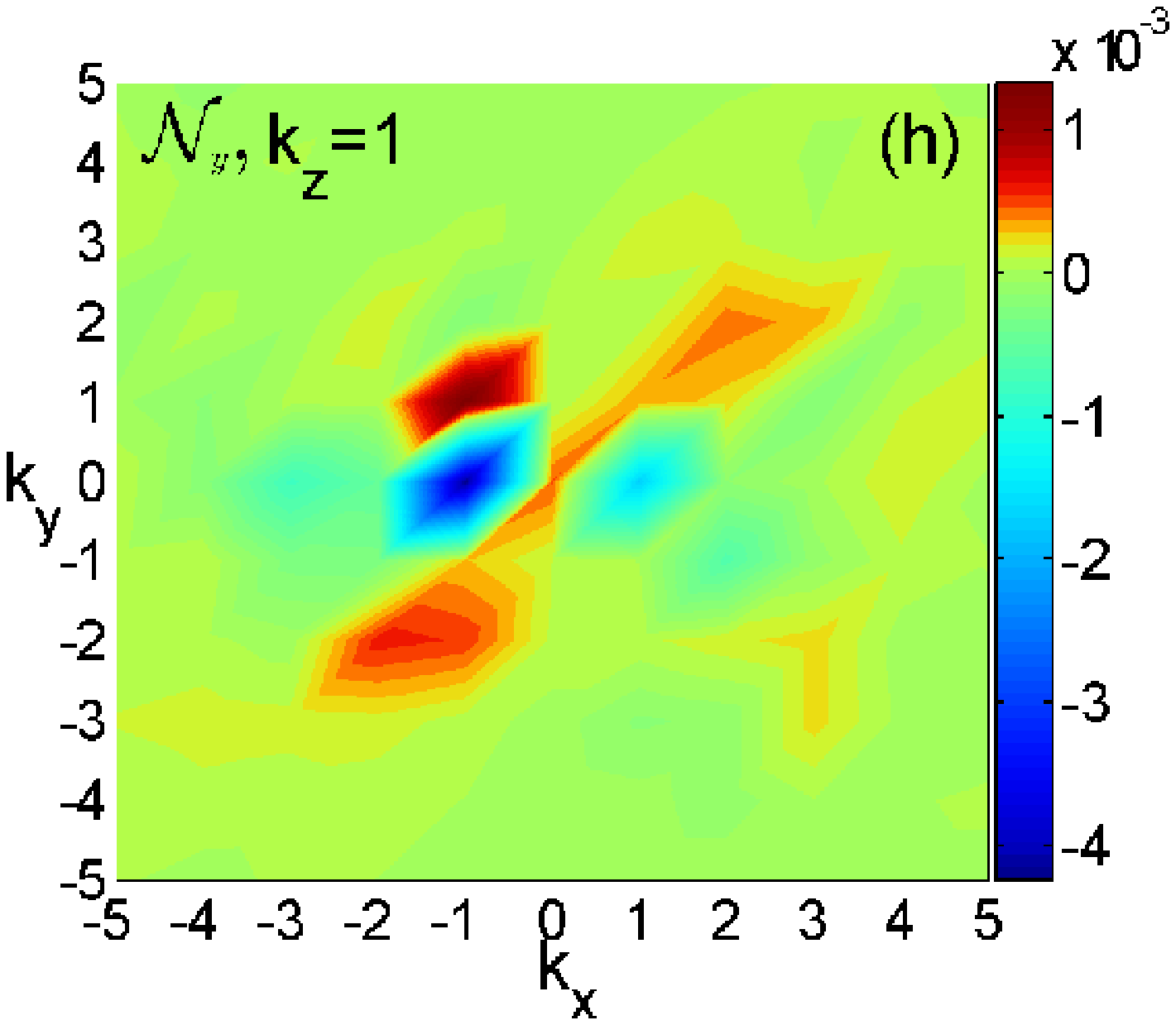}
\includegraphics[width=0.32\textwidth]{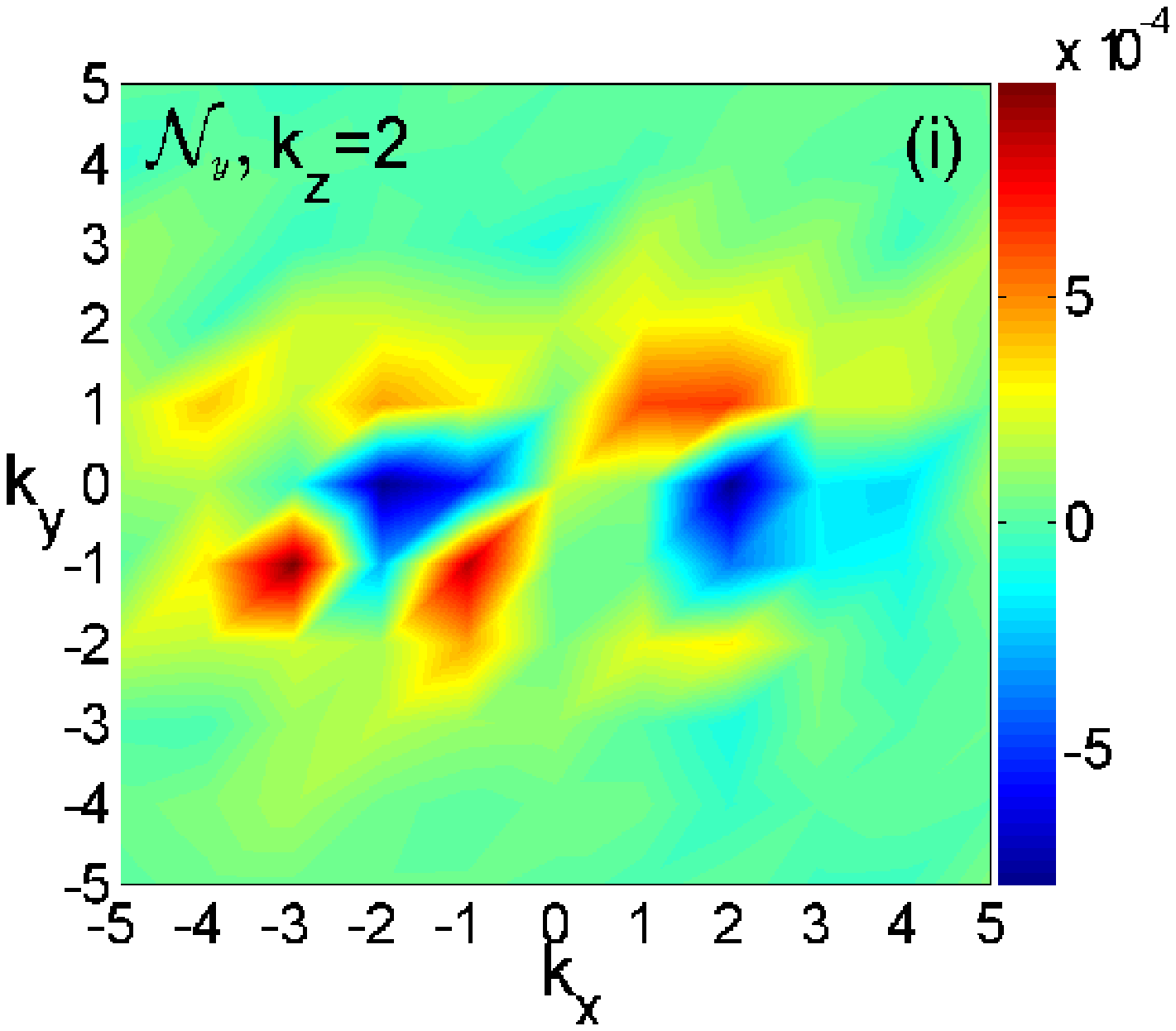}
\includegraphics[width=0.32\textwidth]{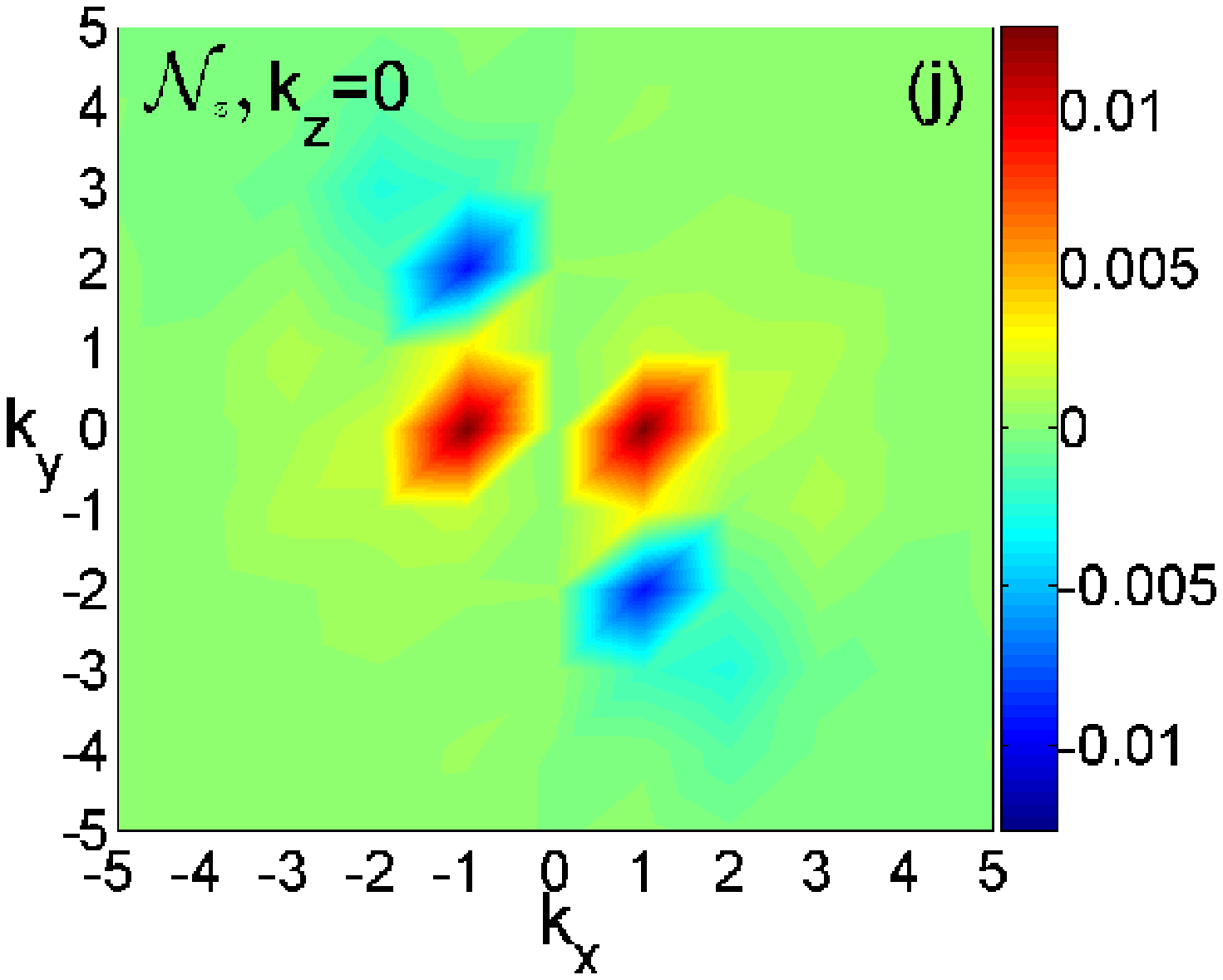}
\includegraphics[width=0.32\textwidth]{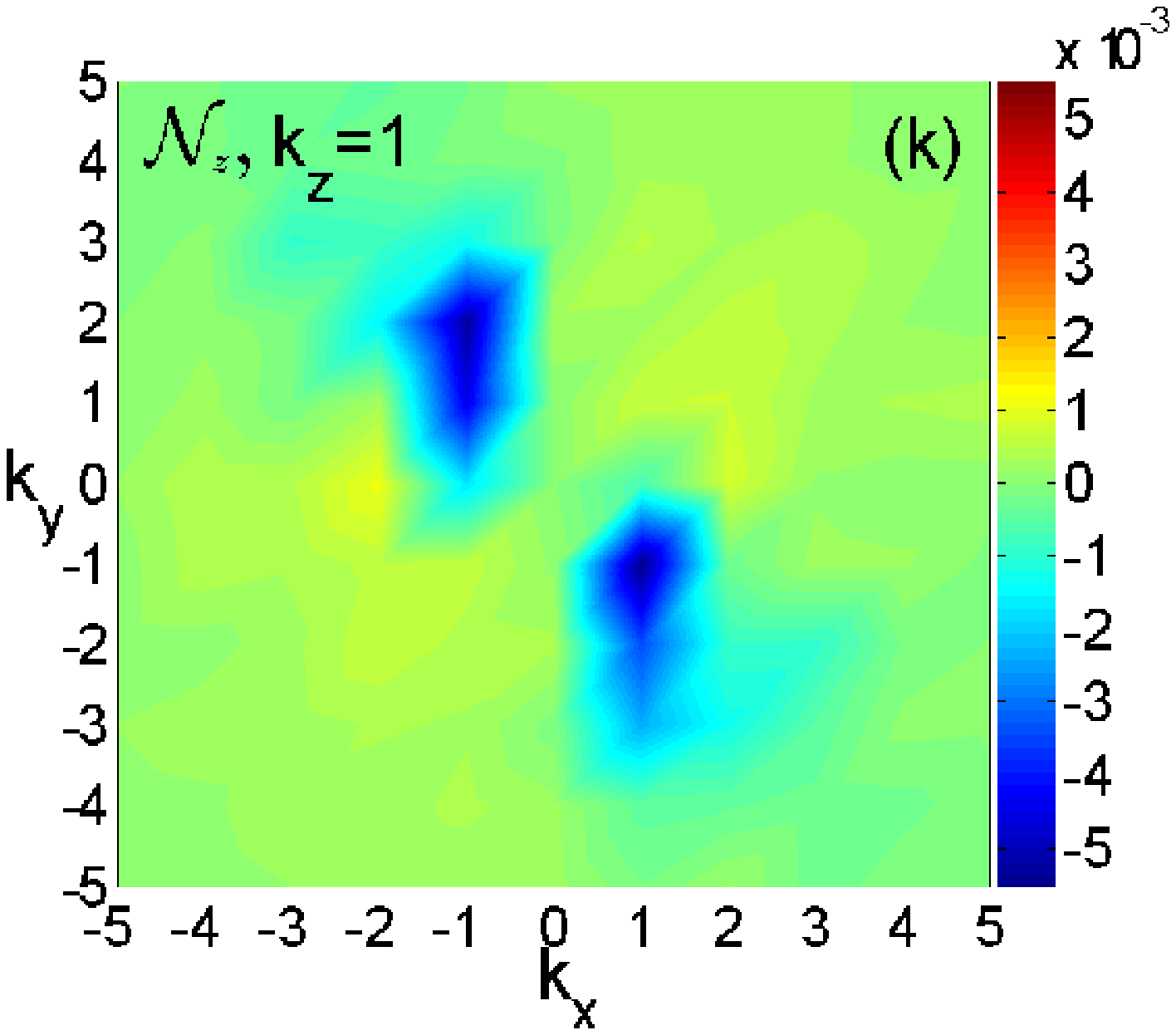}
\includegraphics[width=0.32\textwidth]{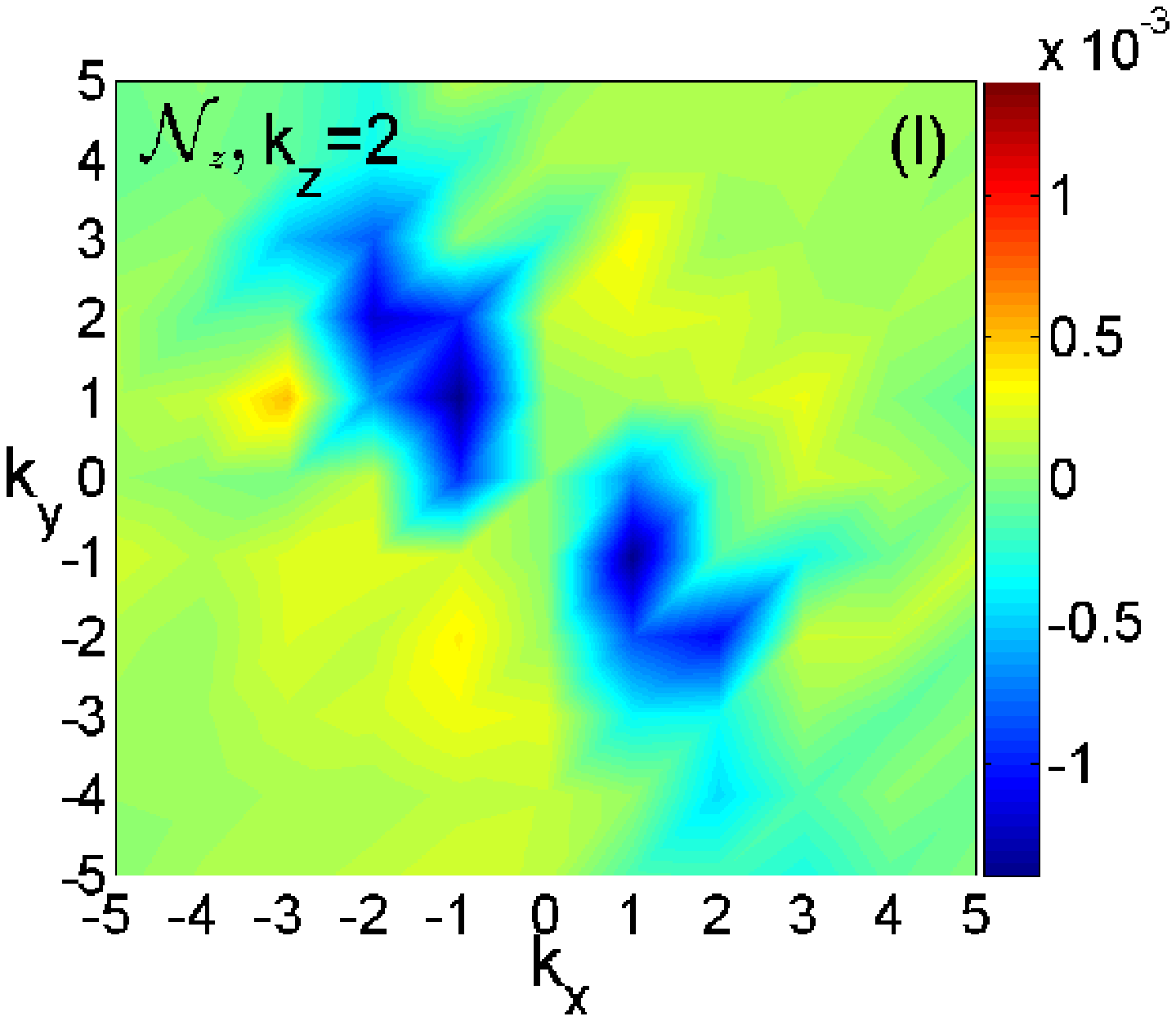}
\caption{(Color online) Maps of the time-averaged energy injecting
stress [(a), (b), (c)], ${\cal H}_k$, and nonlinear transfer, [(d),
(e), (f)] ${\cal N}_x$, [(g), (h), (i)] ${\cal N}_y$, [(j), (k),
(l)] ${\cal N}_z$ functions in {\bf k}-space for the cubic box.
Shown are $(k_x,k_y)$-slices of these quantities at $k_z=0$ (left
column), $k_z=1$ (middle column) and $k_z=2$ (right column). ${\cal
H}_k$ is significant in the vital area $|k_x|\leq 2, |k_y|\leq 3,
|k_z|\leq 2$, on the $k_y/k_x>0$ side. Hence, the energy injection
into turbulence mainly occurs in this region. At $k_z=1$, ${\cal
H}_k$ reaches largest values and consequently energy injection is
most intensive at this wavenumber. ${\cal N}_x$, ${\cal N}_y$ and
${\cal N}_z$ transfer, respectively, the streamwise, shearwise and
spanwise components of the spectral kinetic energy anisotropically,
or transversely over wavevector angles in Fourier space, away from
the regions where they are negative ${\cal N}_x<0,~{\cal
N}_y<0,~{\cal N}_z<0$ (blue) towards the regions where they are
positive ${\cal N}_x>0,~{\cal N}_y>0, ~{\cal N}_z>0$ (yellow and
red).}
\end{figure*}
\begin{figure}
\includegraphics[width=\columnwidth]{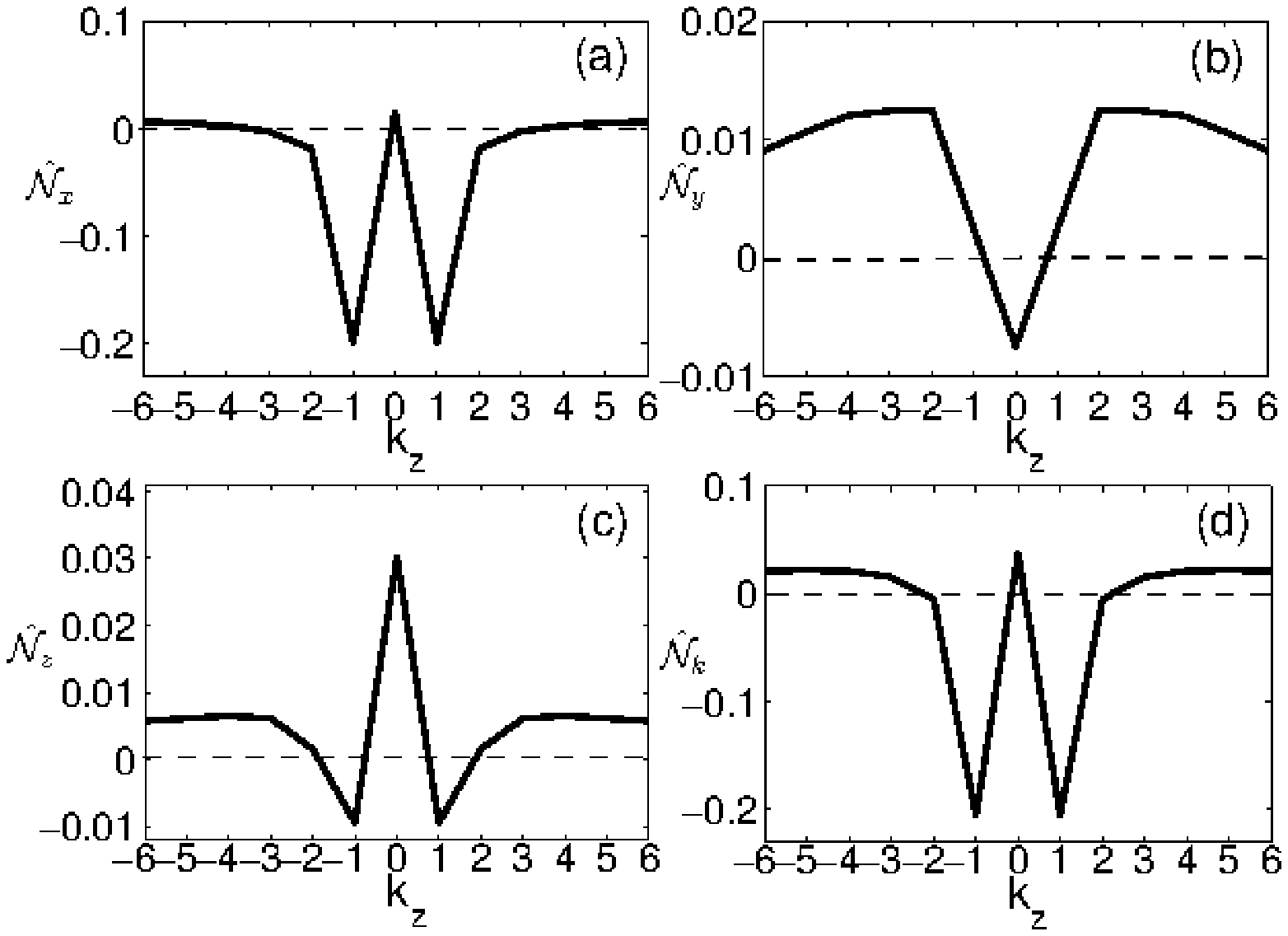}
\caption{The time-averaged nonlinear transfer functions integrated
in $(k_x,k_y)-$plane and represented as a function of $k_z$, (a)
$\hat{\cal N}_x$, (b) $\hat{\cal N}_y$, (c) $\hat{\cal N}_z$ and
their sum (d) $\hat{\cal N}_k$ describing nonlinear transfer of
energy. These functions transfer corresponding quantities from $k_z$
at which they are negative to those $k_z$ at which they are
positive.}
\end{figure}

\subsubsection*{Energy injection, nonlinear transfers and their
interplay in ${\bf k}$-space}

To understand the origin of the anisotropic character of the energy
spectrum, in Fig. 8 we present the distribution of the time-averaged
energy-injecting Reynolds stress spectrum, ${\cal H}_k$, and the
nonlinear transfer functions, ${\cal N}_x, {\cal N}_y$ and ${\cal
N}_z$, in {\bf k}-space. Since these quantities are symmetric with
respect to a change ${\bf k}\rightarrow -{\bf k}$, without loss of
generality, hereafter we concentrate on the upper part ($k_z \geq
0$) of {\bf k}-space. First of all, these plots clearly demonstrate
the anisotropic nature of the linear and nonlinear processes in {\bf
k}-space -- all these quantities exhibit anisotropy over
wavenumbers, that is, depend on wavevector angle, or orientation.
${\cal H}_k$ is significant primarily in the vital area of spectral
space, $|k_x|\leq 2, |k_y|\leq 3, |k_z|\leq 2$, at $k_y/k_x>0$,
where it supplies the turbulence with energy via nonmodal growth of
the dynamically active modes (see below). At $k_z=0$, ${\cal H}_k$
is mainly concentrated around $|k_y|,|k_x| \sim 1$ [Fig. 8(a)], it
is positive at $k_x/k_y>0$, supplying harmonics with energy and
negative at $k_x/k_y<0$, returning energy to the flow. On the other
hand, at $k_z\geq 1$, the distribution of ${\cal H}_k$ is changed --
it is positive and achieves local maximum around $k_x=k_y=0$ at each
$k_z$ [Figs. 8(b) and 8(c)]. The Reynolds stress spectrum is highest
and therefore the energy extraction from the background flow into
turbulence more effectively occurs for modes with $k_{z}=\pm 1$,
reaching the maximum at ${\bf k}_b$ corresponding to the basic mode
[Figs. 8(b), see also Fig. 7(b)]. Outside the vital area, at $|k_x|>
2, |k_y|>3, |k_z|>2$, the stress and hence the energy input into
turbulence rapidly decrease and vanish at large wavenumbers. The
total energy input at $k_z=0$ is also negligible, $\hat{\cal
H}\approx 0$, as seen from Fig. 7(b), since ${\cal H}_k$ changes
sign in $(k_x,k_y)$-plane at this $k_z$ [Fig. 8(a)]. \emph{\it Thus,
the main energy supply for turbulence is due to large-scale modes
with the wavelengths comparable to the box size.} Note that the
energy injection into turbulence mediated by the stress occurs over
a range wavenumbers and in this respect differs from the classical
cases of forced turbulence in flows without shear, where forcing is
applied at a specific wavenumber or a narrow range of wavenumbers
(see e.g., \cite{Biskamp03}).

The nonlinear transfer functions ${\cal N}_x, {\cal N}_y, {\cal
N}_z$ [Figs. 8(d)-8(l)] depend on wavevector angle and,
consequently, redistribute the streamwise, shearwise and spanwise
velocity energies generally not only along wavevector
(direct/inverse cascades), but also transverse to wavevector. The
latter occurs away from the regions in {\bf k}-space where the
respective transfer functions are negative ${\cal N}_x,{\cal
N}_y,{\cal N}_z<0$ [blue areas in Figs. 8(d)-8(l)] to regions where
they are positive ${\cal N}_x,{\cal N}_y,{\cal N}_z>0$ (yellow and
red areas); at all other wavenumbers (light green areas) these
functions are small. It is seen from Figs. 8(d)-8(l) that this
transverse cascade primarily operates within the vital area of the
wavenumber space, $|k_x|\leq 2, |k_y|\leq 3, |k_z|\leq 2$, where the
nonnormality-induced energy exchange between the background flow and
harmonics is dominant. Outside the vital area at large wavenumbers,
the transfer functions become nearly independent of the wavevector
angle and depend mostly on its magnitude, i.e., the cascade is not
transverse anymore but direct. So, outside the vital area the
situation is similar to the classical homogeneous isotropic
turbulence described by Kolmogorov phenomenology. To characterize
transfers along $k_z$, we integrate ${\cal N}_i$ over
$(k_x,k_y)$-plane, $\hat{\cal N}_i(k_z)=\int {\cal N}_i dk_xdk_y,
(i=x,y,z,k)$, and represent them as functions of $k_z$ in Fig. 9.

The interplay between the energy-injecting linear process and the
transverse cascade determines the self-sustaining dynamics of
turbulence and characteristics of the energy spectrum in the
presence of shear. Figures 8(d)-8(f) and 9(a) show that ${\cal N}_x$
is negative and large by absolute value in the vital area, where the
stress, ${\cal H}_k$, is positive and appreciable. Outside the vital
area, i.e., at larger wavenumbers ($|k_x|>2, |k_y|>3, |k_z| \geq
3$), ${\cal N}_x$ is positive but small and the stress is also
negligibly small. In other words, ${\cal N}_x$ provides a sink, or
negative feedback for ${\cal H}_k$ by taking out energy in the
streamwise velocity gained from the background flow and transferring
it to large wavenumbers and part of it to $k_z=0$, as evident from
the dependence of ${\cal {\hat N}}_x$ on $k_z$. By contrast, ${\cal
N}_y$ is positive in those regions of the vital area where ${\cal
H}_k$ is also positive [Figs. 8(g)-8(i) and 9(b)]. As a result, it
provides positive feedback at these wavenumbers by generating the
shearwise energy, $|\bar{u}_y|^2/2$, which, in turn, leads to the
growth of the streamwise energy, $|\bar{u}_x|^2/2$, due to the
linear nonmodal growth mechanism (the first terms of linear origin
on the rhs of Eqs. 13 and 16, while ${\cal N}_x<0$ at these
wavenumbers and cannot contribute to the growth). The streamwise
energy, which in fact appears to be largest among the other two
components, is then transferred to larger wavenumbers by ${\cal
N}_x$. Note that this positive feedback due to ${\cal N}_y$ is
crucial to the self-sustenance of turbulence and is realized by the
transverse cascade, which ensures supply of the shearwise energy for
harmonics with such wavevector orientations ($k_y/k_x>0$) that
correspond to positive stress ${\cal H}_k>0$ and hence with the
capability of nonmodal growth. Figures 8(j)-8(l) and 9(c) show
${\cal N}_z$, which is responsible for the transfer of energy in the
spanwise velocity. However, since the Reynolds stress providing
energy supply for the turbulence is produced by the product
(correlation) of $u_x$ and $u_y$, below we concentrate only on
${\cal N}_x$ and ${\cal N}_y$ that govern nonlinear redistribution
of these velocity components. Figure 9(d) shows the nonlinear
transfer function for the energy, $\hat{\cal N}_k=\hat{\cal
N}_x+\hat{\cal N}_y+\hat{\cal N}_z$. It is dominated by $\hat{\cal
N}_x$ and hence the nonlinear transfer of the total energy over
wavenumbers is similar to that of the streamwise energy.

\emph{\it Thus, the transverse cascade of energy appears to be a
generic feature of nonlinear dynamics of perturbations in spectrally
stable shear flows.} The conventional description of shear flow
turbulence solely in terms of direct and inverse cascades, which
leaves such nonlinear transverse cascade out of consideration, might
be incomplete and misleading. It should be emphasized that revealing
the complete picture of these nonlinear cascade processes has become
largely possible due to carrying out the analysis in Fourier space.
Because of the shear-induced anisotropy of cascade directions, often
used energy spectra and transfer functions integrated, or averaged
over wavevector angles, clearly, are not fully representative of the
actual, more general (transverse) redistribution of the spectral
energies due to nonlinearity in {\bf k}-space.

We have described above the linear and nonlinear process in a
time-average sense in the fully developed turbulence. To gain a
better insight in its self-sustenance and appearance of the bursts
in the total energy evolution (Fig. 2), below we analyze the
temporal evolution of individual harmonics in the vital area for
fixed $k_x, k_y, k_z$\footnote{Although $k_y$ of every harmonic with
$k_x\neq 0$ changes with time as a result of drift due to shear
flow, we can still take Eulerian approach and focus on a fixed $k_y$
in spectral space.}.

First we consider the dynamics of the dominant basic mode with ${\bf
k}_b=(0, 0, \pm 1)$, which corresponds to the maximum energy and
stress in spectral space during the whole evolution in the case of
the cubic box. For this mode, the dynamical Eqs. (16)-(18) are
reduced to the following system ($S=1$ everywhere below):
\begin{equation}
\frac{\partial}{\partial t} \frac{|\bar{u}_x({\bf
k}_b,t)|^2}{2}={\cal H}_k({\bf k}_b,t)+{\cal N}_x({\bf
k}_b,t)-\nu|\bar{u}_x({\bf k}_b,t)|^2,
\end{equation}
\begin{equation}
\frac{\partial}{\partial t}\frac{|\bar{u}_y({\bf
k}_b,t)|^2}{2}={\cal N}_y({\bf k}_b,t)-\nu|\bar{u}_y({\bf
k}_b,t)|^2,
\end{equation}
\begin{equation}
\frac{\partial}{\partial t}\frac{|\bar{u}_z({\bf
k}_b,t)|^2}{2}=-\nu|\bar{u}_z({\bf k}_b,t)|^2.
\end{equation}
Note that ${\cal N}_z({\bf k}_b,t)=0$ and Eq. (23) can be solved
analytically, $\bar{u}_z({\bf k}_b,t)=\bar{u}_z({\bf k}_b,0) \exp
(-2\nu t)$. Thus, the spanwise velocity decays exponentially in time
and the velocity field of the basic harmonic becomes 2D. Figure 10
shows the evolution of the spectral energies of the streamwise and
shearwise velocities, $|\bar{u}_x({\bf k}_b)|^2/2$, $|\bar{u}_y({\bf
k}_b)|^2/2$, and the nonlinear transfer functions ${\cal N}_x({\bf
k}_b)$, ${\cal N}_y({\bf k}_b)$ for the basic mode. Because the
basic mode is of large scale and the Reynolds number is high, the
dissipation terms in these equations are negligible. The velocities
exhibit a sequence of amplifications (bursts) and decays, with the
streamwise velocity being always much larger than the shearwise one.
${\cal N}_x({\bf k}_b)$ is always negative and therefore acts as a
sink in the streamwise velocity evolution Eq. (21), while ${\cal
N}_y({\bf k}_b)$ oscillates irregularly with alternating sign. When
${\cal N}_y({\bf k}_b)$ is positive, it amplifies the shearwise
velocity, while when negative causes its decrease, but the
time-average of ${\cal N}_y({\bf k}_b)$ is positive, though small
[see also Fig. 8(h)], indicating the growth trend of the shearwise
velocity on average. The shearwise velocity generated by ${\cal
N}_y({\bf k}_b)$, in turn, leads to the amplification of the
streamwise velocity due to the shear-related first linear term on
the rhs of dynamical Eq. (13), or equivalently, to the amplification
of the streamwise energy due to the stress term ${\cal H}_k({\bf
k}_b)$ in Eq. (21). Note that this is the only cause of
amplification for $|\bar{u}_x({\bf k}_b)|$, since the corresponding
nonlinear term, ${\cal N}_x({\bf k}_b)$, is always negative and
cannot amplify it. Then, $\bar{u}_y({\bf k}_b)$ and the amplified
$\bar{u}_x({\bf k}_b)$ jointly give rise to positive stress, ${\cal
H}_k({\bf k}_b)>0$ [Figs. 2(b) and 11(b)]. As a result, the stress
production rate, $d{\cal H}_k({\bf k}_b)/dt$ (calculated by
combining Eqs. 13 and 14), closely follows, or correlates with
${\cal N}_y({\bf k}_b)$ [Fig. 11(a), the correlation coefficient is
$R(d{\cal H}_k/dt,{\cal N}_y)=0.93$,\footnote{The correlation
coefficient $R(a,b)$ between two time-dependent functions $a$ and
$b$ is defined in a usual way as Pearson's product-moment
correlation coefficient. The time average is done over an entire
duration of the saturated turbulent state in the simulations.}] and,
consequently, the stress itself follows $|\bar{u}_y({\bf k}_b)|$
[Fig. 11(b), $R({\cal H}_k,|\bar{u}_y|)=0.97$], i.e., the peaks, or
bursts and dips in the temporal evolution of these quantities
coincide. So, the transfer function ${\cal N}_y({\bf k}_b)$ plays a
central role in the dynamics of the basic mode by providing a
positive feedback to the linear nonmodal growth: when ${\cal
N}_y({\bf k}_b)>0$, it causes the amplification of $\bar{u}_y({\bf
k}_b)$, which via nonmodal growth mechanism leads in turn to the
amplification of $\bar{u}_x({\bf k}_b)$ and hence of the
corresponding stress. These growth intervals in the evolution of
these quantities alternate with the decaying intervals where ${\cal
N}_y({\bf k}_b)<0$. Since for the aspect ratio $(1,1)$, the basic
mode is the dominant one among all the active modes, these bursts
also manifest themselves in the total stress evolution, as is seen
in Fig. 2(b).

\begin{figure}
\includegraphics[width=\columnwidth]{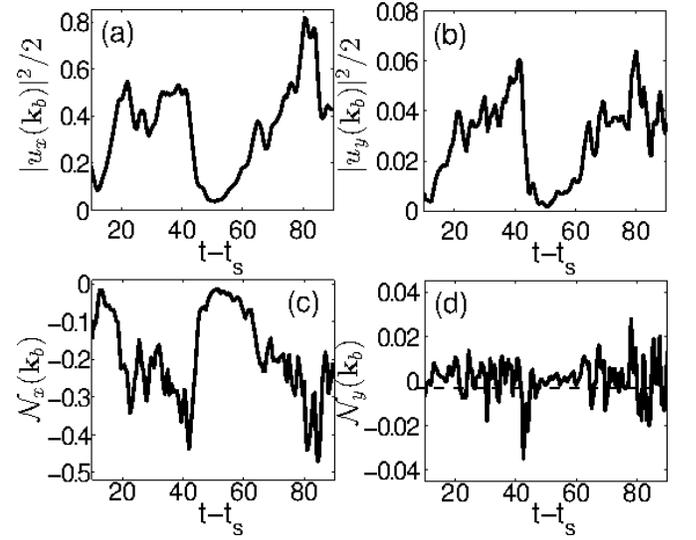}
\caption{Evolution of (a) the streamwise, $|\bar{u}_x({\bf
k}_b)|^2/2$, and (b) shearwise, $|\bar{u}_y({\bf k}_b)|^2/2$,
spectral energies as well as the nonlinear transfer functions (c)
${\cal N}_x({\bf k}_b)$ and (d) ${\cal N}_y({\bf k}_b)$ for the
basic mode with ${\bf k}_b=(0,0,\pm 1)$ in the cubic box. The
streamwise energy is much larger than the shearwise one due to the
linear nonmodal growth process. ${\cal N}_x({\bf k}_b)$ is negative
during the whole evolution, acting as a sink for the streamwise
energy, while ${\cal N}_y({\bf k}_b)$ oscillates and, when positive,
causes increase in $|\bar{u}_y|$; its time-averaged value over the
evolution is positive $4.78\cdot 10^{-4}$.}
\end{figure}
\begin{figure}
\includegraphics[width=\columnwidth]{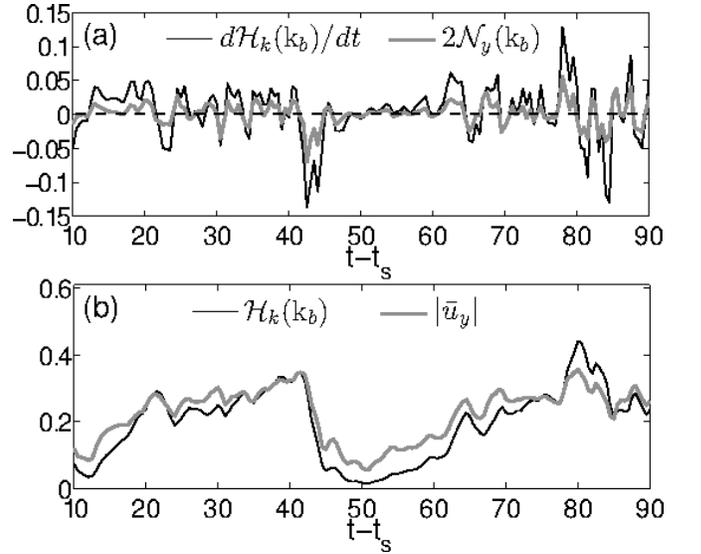}
\caption{Evolution of (a) the stress production rate, $d{\cal
H}_k({\bf k}_b)/dt$, and nonlinear transfer term ${\cal N}_y({\bf
k}_b)$ as well as (b) the spectral Reynolds stress, ${\cal H}_k({\bf
k}_b)$, and shearwise velocity, $|\bar{u}_y({\bf k}_b)|$ of the
basic mode in the cubic box. The stress production rate closely
follows ${\cal N}_y({\bf k}_b)$ (which is scaled by a factor of 2 to
make comparison easier) and, consequently, the amplifications
(bursts) of both shearwise velocity and stress occur when ${\cal
N}_y({\bf k}_b)$ is positive and decrease when this term is
negative.}
\end{figure}
\begin{figure}
\includegraphics[width=\columnwidth, height=8cm]{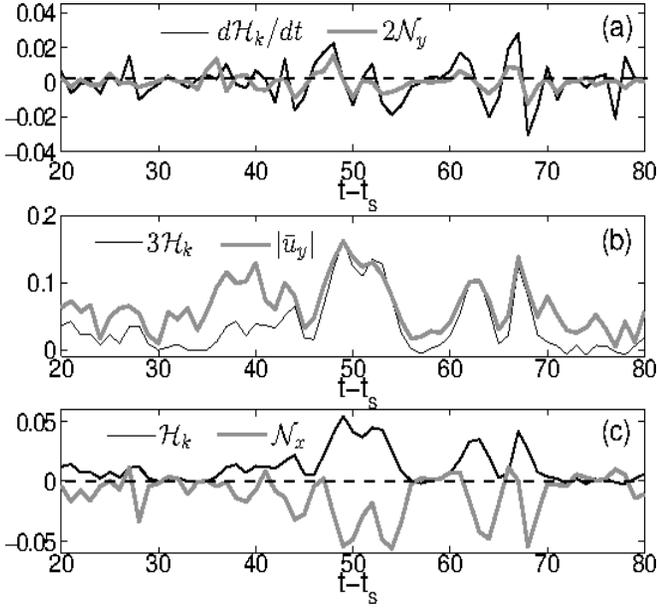}
\caption{Evolution of (a) the stress production rate and nonlinear
transfer term ${\cal N}_y$, (b) the stress ${\cal H}_k$ and
shearwise velocity $|\bar{u}_y|$ and (c) the stress and ${\cal N}_x$
for the mode ${\bf k}_1=(0, 1, 1)$. The nonlinear function in (a)
and the stress in (b) are scaled by 2 and 3, respectively, for the
sake of comparison.}
\end{figure}
\begin{figure}
\includegraphics[width=\columnwidth, height=8cm]{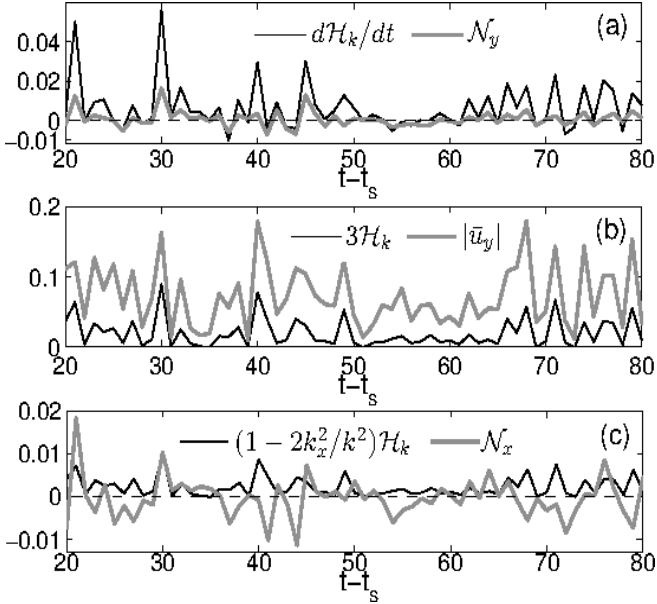}
\caption{Same as in Fig. 12, but for the non-symmetric mode ${\bf
k}_2=(1,1,1)$. The stress in (b) is scaled by a factor of 3. A high
degree of correlation exists between the stress production rate and
the nonlinear term ${\cal N}_y$ (a) as well as the stress and
shearwise velocity (b), implying that the stress amplification is
due to the nonmodal growth of $\bar{u}_x$ from $\bar{u}_y$.}
\end{figure}
\begin{figure}
\includegraphics[width=\columnwidth, height=8cm]{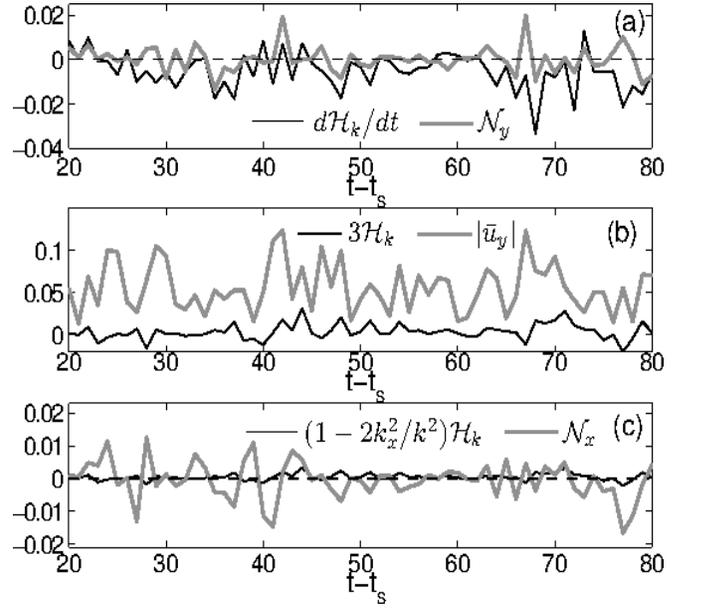}
\caption{Same as in Fig. 13, but for the mode ${\bf k}_3=(1, -1,
1)$. The stress in (b) is scaled by a factor of 3. Unlike the modes
in Figs. 12 and 13, ${\cal N}_x$ dominates the
nonnormality/shear-induced linear term and consequently the stress
and shearwise velocity do not correlate for this harmonic. The
corresponding stress is smaller than those of the other two modes.}
\end{figure}

The same nonmodal linear growth and nonlinear processes underlie the
dynamics of other symmetric ($k_x=0$) and non-symmetric ($k_x\neq
0$) active modes in the vicinity of ${\bf k}_b$ within the vital
area (Fig. 5), however, their quantitative character depends on mode
wavenumbers. To demonstrate this, in Figs. 12-14 we show the
dynamics for such nearby symmetric mode with ${\bf k}_1=(0,1,1)$ and
non-symmetric modes with ${\bf k}_2=(1,1,1)$ and ${\bf
k}_3=(1,-1,1)$. These figures display the time-history of (a) the
stress production rate $d{\cal H}_k/dt$ and the nonlinear transfer
term ${\cal N}_y$, (b) stress ${\cal H}_k$ and the absolute value of
the shearwise velocity, $|\bar{u}_y|$, (c) the first shear-induced
term of linear origin, $(1-2k_x^2/k^2){\cal H}_k$, on the rhs of Eq.
(16) and the nonlinear transfer term ${\cal N}_x$. We focus on these
harmonics with $k_z=1$, because, as noted above, they correspond to
higher levels of stress and energy than those at other values of
$k_z$ (Fig. 7), and hence contribute most to the turbulence
dynamics. The time-development of the symmetric harmonic with ${\bf
k}_1$ (Fig. 12) is similar to that of the basic mode except that it
has much less energy and stress. It is again governed by Eqs. (21)
and (22) (with ${\bf k}_b$ replaced by ${\bf k}_1$), but since
${\cal N}_z({\bf k}_1)$ is no longer zero, it generates the spanwise
velocity, which, although does not contribute to the stress, takes
up part of the mode's energy. The stress production rate and ${\cal
N}_y({\bf k}_1)$ evolve nearly similarly with time [Fig. 12(a)],
though with a bit lesser correlation $R(d{\cal H}_k/dt,{\cal
N}_y)=0.78$ than that in the case of the basic mode. Therefore, the
stress ${\cal H}_k({\bf k}_1)$ and the shearwise velocity,
$|\bar{u}_y({\bf k}_1)|$, also display similar behavior in time with
coinciding amplification and decay intervals [Fig. 12(b), $R({\cal
H}_k,|\bar{u}_y|)=0.86$]. Figure 12(c) shows that the stress remains
positive at all times and supplies the mode with energy, while
${\cal N}_x({\bf k}_1)$ is negative and acts as a sink for the
energy of the streamwise velocity. This indicates again that ${\cal
N}_y$ represents the only source maintaining the shearwise velocity
and hence the stress.

The linear nonmodal dynamics for the non-symmetric harmonics is
described by the terms of linear origin related to shear in the
above dynamical equations in Fourier space. First consider the
non-symmetric mode with ${\bf k}_2$ (Fig. 13). For this mode, ${\cal
N}_y({\bf k}_2)$ acts as a source and generates the shearwise
velocity, $|\bar{u}_y({\bf k}_2)|$. The latter is then nonmodally
amplified due to the first linear shear-induced term on the right
hand side of Eq. (17), $k_{x2}k_{y2}|\bar{u}_y({\bf
k}_2)|^2/k_{2}^2$, which is always positive for this mode since
$k_{x2}k_{y2}>0$. Because ${\cal N}_y({\bf k}_2)$ is also positive
on average in time [see Fig. 8(h)], the linear and nonlinear terms
in this equation cooperate with each other in the growth of
$|\bar{u}_y({\bf k}_2)|$. During its amplification, the shearwise
velocity generates and nonmodally amplifies the streamwise velocity,
$\bar{u}_x({\bf k}_2)$, due to the first shear-related linear term
on the rhs of Eq. (13), or equivalently, by the
$(1-2k_{x2}^2/k_2^2){\cal H}_k({\bf k}_2)$ term on the rhs of Eq.
(16). As seen from Fig. 13(c), ${\cal N}_x({\bf k}_2)$ is negative
most of the time (and hence on average) for this harmonic and
therefore acts as a sink for $|\bar{u}_x({\bf k}_2)|$, opposing its
growth. $\bar{u}_x$ and $\bar{u}_y$ velocity components give rise to
the stress ${\cal H}_k({\bf k}_2)$, whose production rate, as a
result, correlates with ${\cal N}_y({\bf k}_2)$ in time [Fig. 13(a),
$R(d{\cal H}_k/dt,{\cal N}_y)=0.84$]. Due to this, the temporal
evolution of ${\cal H}_k({\bf k_2})$ itself closely correlates with
$|\bar{u}_y({\bf k}_2)|$ [Fig. 13(b), $R({\cal
H}_k,|\bar{u}_y|)=0.9$]. So, increase (decrease) of ${\cal N}_y({\bf
k}_2)$ results in the increase (decrease) of the stress production
rate. Consequently, the increase (decrease) in the shearwise
velocity leads to the increase (decrease) in the stress and hence in
the energy extraction from the flow. It is clear from Figs. 13(a)
and 13(b) that the peaks, respectively, in the evolution of ${\cal
N}_y({\bf k}_2)$ and $d{\cal H}_k({\bf k}_2)/dt$ and in the
evolution of $|\bar{u}_y({\bf k}_2)|$ and ${\cal H}_k({\bf k}_2)$
coincide. We note that these high correlations, leading to the
growth of the Reynolds stress, are related to the fact that
$\bar{u}_x({\bf k}_2)$ grows from $\bar{u}_y({\bf k}_2)$ due to
nonnormality/shear (first term on the rhs of Eq. 13), whereas
$\bar{u}_y({\bf k}_2)$ is produced by the nonlinear transfer term
${\cal N}_y({\bf k}_2)$. The nonmodal growth mechanism, as for the
basic mode, also for this non-symmetric mode ensures that streamwise
and shearwise velocities correlate such that the resulting Reynolds
stress remains positive, ${\cal H}_k({\bf k}_2)>0$, at all times
[Fig. 13(b)], which is necessary for the continual extraction of
energy from the background flow into the harmonic. This is the
essence of the interplay between nonlinearity, which generates seed
shearwise velocity, and the nonmodal growth process that generates
and amplifies streamwise velocity from the shearwise one and
ultimately the positive stress.

The situation is different for the mode with ${\bf k}_3=(1,-1,1)$
(Fig. 14). The stress and shearwise velocity are not related -- the
corresponding correlation coefficients are much smaller than those
in the previous cases, $R(d{\cal H}_k/dt, {\cal N}_y)=0.17$,
$R({\cal H}_k,|\bar{u}_y|)=0.24$. The first shear-induced linear
term on the rhs of Eq. (16), responsible for the nonmodal growth of
the streamwise velocity, is dominated by ${\cal N}_x({\bf k}_3)$
[Fig. 14(c)]. As a result, the evolution of $|\bar{u}_x({\bf k}_3)|$
is governed mainly by ${\cal N}_x({\bf k}_3)$, and the effect of the
nonnormality is small. Moreover, in contrast to the harmonic with
${\bf k}_2$, the term responsible for the nonmodal growth of the
shearwise velocity is always negative, $k_{x3}k_{y3}|\bar{u}_y({\bf
k}_3)|^2/k_{3}^2<0$, and opposes ${\cal N}_y({\bf k}_3)$ in the
amplification of $|\bar{u}_y({\bf k}_3)|$. In other words, for the
shearwise velocity, there is no constructive interplay between
nonnormality-induced linear dynamics and the nonlinear term. So, the
production rate of the streamwise velocity at ${\bf k}_3$ is much
reduced compared to that for ${\bf k}_2$. This leads to the absence
of correlation between the temporal evolution of the stress
production rate and ${\cal N}_y({\bf k}_3)$ [Fig. 14(a)] and hence
between the stress ${\cal H}_k({\bf k}_3)$ and shearwise velocity
[Fig. 14(b)]. Since $|\bar{u}_x({\bf k}_3)|$ is determined mainly by
the nonlinear transfer term ${\cal N}_x({\bf k}_3)$ rather than by
the nonmodal dynamics, it is not correlated with $\bar{u}_y({\bf
k}_3)$. As a result, the Reynolds stress produced by these two
velocity components is irregularly oscillating between positive and
negative values [Fig. 14(b)] in contrast to the basic mode and the
${\bf k}_1$ and ${\bf k}_2$ modes, whose corresponding stresses are
always positive owing to the linear nonmodal growth mechanism [see
for comparison Figs. 12(b) and 13(b)]. Because of this, the
time-averaged value of the stress corresponding to this mode, being
equal to $0.0011$, is smaller than that for the ${\bf k}_2$ mode,
which is 0.006. So, due to the constructive interplay between the
nonlinear and linear terms in the case of the basic mode and ${\bf
k}_1$ and ${\bf k}_2$ modes, their effectiveness in the energy
extraction from the background flow is higher than that of the ${\bf
k}_3$ mode, for which such an interplay is practically absent.

Although the dynamics of only three modes, one symmetric and other
two non-symmetric ones with $k_y/k_x>0$ and $k_y/k_x<0$ at $k_z=1$,
were described above, the other neighboring modes in these regions
of {\bf k}-space exhibit a similar dynamics too. The time-averaged
plots of Figs. 8(a)-8(c) clearly show that in the vital area, the
Reynolds stress of active modes with $k_y/k_x<0$ on average is
smaller than that of ones with $k_y/k_x>0$, which thus contribute
most to the positive stress. They play a key role in supplying
turbulence with energy, together with the basic and nearby symmetric
modes. These modes share a similar dynamics with the ${\bf k}_2$
mode and therefore the higher values of the associated stress are
due to a constructive interplay between nonmodal growth and
nonlinear feedback discussed above. From these plots we see that on
average ${\cal N}_y
> 0$ at $k_y/k_x
> 0$ and $k_x=0$, indicating that the transverse cascade continually regenerates
just these modes, thereby ensuring the self-sustenance of the
turbulence.

\begin{figure}
\includegraphics[width=\columnwidth]{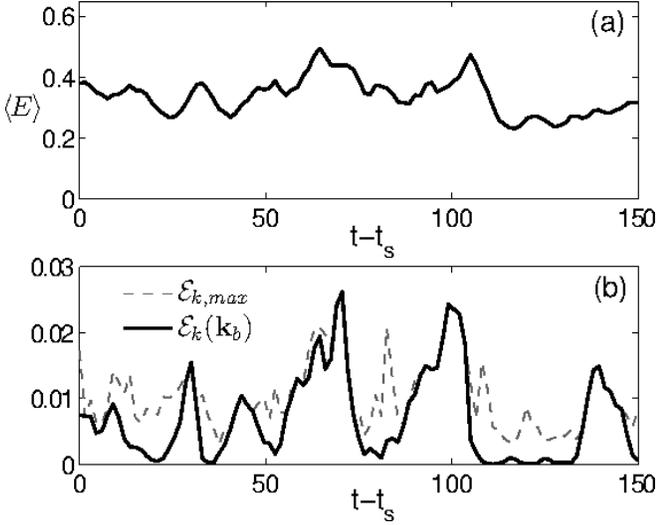}
\caption{Evolution of the volume-averaged (a) perturbed total
kinetic energy, $\langle E\rangle$, and (b) the energy ${\cal
E}_k({\bf k}_b)$ of the basic mode with ${\bf k}_b=(0,0,\pm 1)$ as
well as the maximum energy in {\bf k}-space, ${\cal E}_{k,max}$
(dashed line) for the aspect ratio $(3,2)$. The basic mode
corresponds to the maximum energy only during large bursts of the
spectral energy, while at other times this maximum is achieved by
other neighboring active modes [bigger black dots in Figs. 16(a) and
16(b)]. $\langle E \rangle$ approximately follows the maximum
spectral energy and therefore the highest peaks are associated with
the dominance of the basic mode at these times.}
\end{figure}
\begin{figure*}
\includegraphics[width=5.9cm]{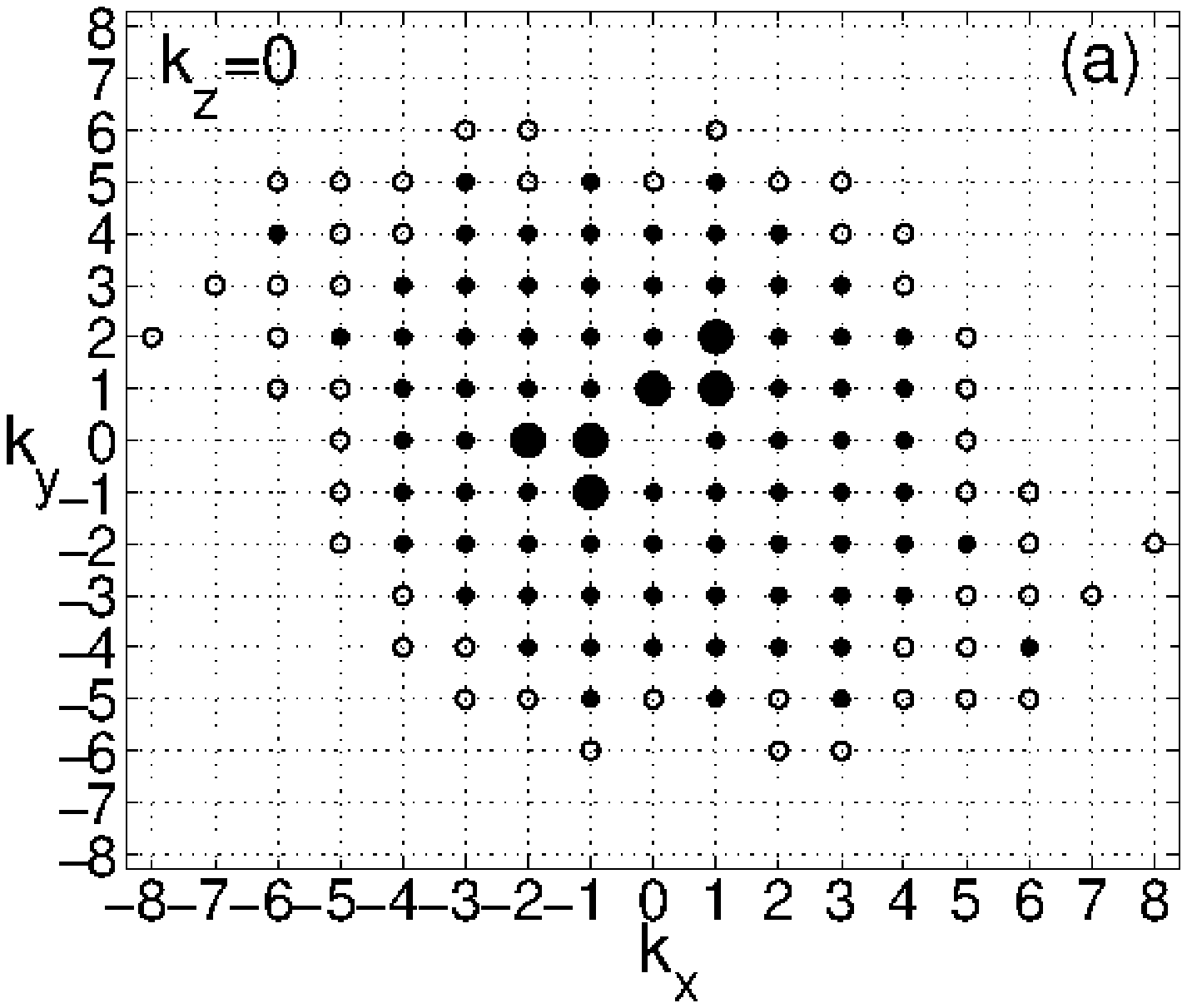}
\includegraphics[width=5.9cm]{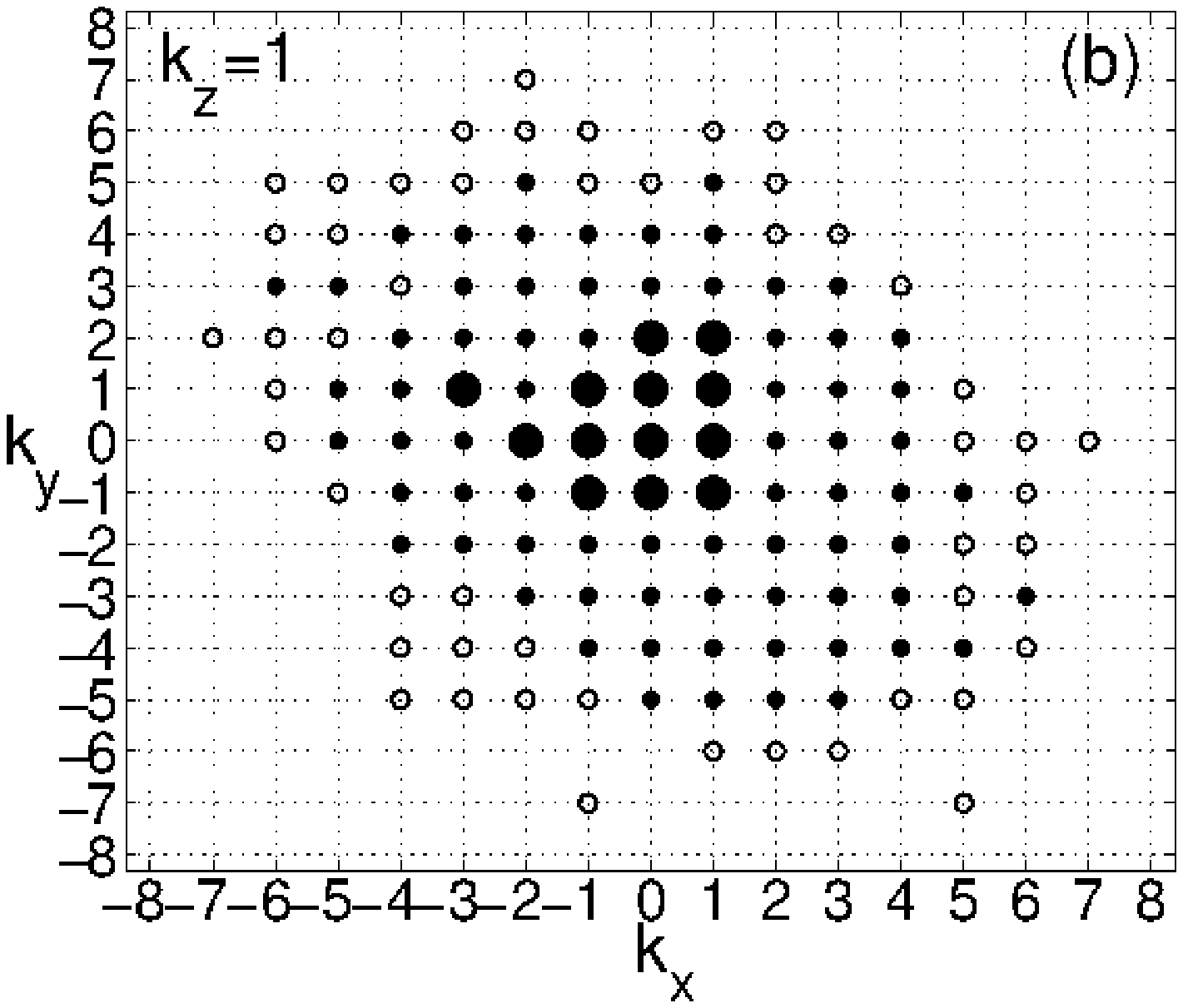}
\includegraphics[width=5.9cm]{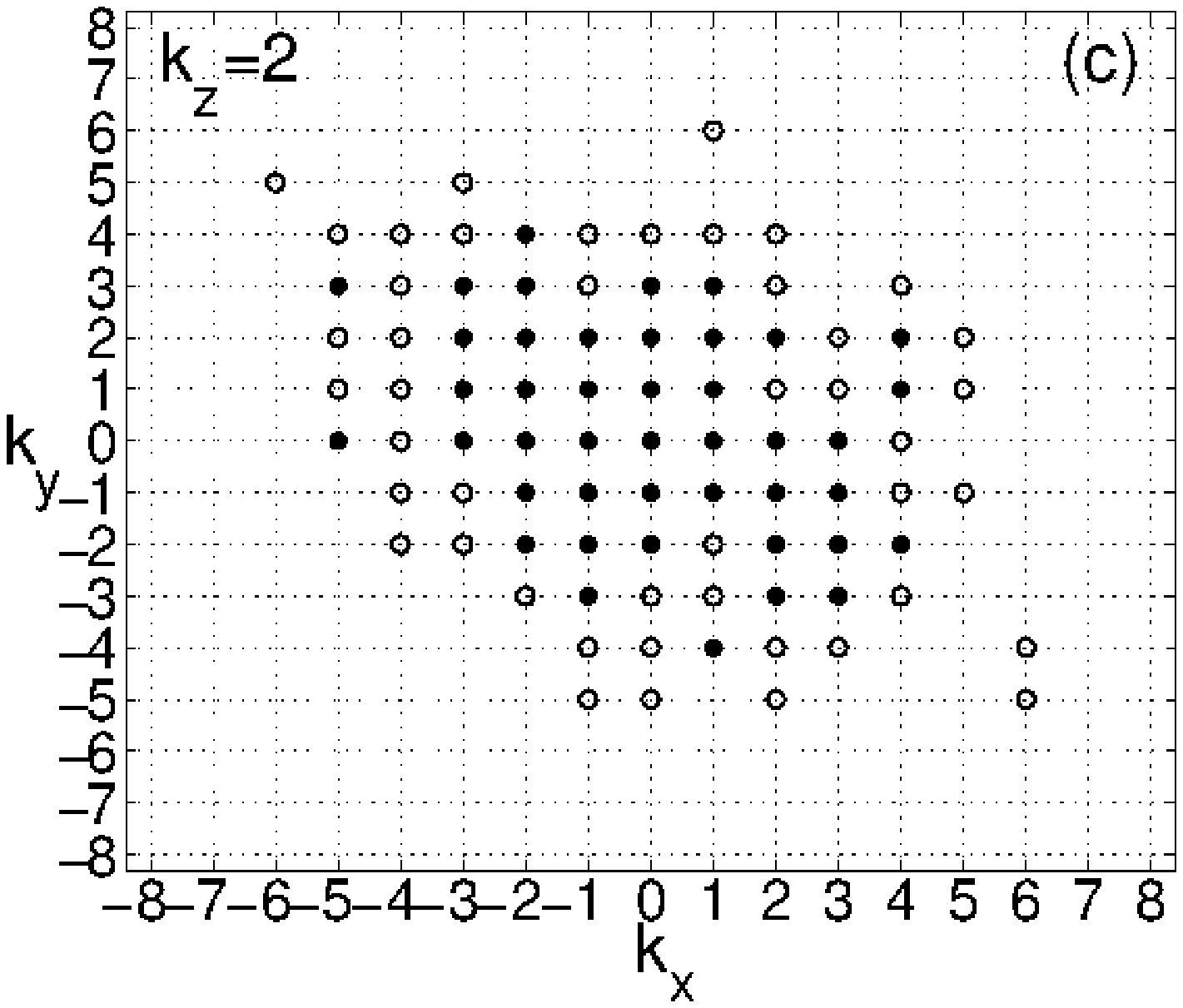}
\caption{Distribution of the dominant energy-carrying modes in {\bf
k}-space at (a) $k_z=0$, (b) $k_z=1$ and (c) $k_z=2$ during the
whole run, for the aspect ratio $(3,2)$. The symbols have the same
meaning as in Fig. 5. The bigger black dots represent modes that
correspond to the maximum of ${\cal E}_k$ at least once during the
evolution, including the basic mode at $k_x=k_y=0, k_z=1$.}
\end{figure*}

\subsection{Aspect ratio $(A_{xz}, A_{yz})=(3,2)$}

The evolution of the volume-averaged energy in the saturated
turbulence for the aspect ratio $(3,2)$ is shown in Fig. 15(a). It
is different from that in the previous case, having weaker and less
pronounced bursts. This difference is due to the specific changes of
the dynamics in spectral space as the aspect ratio varies. To make a
detailed comparison with the previous case, we performed a similar
analysis of energy spectra, linear energy-injecting stress and
nonlinear transfer terms.

Figure 16 shows the dynamically active, energy-carrying modes during
the evolution, as defined above. The number of these modes (209
black dots in Fig. 16) is much larger compared to that in the cubic
box. The vital area of turbulence containing these harmonics has
become wider in {\bf k}-space and spans the range $|k_x|\leq 8,
|k_y| \leq 7, |k_z|\leq 4$. The basic mode ${\bf k}_b=(0,0,\pm 1)$
no longer dominates over the other active modes during the whole
evolution, but corresponds to the maximum energy in spectral space
only during the highest bursts [Fig. 15(b)]. The lower peaks in the
maximum spectral energy curve in Fig. 15(b) are due to symmetric and
non-symmetric modes that lie in the vicinity of the basic one in the
vital area; these modes are marked by bigger black dots in Figs.
16(a) and 16(b) and have $k_z=0,1$. The modes with larger $k_z\geq
2$ in the vital area, although being active, never yield maximum of
the spectral energy. One can say that the dominance of the basic
mode is still retained during the strongest bursts, but because
other active modes also reach comparable maximum energies at
different times, the bursts in the volume-averaged energy do not
appear to be as pronounced as in the cubic box, where only the basic
mode prevails over the other active modes at all times.

\begin{figure*}
\includegraphics[width=5.9cm]{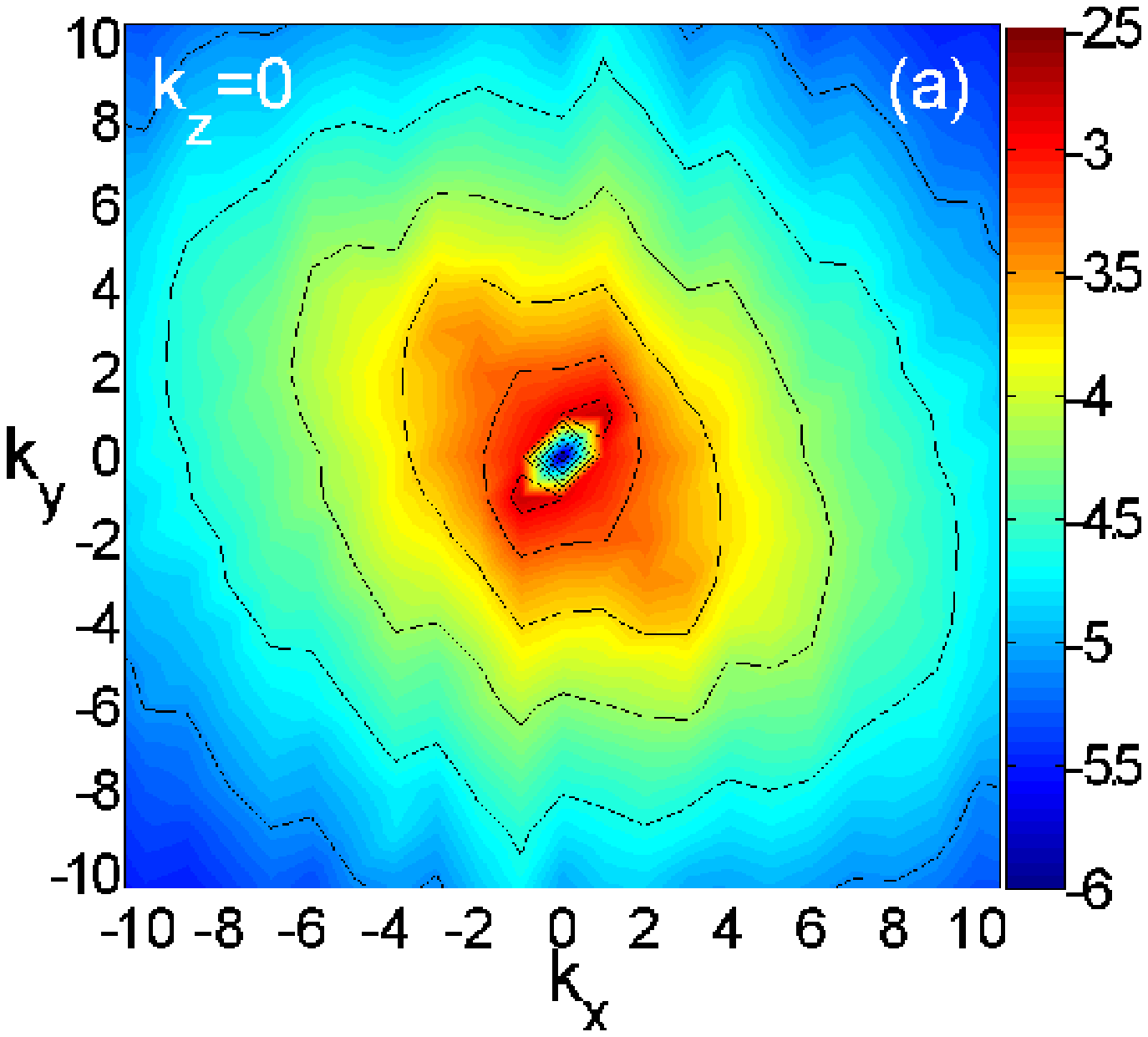}
\includegraphics[width=5.9cm]{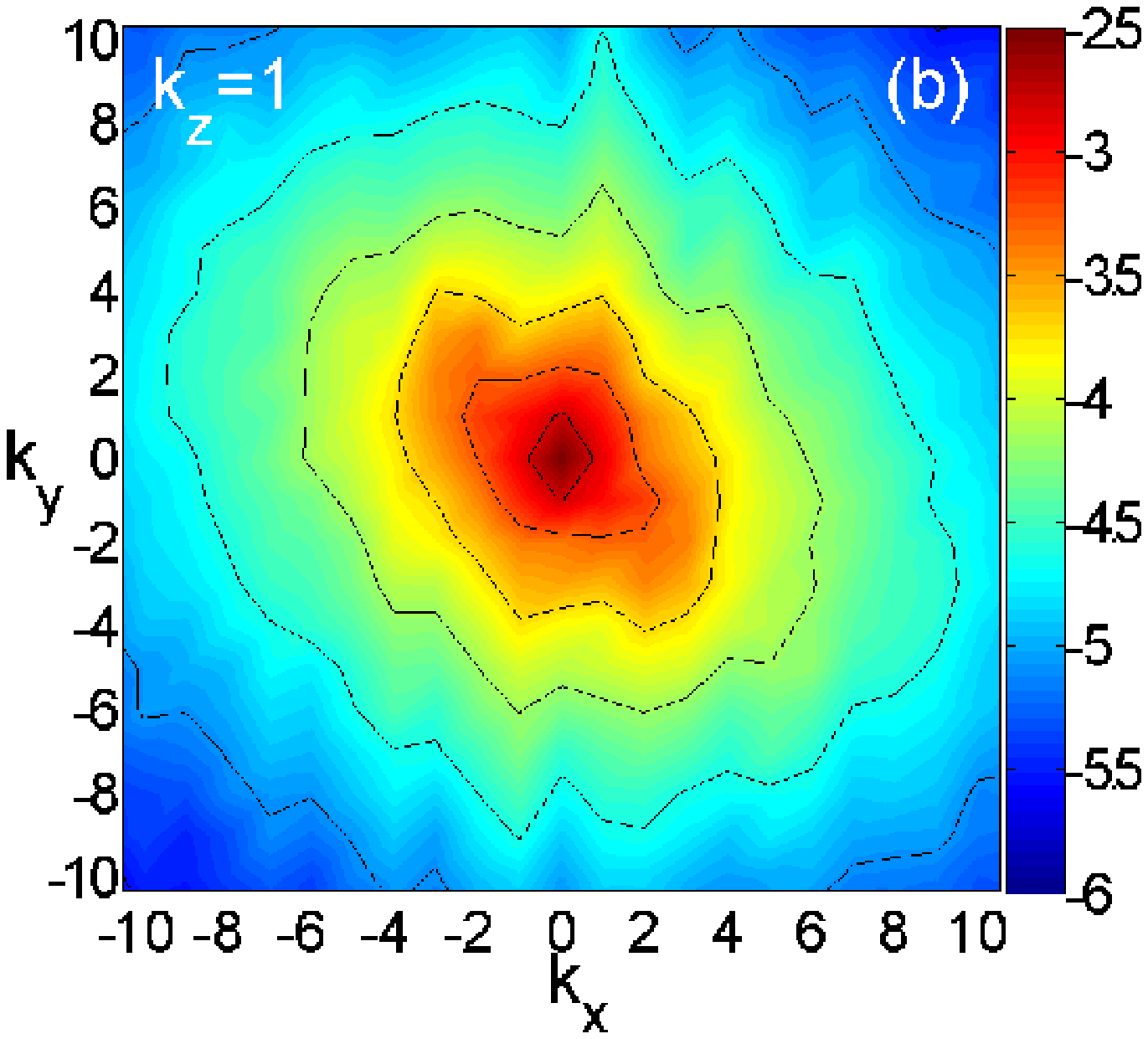}
\includegraphics[width=5.9cm]{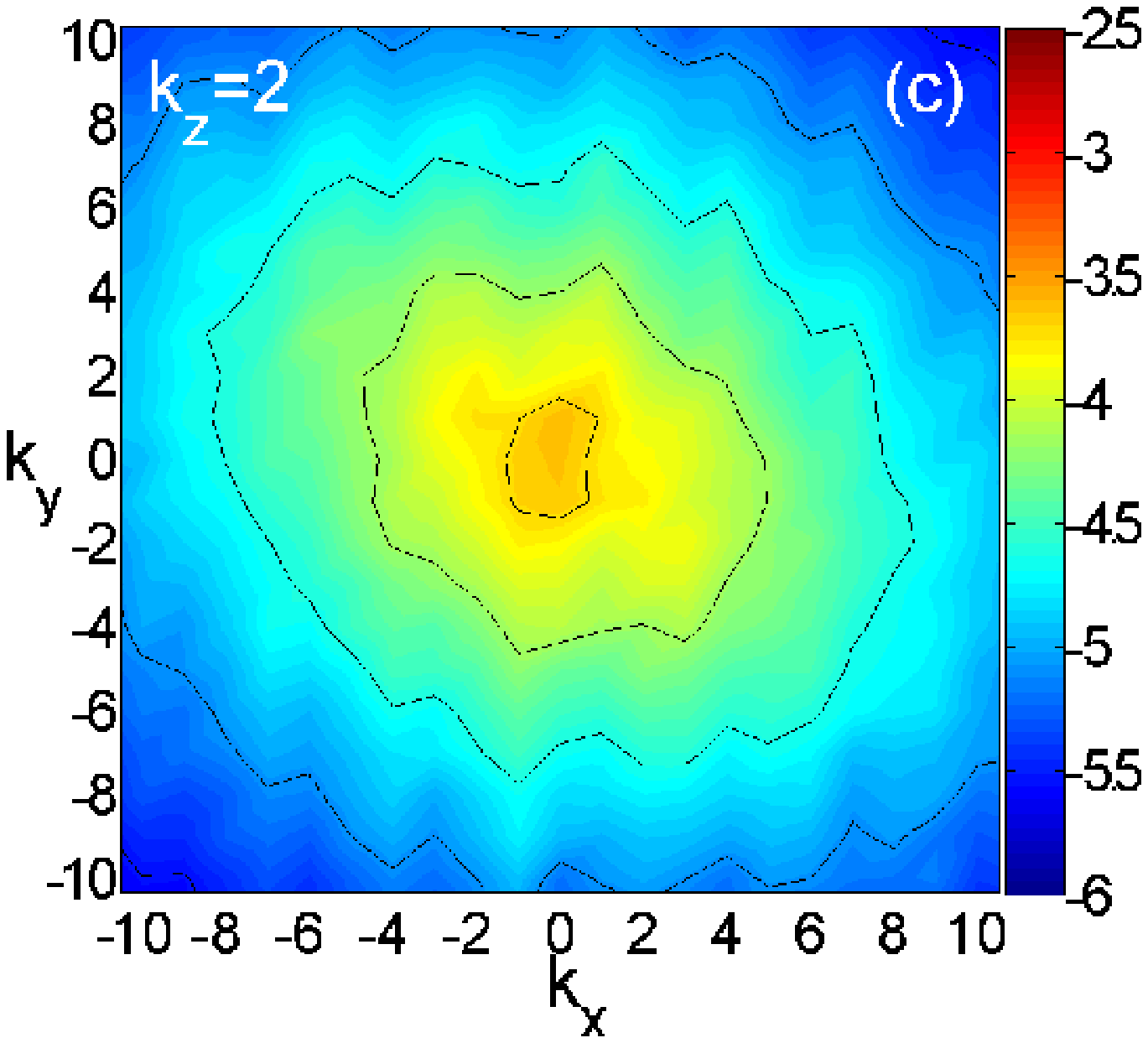}
\caption{(Color online) Logarithm of the time-averaged kinetic
energy spectrum, $log_{10}({\cal E}_k)$, for the aspect ratio
$(3,2)$. The slices in $(k_x,k_y)$-plane are at (a) $k_z=0$, (b)
$k_z=1$ and (c) $k_z=2$. The spectrum is anisotropic, with larger
power at the $k_y/k_x<0$ side.}
\end{figure*}
\begin{figure}
\includegraphics[width=\columnwidth]{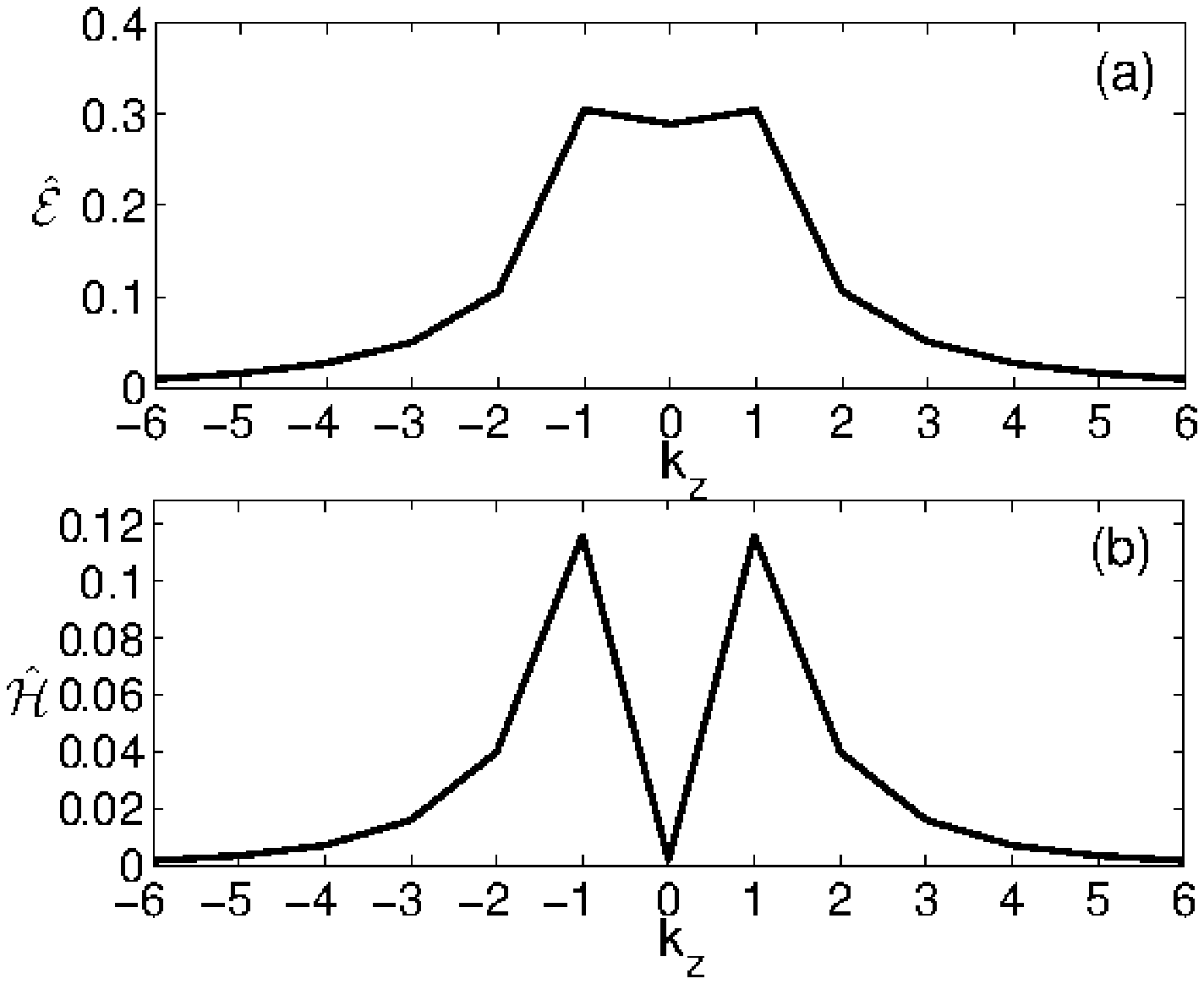}
\caption{Integrated in $(k_x, k_y)$-plane time-averaged (a) kinetic
energy spectrum, $\hat{\cal E}$, and (b) spectral Reynolds stress,
$\hat{\cal H}$, vs. $k_z$ for the aspect ratio $(3,2)$. Both the
energy and stress achieve their maximum at $k_z=\pm 1$, although the
energy at $k_z=0$ is close to that at $k_z=\pm 1$.}
\end{figure}

Figure 17 shows the time-averaged energy spectrum in
$(k_x,k_y)$-plane at $k_z=0, 1, 2$. This spectrum is more extended
over the wavenumbers compared to that in the cubic box, but has a
similar shape and type of anisotropy. Although the maximum of energy
is again at $k_z=1$, modes with $k_z=0$ also have comparable
energies, unlike that in the cubic box. As seen from Fig. 16, modes
with these two spanwise wavenumbers correspond to the maximum of
spectral energy at certain times during evolution and hence play a
main role in the turbulence dynamics. The spectral energy then
rapidly decreases with increasing $k_z$. This behavior is also seen
in Fig. 18 showing the time-averaged spectra of the energy and
stress integrated in $(k_x,k_y)$-plane, $\hat{\cal E}$ and
$\hat{\cal H}$. Both these quantities reach their maximum at
$k_z=\pm 1$. $\hat{\cal E}$ at $k_z=0$ is close to its maximum
value, while $\hat{\cal H}$ at $k_z=0$ is small, because ${\cal
H}_k(k_x,k_y,0)$ changes sign in $(k_x,k_y)$-plane, as in the case
of the cubic box [see Fig. 8(a)]. So, although harmonics with
$k_z=0$ can reach energies comparable to those with $k_z=1$, in
fact, they are not as effective in the net energy extraction from
the flow as the latter ones, for which ${\cal H}_k$ is positive over
the whole $(k_x,k_y)$-plane at $k_z=1$ [Fig. 19(a)].

\begin{figure*}
\includegraphics[width=5.9cm]{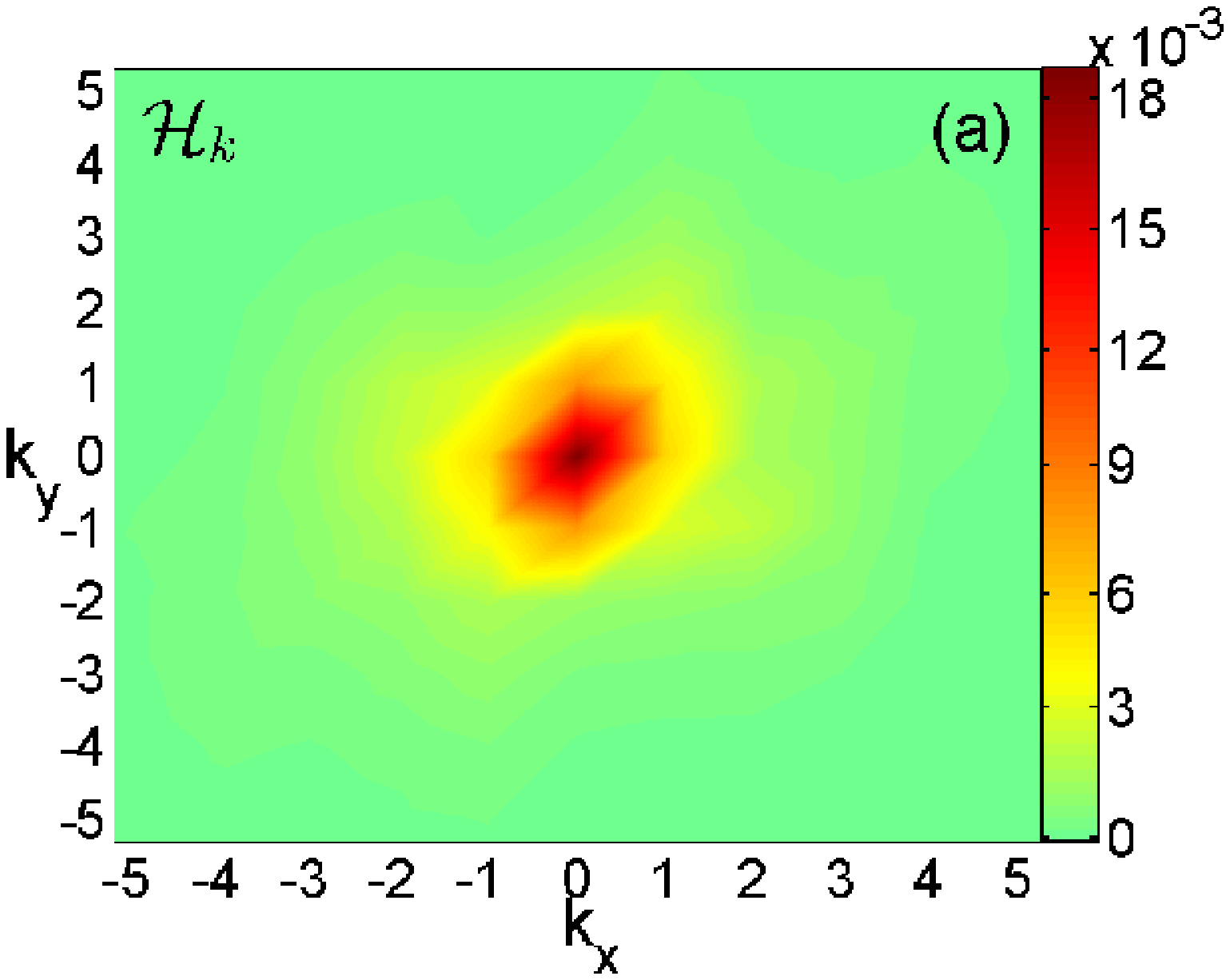}
\includegraphics[width=5.9cm]{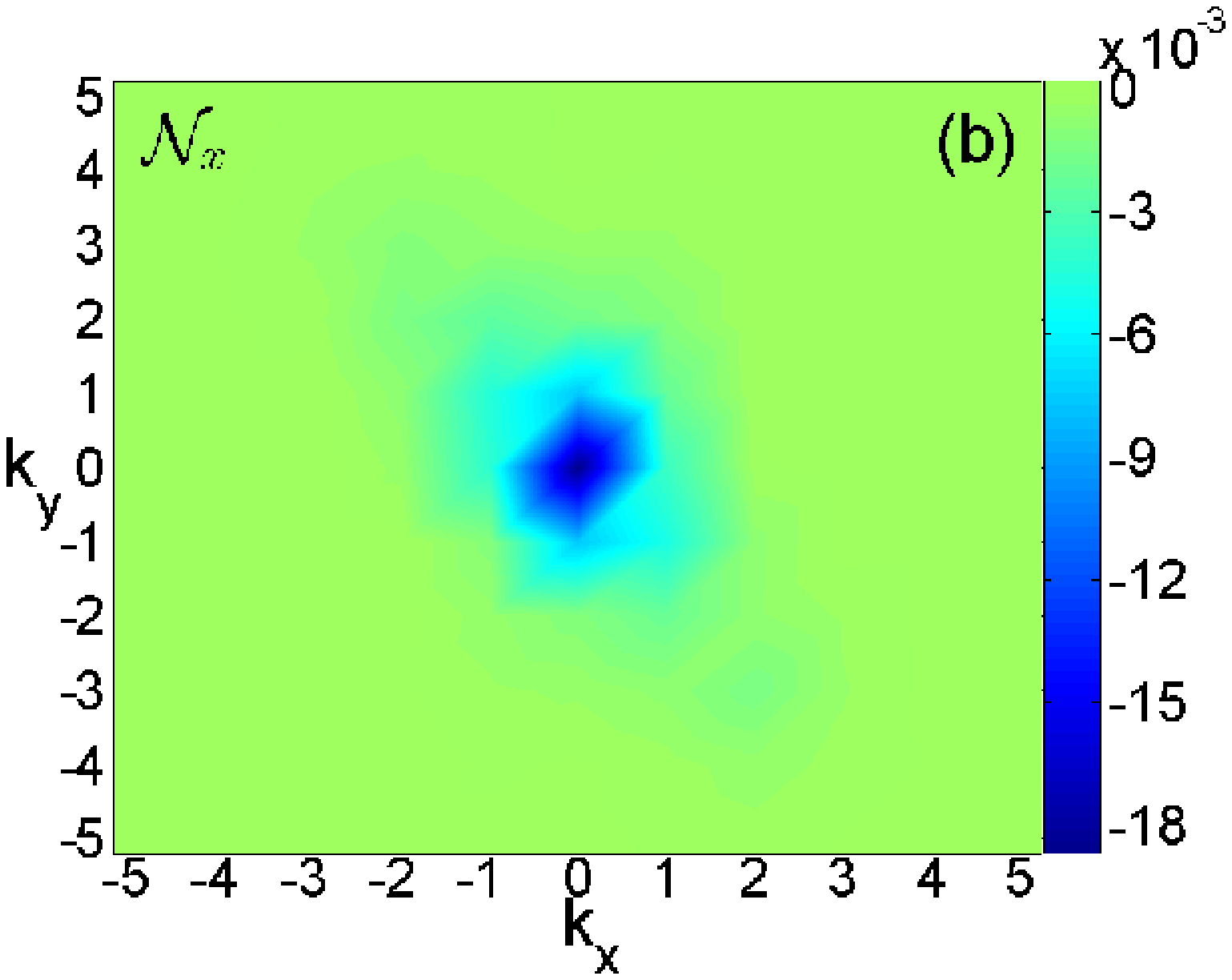}
\includegraphics[width=5.9cm]{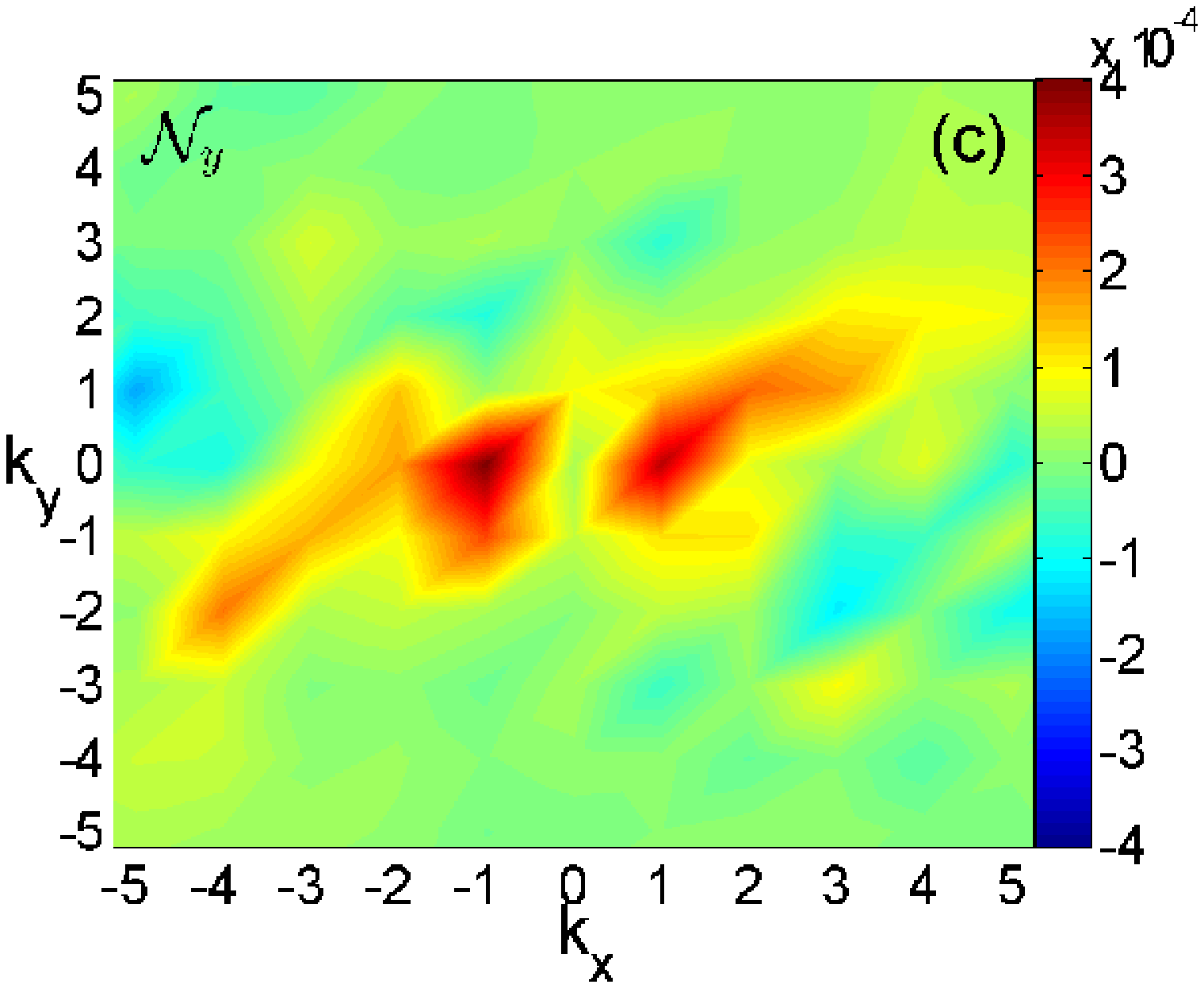}
\caption{(Color online) Maps of the time-averaged (a)
energy-injecting stress, ${\cal H}_k$, and nonlinear transfer terms,
(b) ${\cal N}_x$ and (c) ${\cal N}_y$ in {\bf k}-space for the
aspect ratio $(3,2)$. Shown are slices in $(k_x,k_y)$-plane at
$k_z=1$.}
\end{figure*}
\begin{figure}
\includegraphics[width=\columnwidth]{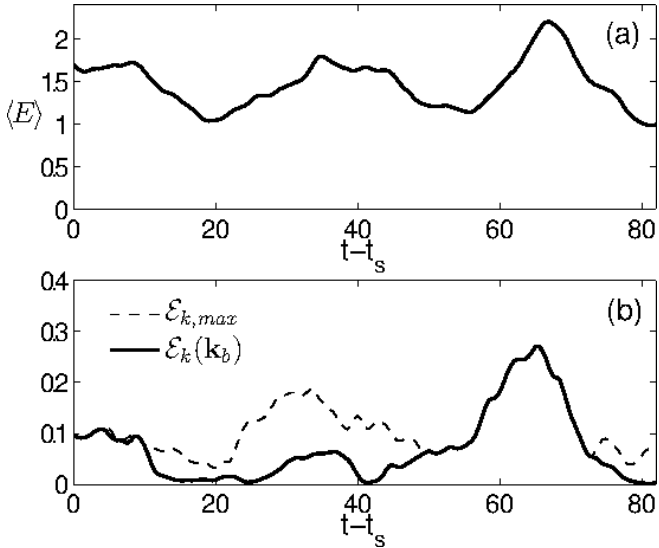}
\caption{Same as in Fig. 15, but for the aspect ratio $(1,2)$. The
basic mode corresponds to the maximum energy only during largest
bursts, while at other times maximum energy is achieved by other
nearby symmetric active modes [bigger black dots in Figs. 21(a) and
21(b)]. The volume-averaged energy follows the maximum spectral
energy and therefore the highest peaks are manifestation of the
basic mode dynamics (growth) at these times.}
\end{figure}\begin{figure*}
\includegraphics[width=5.9cm]{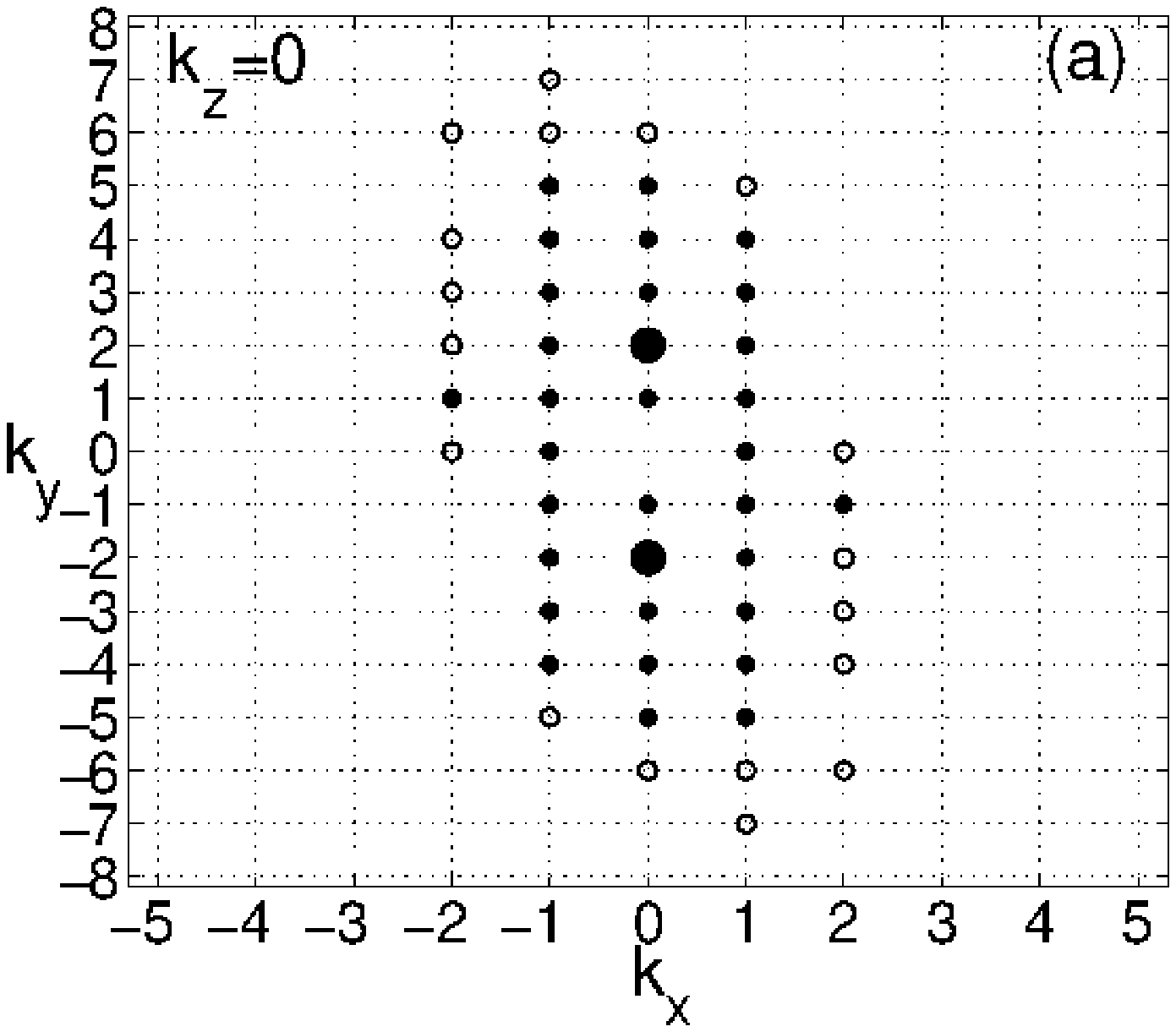}
\includegraphics[width=5.9cm]{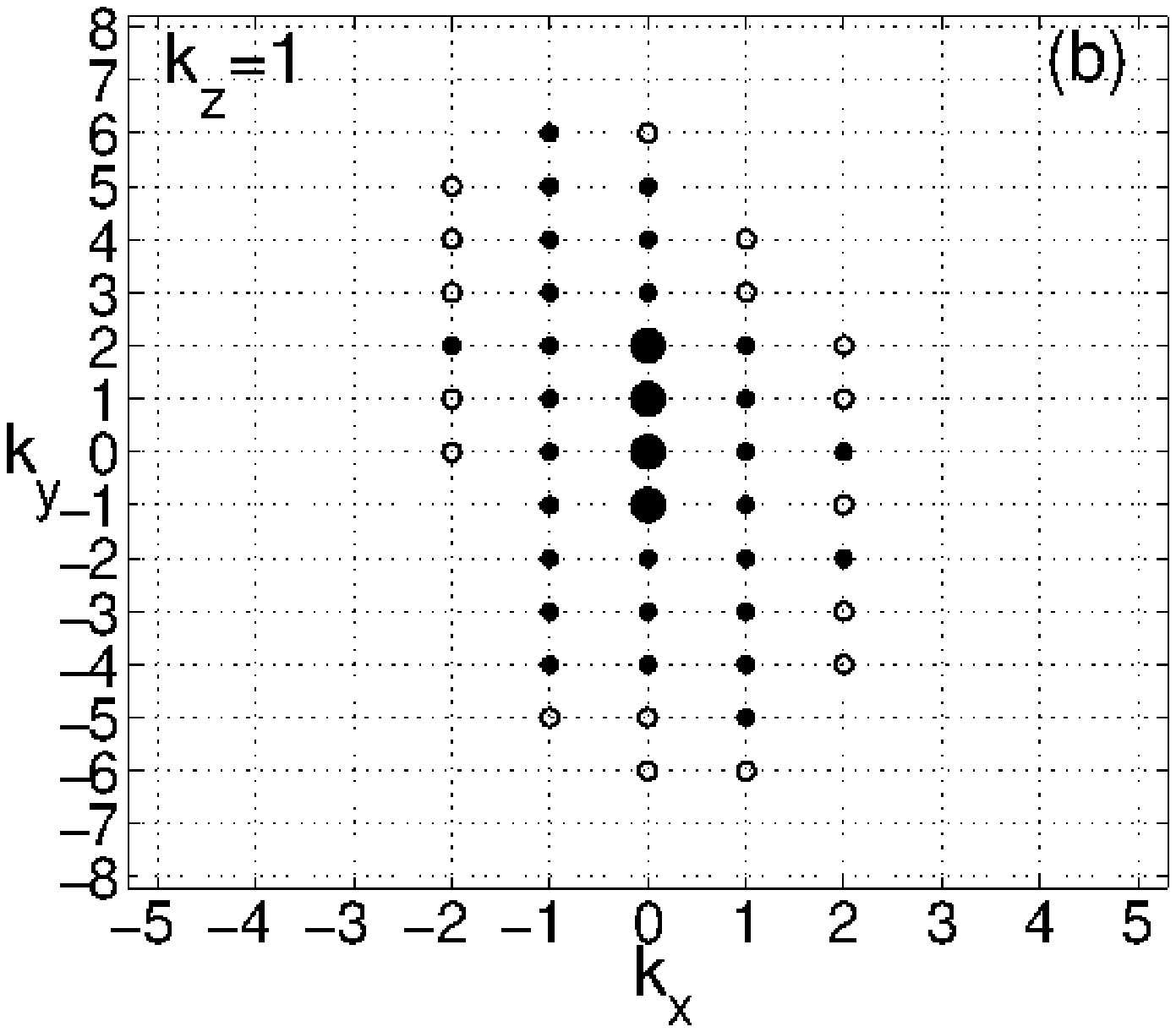}
\includegraphics[width=5.9cm]{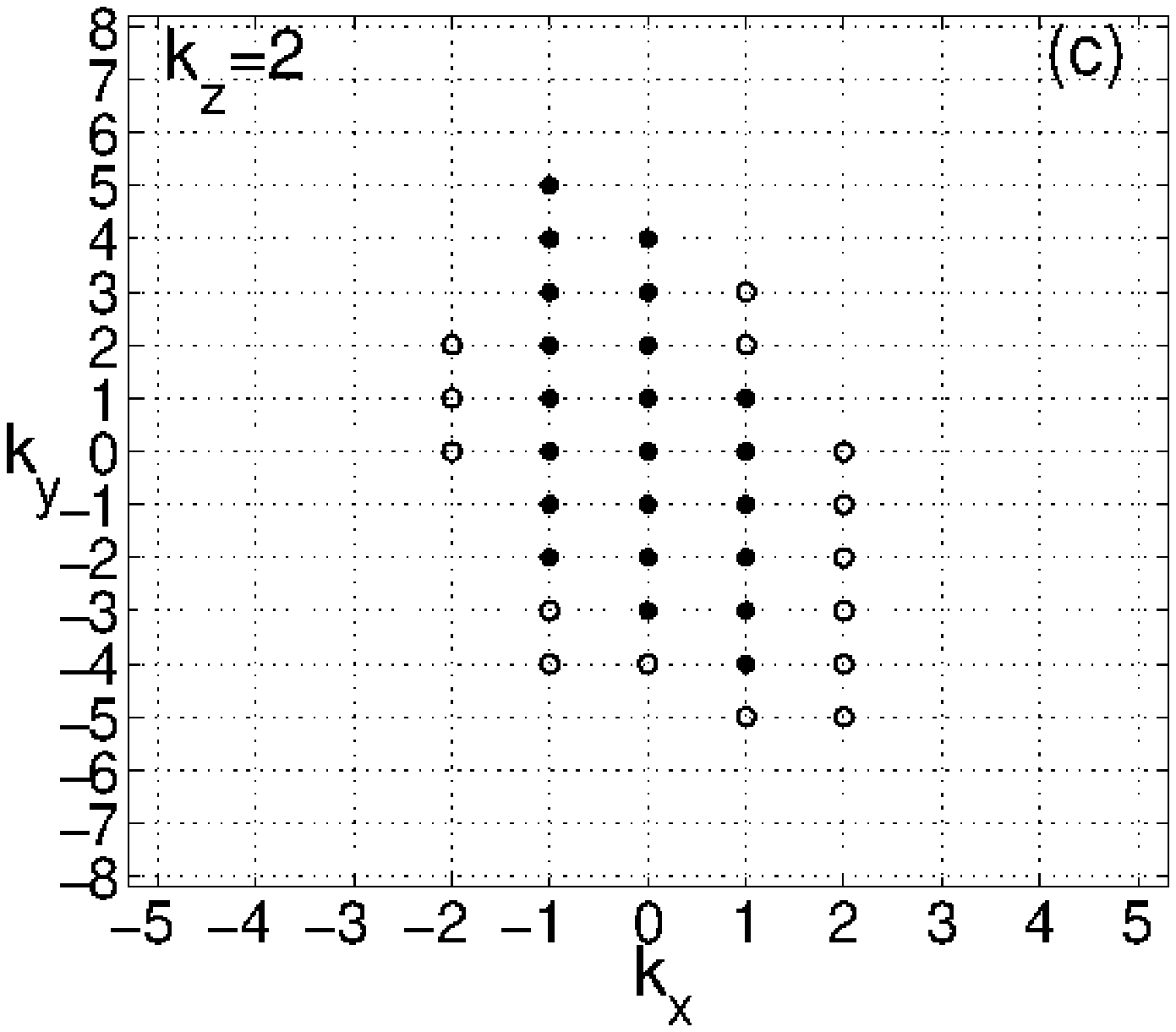}
\caption{Distribution of the dominant energy-carrying modes in {\bf
k}-space at (a) $k_z=0$, (b) $k_z=1$ and (c) $k_z=2$ for the aspect
ratio $(1,2)$. The symbols have the same meaning as in Fig. 16. The
most active modes that contribute to the maximum of the spectral
energy at least once during the evolution are only symmetric ones,
including the basic mode at $k_x=k_y=0, k_z=1$.}
\end{figure*}
\begin{figure*}
\includegraphics[width=5.9cm]{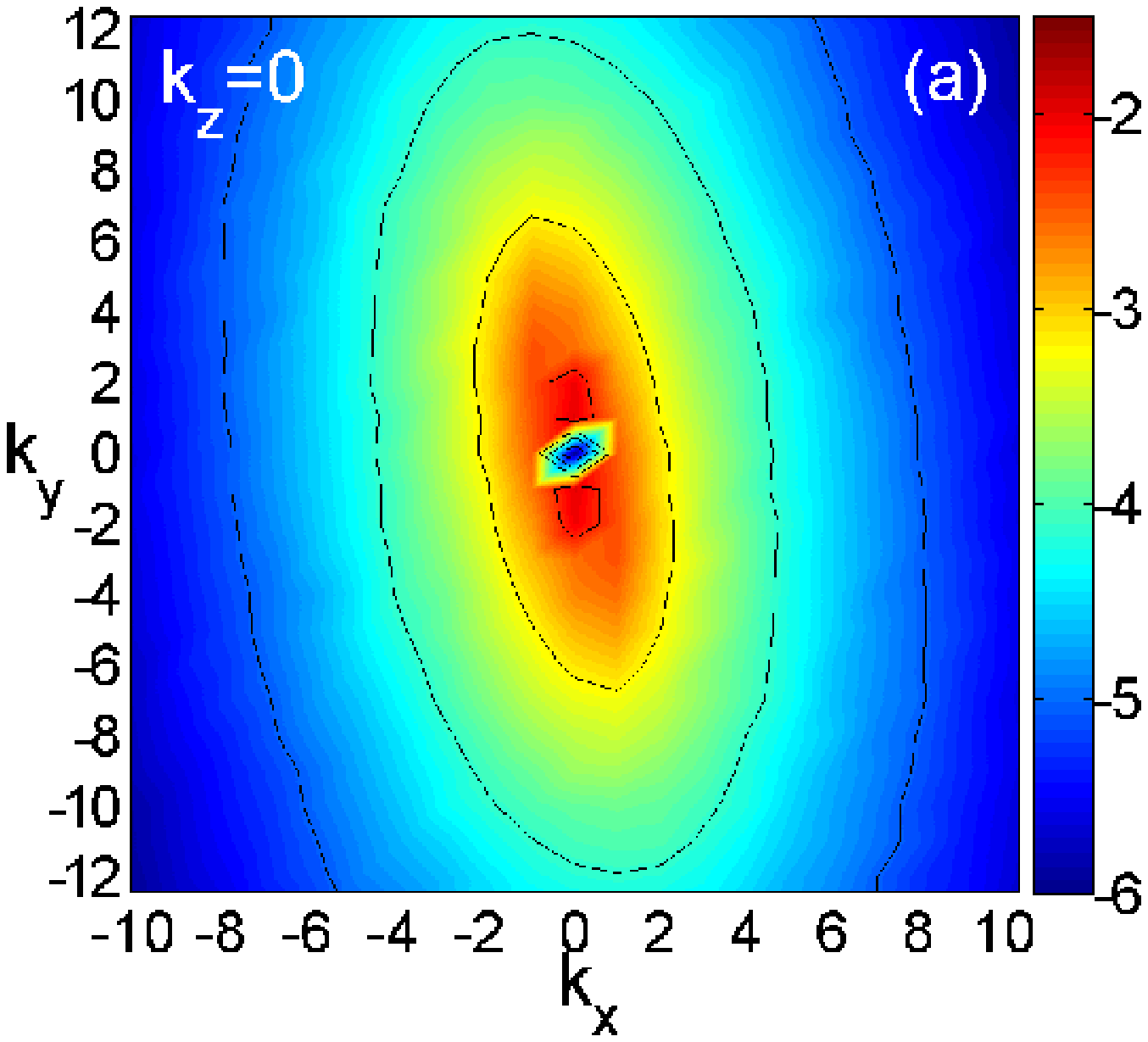}
\includegraphics[width=5.9cm]{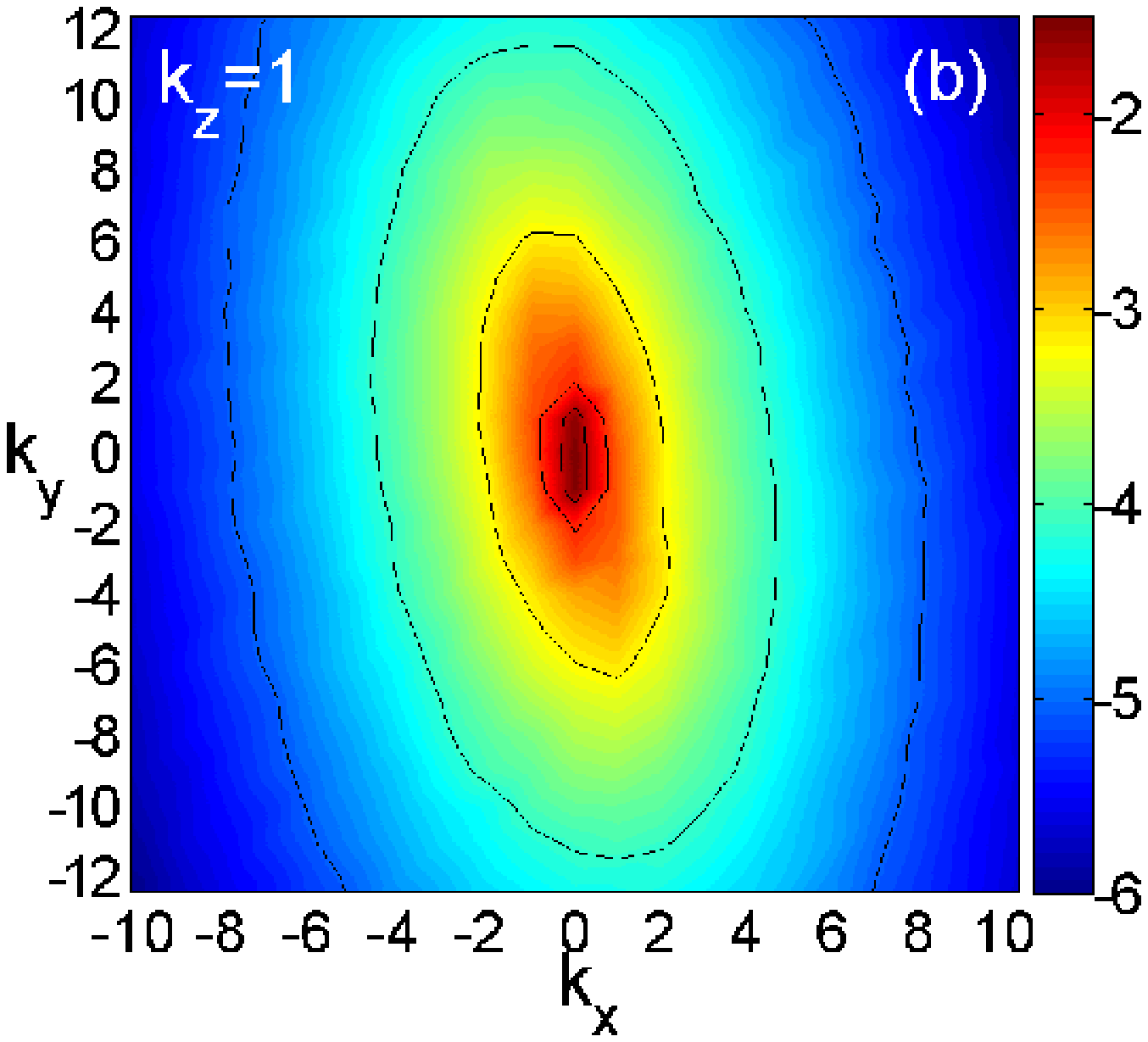}
\includegraphics[width=5.9cm]{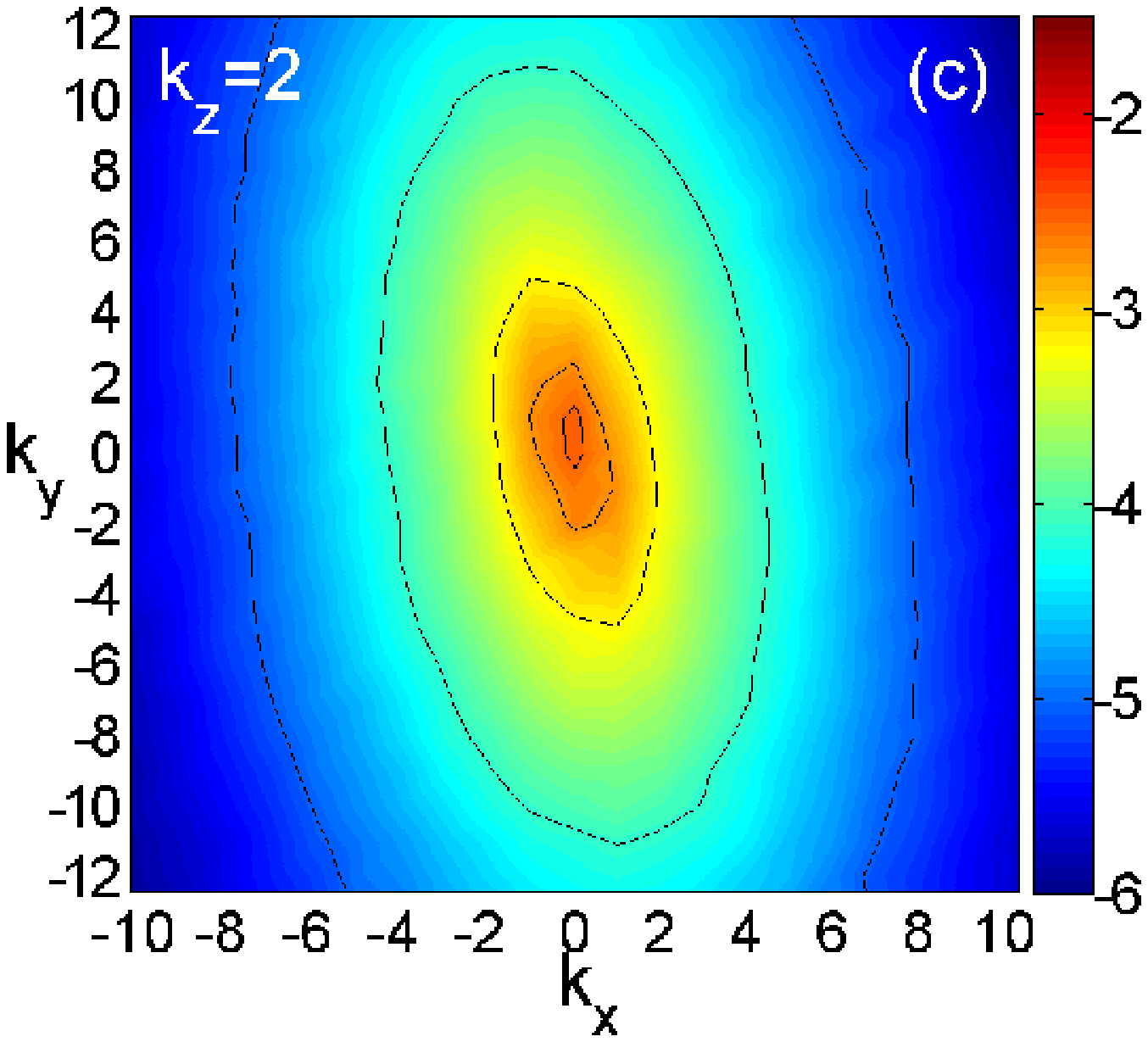}
\caption{(Color online) Logarithm of the time-averaged kinetic
energy spectrum, $log_{10}({\cal E}_k)$ for the aspect ratio
$(1,2)$. The slices in $(k_x,k_y)$-plane at (a) $k_z=0$, (b) $k_z=1$
and (c) $k_z=2$ are presented.}
\end{figure*}
\begin{figure}
\includegraphics[width=\columnwidth]{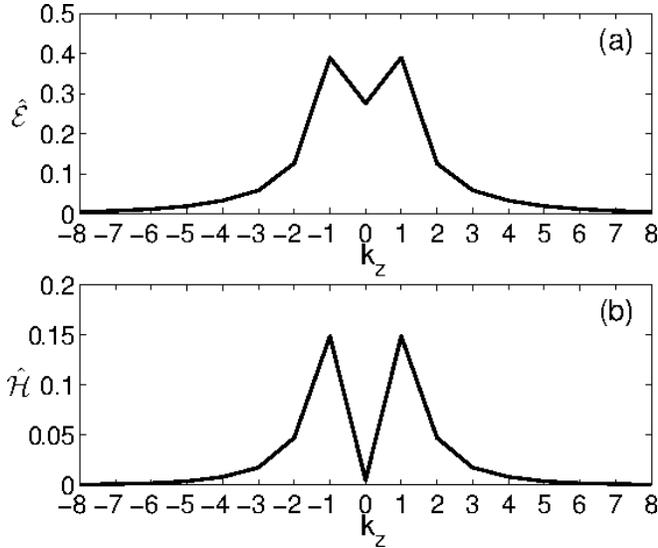}
\caption{Same as in Fig. 18, but for the aspect ratio $(1,2)$. Both
the energy and stress achieve their maximum again at $k_z=\pm 1$.}
\end{figure}
\begin{figure*}
\includegraphics[width=5.9cm]{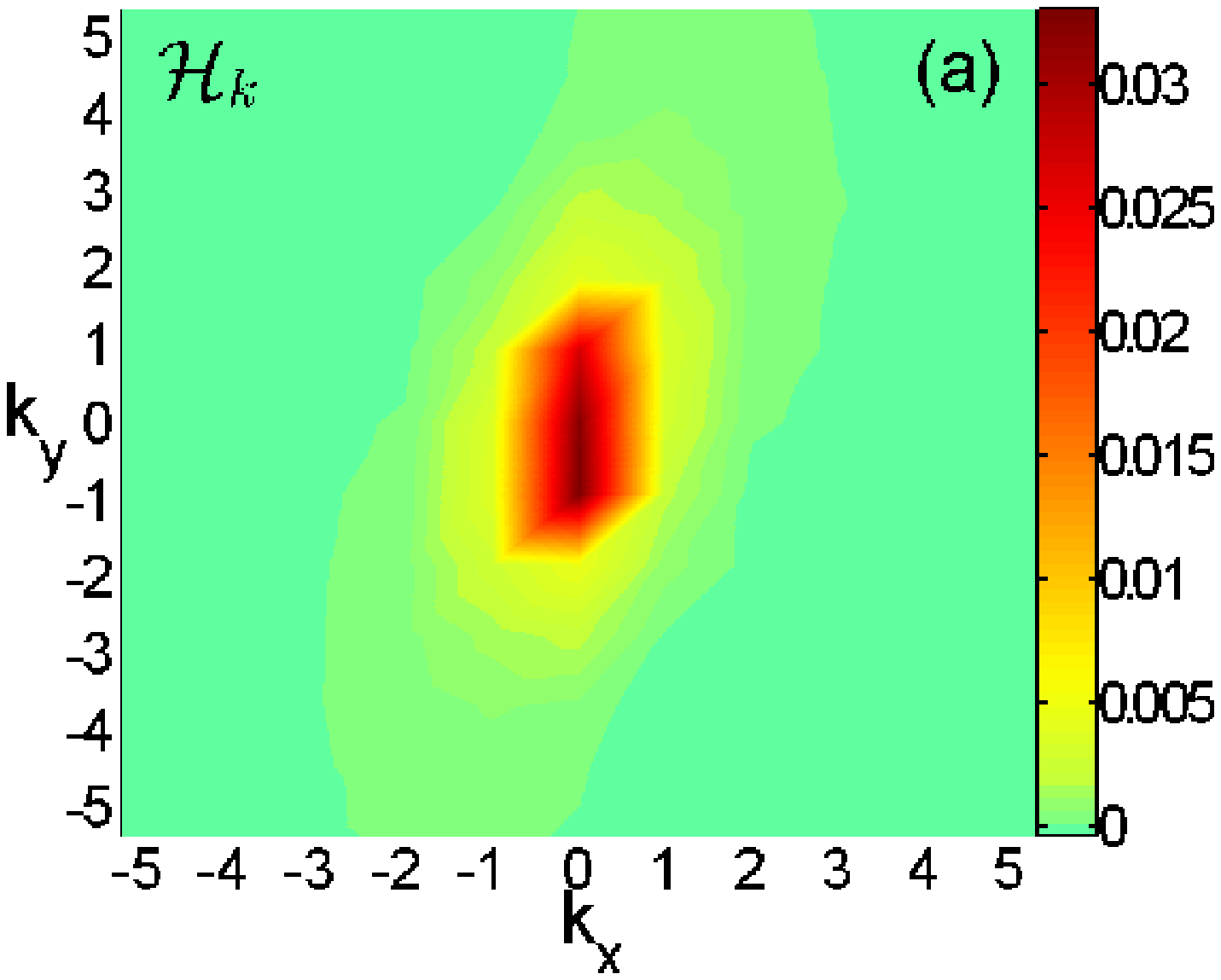}
\includegraphics[width=5.9cm]{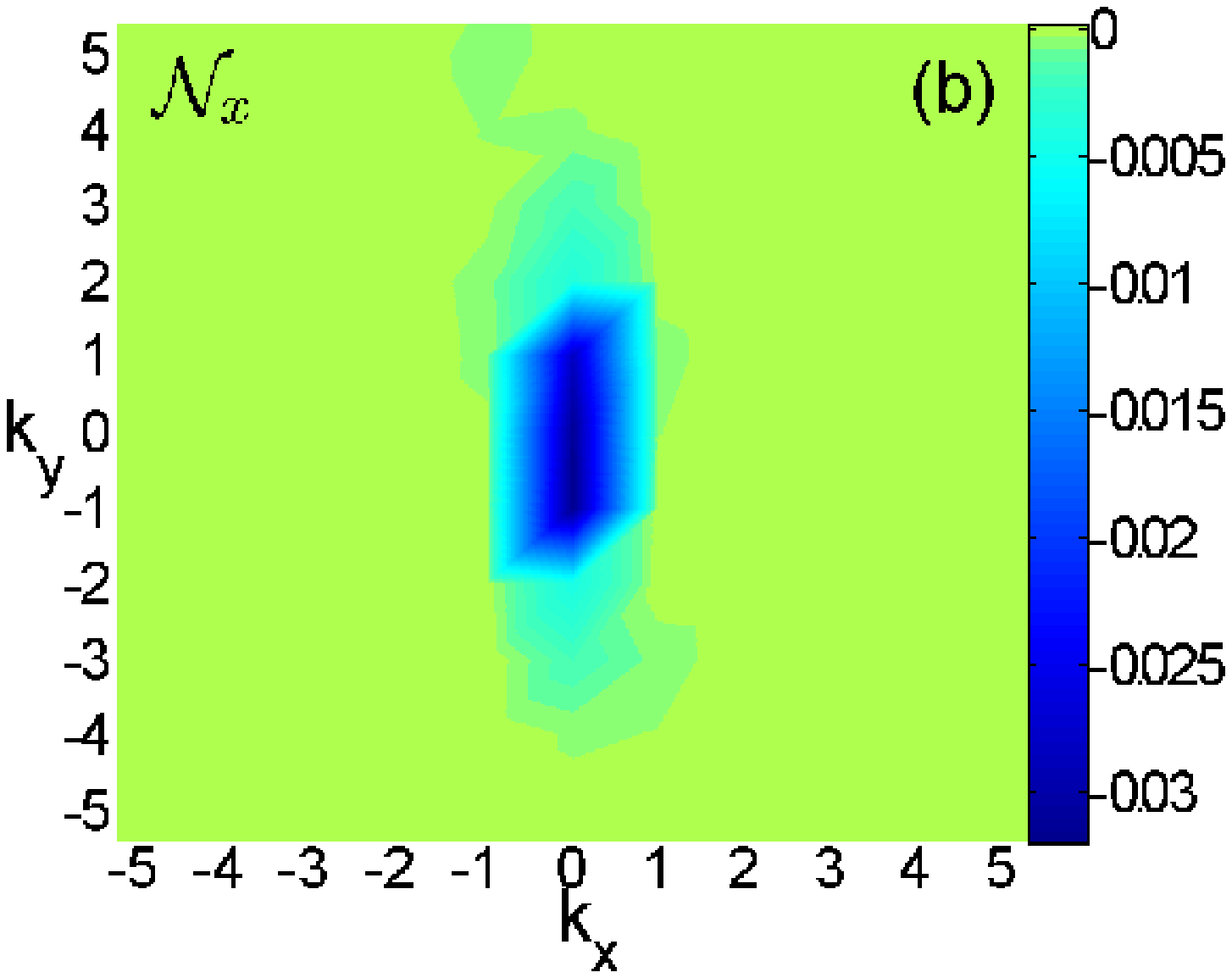}
\includegraphics[width=5.9cm]{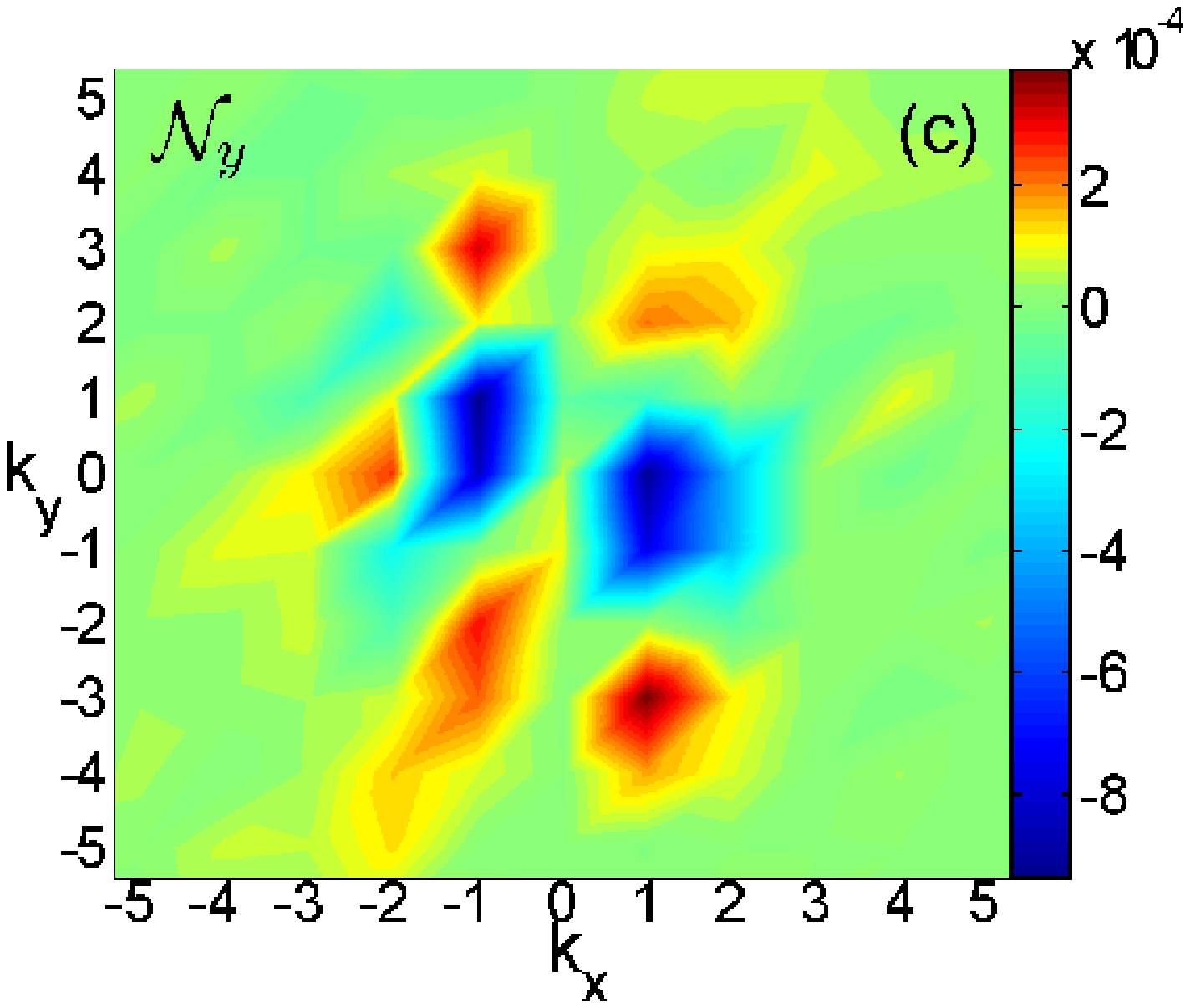}
\caption{(Color online) Maps of the time-averaged (a)
energy-injecting stress ${\cal H}_k$, and nonlinear transfer terms,
(b) ${\cal N}_x$ and (c) ${\cal N}_y$ in {\bf k}-space for the
aspect ratio $(1,2)$. Shown are sections of these quantities in
$(k_x,k_y)$-plane at $k_z=1$.}
\end{figure*}

Figure 19 shows the time-averaged stress and the nonlinear transfer
functions in spectral space at slices $k_z=1$: (a) ${\cal H}_k$, (b)
${\cal N}_x$ and (c) ${\cal N}_y$. Note that all these spectra
exhibit qualitatively a similar shape and anisotropy as are in the
cubic box. The maximum of the time-averaged stress comes again at
the basic mode with ${\bf k}_b=(0,0, \pm 1)$, indicating that in
time-average sense it is dominant, however, as mentioned above, the
basic mode in fact prevails over other modes only during the high
burst events [Fig. 15(b)]. The nonlinear transfer term ${\cal N}_x$,
is negative in the vital area and therefore acts as a sink for
streamwise energy there, but transfers it to intermediate and larger
wavenumbers where this term is positive. For these modes lying
outside the vital area, the stress is negligible and therefore they
get energy mainly due to the nonlinear transfer by ${\cal N}_x$
rather than from the flow. On the other hand, ${\cal N}_y$ exhibits
a noticeable dependence on the wavevector angle (i.e., the
transverse cascade): ${\cal N}_y$ is positive in the first and third
quadrants [$k_y/k_x>0$, yellow and red areas in Fig. 19(c)], where
it creates the shearwise velocity, like that in the the cubic box
[see Fig. 8(h)]. It is then further amplified jointly by ${\cal
N}_y$ and the linear shear-induced term on the rhs of Eq. (17),
which is also positive in this region of spectral space. The
shearwise velocity, undergoing amplification, induces growth of
streamwise velocity due to the nonnormality and ultimately stress.
As a result, as seen from Fig. 19(a), majority of the stress occurs
just in this quadrants of spectral space (at a given $k_z$). So, a
basic self-sustaining scheme discussed above in the case of the
cubic box, extends naturally to this aspect ratio.

Finally, we note that the main distinction between the spectral
dynamics of turbulence in the cubic and $(A_{xz},A_{yz})=(3,2)$
boxes is the difference between the number of active,
energy-carrying modes and, consequently, the size of the vital area
in spectral space. Increasing $A_{xz}$ with respect to $A_{yz}$
introduces more active symmetric and non-symmetric modes into the
turbulence dynamics, widening the vital area. Together with the
basic mode, more active harmonics with $|k_z|=1$ and $k_z=0$ reach
the maximum spectral energy in succession in short time intervals.
Due to this the bursts in the volume-averaged energy evolution are
not as pronounced, or strong as for the cubic box, however, largest
peaks can still be attributed to the dynamics of the basic mode.

\subsection{Aspect ratio $(A_{xz},A_{yz})=(1,2)$}

Evolution of the volume-averaged energy in the case of the aspect
ratio $(1,2)$ is shown in Fig. 20. In this figure, we also show the
time-evolution of the maximum value of the spectral energy and the
energy of the basic mode. Note that the volume-averaged energy well
follows the maximum spectral energy: the bursts (peaks) in the
maximum spectral energy are related to the basic mode dynamics
(amplification). Figure 21 displays the active modes (black dots)
during the evolution that form the self-sustaining dynamics of
turbulence for this aspect ratio. Their total number is 86, more
than that in the cubic box but less than that for the aspect ratio
$(3,2)$. The vital area, where these modes lie, spans the range
$|k_x|\leq 2, |k_y|\leq 7, |k_z|\leq 2$. Note that compared to the
cubic box (Fig. 5), increasing $A_{yz}$ with respect to $A_{xz}$
introduces more active modes with larger $k_y$, but with the same
$k_x$ and $k_z$, resulting in the extension of the vital area along
the $k_y$-axis. Together with the basic mode with ${\bf
k}_b=(0,0,\pm 1)$, the five symmetric modes -- two with $k_z=0$ and
three with $k_z=1$ [shown with bigger black dots in Figs. 21(a) and
21(b)], contribute to the maximum spectral energy, ${\cal
E}_{k,max}$. As a result, the peaks in the volume-averaged energy
appear more pronounced than that in the case $(3,2)$, where larger
number of modes reach comparable energy maxima. However, as evident
from Fig. 20(b), during the largest bursts, the maximum of the
spectral energy is still due to the basic mode, as in the case of
the above two aspect ratios. So, the conclusion drawn above that the
stronger bursts in the total energy are associated with the basic
mode dynamics holds also for the aspect ratio $(1,2)$.

The time-averaged energy spectrum at three different $k_z=0, 1, 2$
is shown in Fig. 22. It exhibits the same type of anisotropy due to
shear as those in the above two cases. The time-averaged energy and
stress spectra integrated in $(k_x,k_y)$-plane is represented as a
function of $k_z$ in Fig. 23. Both spectra reach maximum again at
$k_z=\pm 1$ and decrease with $|k_z|$. Compared to the aspect ratio
$(3,2)$, at $k_z=0$, the integrated spectral energy is smaller than
that at $k_z=\pm 1$ and the integrated stress is again small,
implying that a net energy extraction by 2D modes is much less
effective compared to that by 3D modes. Figure 24 shows the
time-averaged spectra of the stress and nonlinear transfer functions
at $k_z=1$, which are qualitatively similar to those for the other
two aspect ratios considered above and hence the self-sustaining
process works in the same way.

\section{summary and discussion}

We considered a spectrally stable flow with constant shear of
velocity, where the only energy supply of (subcritical) turbulence
is due to the flow nonnormality-induced linear growth of
perturbations. To understand the underlying self-sustenance
mechanism of homogeneous turbulence in this flow, we performed DNS
and analyzed the dynamical processes in Fourier space. From the
Navier-Stokes equations we derived the evolution equations for the
perturbed velocity components in wavenumber space using Fourier
transformation. We explicitly calculated and visualized individual
linear and nonlinear terms in these spectral equations and analyzed
them in detail qualitatively as well as quantitatively based on the
DNS data.

The main results corresponding to the five objectives outlined in
Introduction are summarized below:


-- \emph{\it We first calculated the dynamically important optimal
harmonics (Fig. 1) using the linear nonmodal analysis and identified
the range of aspect ratios, $A_{xz}\leq A_{yz}$ and $A_{xz}\geq 1$,
for which as many of them as possible are included in the simulation
domain.} Afterwards, based on this, we considered three aspect
ratios of the flow domain: cubic $(A_{xz},A_{yz})=(1,1)$, streamwise
elongated $(A_{xz},A_{yz})=(3,2)$ and shearwise elongated
$(A_{xz},A_{yz})=(1,2)$ boxes to understand the dependence of the
turbulence dynamics on the latter.


-- The Reynolds stress, which is of linear origin, is responsible
for the energy extraction from the flow into turbulence via the
nonmodal growth of Fourier harmonics of perturbations. In the
nonlinear regime, we identified a range of wavenumbers in which this
growth is significant and hence the dynamically active modes that
determine the self-sustenance process of the turbulence, carrying
high values of energy and Reynolds stress (Figs. 5, 16, 21). They
have length scales comparable to the box size and hence are mainly
concentrated at small wavenumbers in Fourier space, which is
suitably labeled as \emph{the vital area of the turbulence}.


-- From these active harmonics, those contributing to the maximum of
the spectral energy and stress play a key role in the
self-sustenance process (bigger black dots in Figs. 5, 16, 21). They
have smallest spanwise wavenumbers $k_z=0,2\pi/L_z$ and their number
generally depends on the box aspect ratio. It was shown that for all
the three aspect ratios considered, \emph{\it the basic large scale
mode with the wavenumber ${\bf k}_b = (0, 0, \pm 2\pi/L_z)$ plays a
special role in the homogeneous shear turbulence dynamics},
consistent with previous study by Pumir \cite{Pumir96}. For the
cubic box, it is the only mode that yields the maximum spectral
energy and Reynolds stress in Fourier space during the entire
evolution, much larger than that of the nearby active modes, and
undergoes the largest nonmodal amplification. This amplification
occurs in bursts, alternating with decaying intervals (Fig. 2).
During these bursts, it contributes up to $70\%$ of the total
Reynolds stress. \emph{\it As a result, the basic mode is
responsible for the production of a major fraction of the
turbulence's total energy and largely determines its behavior -- the
total energy also exhibits bursts closely following those of the
basic mode.} Such bursting events in the energy are also observed in
turbulent boundary layers \cite{Robinson91,Jimenez_etal05} and the
homogeneous shear turbulence model is often invoked to understand
their nature \cite{Pumir96,Sekimoto_etal15}. The prevalence of the
basic mode is somewhat reduced for other two aspect ratios. We found
that increasing streamwise or/and shearwise dimensions of the box
introduces more modes, respectively, along these directions in the
vital area, which grow comparably with the basic mode, also
corresponding to the maximum of the spectral energy at certain times
during evolution. As a result, their contribution to the total
energy increases. The evolution of the latter is now determined by
these active modes, rather than by the basic mode only. Due to this,
the total energy is characterized by weaker bursts, especially for
the aspect ratio $(A_{xz},A_{yz})=(3,2)$, with the largest number of
the active modes in the vital area. Nevertheless, the appearance of
highest bursts in the total energy and stress appears to be related
to the basic mode dynamics for these aspect ratios too (Figs. 15,
20).


 -- The nonmodal growth process and, therefore the resulting energy
and Reynolds stress spectra, are anisotropic in Fourier ($\bf k$-)
space, i.e., mainly depend on the orientation of wavevector (Figs.
8, 19, 24). This anisotropy of the linear processes in shear flows,
in turn, leads to anisotropy of nonlinear processes in Fourier
space: the nonlinear terms, which do not directly draw the
background flow energy, transversely redistribute this energy over
wavevector angles. \emph{It was found that this new -- transversal
-- type of nonlinear redistribution of harmonics, referred to as the
transverse cascade, is a generic feature of nonlinear dynamics of
perturbations in spectrally stable shear flows. It differs from and
represents an alternative to the classical energy cascade processes
(direct/inverse) in shear flow turbulence.} Previously, the
transverse cascade was also demonstrated to take place and play an
important role in 2D HD and MHD shear flows
\cite{Horton_etal10,Mamatsashvili_etal14}. This cooperation of
anisotropic linear and nonlinear processes, in turn, gives rise to
an anisotropic energy spectrum in shear flows, which, in general,
differs from the classical Kolmogorov spectrum in homogeneous
turbulence without background shear, especially at small and
intermediate wavenumbers. As a result, the transverse cascade may
naturally appear to be a keystone of the \emph{bypass} concept of
subcritical turbulence in spectrally stable HD shear flows, which is
being actively discussed among the hydrodynamical community.
Therefore, the conventional description of shear flow turbulence
solely in terms of direct and inverse cascades, which leaves out the
transverse cascade, might be incomplete and misleading.


-- The vital area in wavenumber space is the primary site of
activity of the energy-supplying linear nonmodal growth and the
nonlinear transverse cascade that lie at the heart of the
self-sustaining dynamics of turbulence. The course of events
ensuring the self-sustenance of the turbulence is as follows. The
transverse cascade continually transfers harmonics to those
quadrants of the vital area where the shear flow causes linear
nonmodal growth, i.e., regenerates harmonics that can undergo
amplification due to the flow nonnormality, and thereby closes
feedback loop ensuring sustenance of the turbulence. This general
picture comprises refined, somewhat differing interplay of linear
and nonlinear processes for the basic mode and active symmetric and
non-symmetric modes. These, qualitatively quite simple but
quantitatively a bit involved schemes are described in detail for
different aspect ratios in Section III (see also Figs. 10-14).
\emph{\it Thus, we have demonstrated that in the considered constant
shear flow, the turbulence is maintained by a subtle interplay
between linear and nonlinear processes: the first supplies energy
for turbulence via shear-induced nonmodal growth process
(characterized by the Reynolds stresses) and the second plays an
important role of providing a positive feedback that makes this
growth process long-lived against viscous dissipation, which would
otherwise be transient in the absence of the latter.}

The number of the active harmonics (defined as the harmonics whose
energies grow more than $10\%$ of the maximum spectral energy at
least once during the evolution) in the vital area, which are the
main participants in the turbulence dynamics, generally depends on
the box aspect ratio and is quite large. In the considered here
aspect ratios, it exceeds 35 for the cubic box and can be more than
200 for longer boxes in the streamwise or shearwise directions (see
Figs. 5, 16 and 21), indicating fairly complex nature of
self-sustaining dynamics of turbulence, which apparently cannot be
described by low-order models. In this paper, performing a full DNS,
the realization of the nonlinear transverse cascade and its role in
the self-sustenance of homogeneous shear turbulence are demonstrated
in the most general manner, without truncating the number of modes
and using other simplifying assumptions usually made in low-order
models of self-sustaining processes (e.g., Refs.
\cite{Waleffe95,Waleffe97,Mohelis_etal04}). Nevertheless, one can
relate the described here self-sustaining scheme of homogeneous
shear turbulence in the constant shear flow without (physical)
boundaries to the self-sustenance scenario in low-order models
which, however, commonly include walls affecting the dynamics (see
e.g., Refs.~\cite{Farrell_Ioannou12,Thomas_etal14,Thomas_etal15}).
Namely, the nonlinear regeneration of streamwise rolls is analogous
to the production of shearwise velocity due to the nonlinear
transfer terms and the generation and nonmodal amplification of
streamwise streaks from the former due to shear (nonnormality) is
similar to the amplification of streamwise velocity. One of the main
results of the present study is that the self-sustaining process of
turbulence in spectrally stable shear flows in fact does not require
the presence of walls and is intrinsic to the flow system itself,
being governed by the interplay between linear nonnormality-induced
and the nonlinear transverse cascade processes.

Finally, we would like to emphasize that revealing the new aspects
of the homogeneous shear turbulence dynamics -- anisotropic energy
spectra, the nonlinear transverse cascade and the notion of the
vital area, where the self-sustaining process is concentrated, have
become largely possible owing to analysis in Fourier space. This
allows us to gain a deeper insight into turbulence dynamics and its
underlying sustenance mechanism. At the same time, the performed
simulations show that the basic mode, which is central in the energy
exchange between the shear flow and turbulence, has a large scale
spanwise variation and is uniform in the streamwise and shearwise
directions, $k_x= k_y = 0, k_z = \pm 2\pi/L_z$; it has streamwise
and shearwise velocities, but zero spanwise velocity. This simple
configuration (geometry) of the basic mode makes it easily
describable in physical space too -- its amplification due to the
nonnormality can be readily understood from Eqs. (3) and (4) in
physical space. The streamwise velocity $u_x$ is generated from the
shearwise velocity due to shear (linear term -$Su_y$ in Eq. 3). As a
result, this two velocities correlate and produce positive stress,
extracting flow energy. The role of nonlinearity in Eq. (4) is then
to regenerate the shearwise velocity $u_y$.

\begin{acknowledgments}
GM, GK and GC are thankful for hospitality at the School of
Aeronautics, Universidad Polit{\'e}cnica de Madrid, and for
financial support within the Second Multiflow Summer Workshop, 25
May-26 June, 2015, funded in part by the Multiflow Program of the
European Research Council. GM is supported by the Georg Forster
Postdoctoral Research Fellowship from the Alexander von Humboldt
Foundation. SD is supported by the China Scholarship Council. We
thank the Referees for useful comments that improved the
presentation of our results.
\end{acknowledgments}

\appendix
\section{Fourier transform in the presence of shear}

As discussed in the text, $k_y$ of an individual harmonic linearly
changes with time, $k_y(t)=k_y(0)-Sk_xt$, due to the background
shear flow, that is also consistent with the shear-periodic boundary
conditions (see e.g., Refs.
\cite{Farrell87,Chagelishvili_etal03,Jimenez13}). For this reason, a
standard FFT technique cannot be applied for calculating Fourier
transforms along this direction during post-processing. To
circumvent this problem, we used the approach of Refs.
\cite{Mamatsashvili_Rice09,Heinemann_Papaloizou09}, which we recap
below.

In a finite computational domain, streamwise $k_x$ and spanwise
$k_z$ wavenumbers are both discrete: $k_x = 2\pi n_x/L_x, k_z = 2\pi
n_z/L_z$, while the shearwise wavenumber is $k_y(t)=2\pi
n_y/L_y-St(2\pi n_x/L_x)$, where $n_x, n_y, n_z$ are integers,
$-N_x/2\leq n_x \leq N_x/2, -N_y/2\leq n_y \leq N_y/2, -N_z/2\leq
n_z \leq N_z/2$. In this case, any flow field at time $t$ can be
decomposed in Fourier expansion
\begin{equation*}
f(x,y,z,t)=\sum_{n_x,n_y,n_z}\bar{f}_{n_x,n_y,n_z}(t){\rm exp}[{\rm
i}k_xx+{\rm i}k_y(t)y+{\rm i}k_zz],
\end{equation*}
where the Fourier coefficients can be computed from the inverse
transform
\begin{multline*}
\bar{f}_{n_x,n_y,n_z}(t)=\frac{1}{L_xL_yL_z}\times \\ \int\int\int
f(x,y,z,t){\rm exp}[-{\rm i}k_xx-{\rm i}k_y(t)y-{\rm i}k_zz]dxdydz,
\end{multline*}
or more explicitly, after separating out the time-dependent part in
the exponent and rearranging, we get
\begin{multline*}
\bar{f}_{n_x,n_y,n_z}(t)=\\\frac{1}{L_y}\int{\rm exp}\left({\rm
i}St\frac{2\pi n_x}{L_x}y\right)\left[\frac{1}{L_xL_z}\int\int
f(x,y,z,t)\times\right.\\\left. {\rm exp}\left(-{\rm i}\frac{2\pi
n_x}{L_x}x-{\rm i}\frac{2\pi n_z}{L_z}z\right)dxdz\right]{\rm
exp}\left(-{\rm i}\frac{2\pi n_y}{L_y}y \right)dy.
\end{multline*}
This last expression suggest the way to compute the Fourier
transform of a real quantity from the simulation in the presence of
shear. So, we first do a standard FFT in $x$- and $z$-directions,
multiply the result by ${\rm exp}[{\rm i}St(2\pi n_x/L_x)y]$, which
accounts for the effect of shear, and finally do again FFT in the
$y$-direction. The time-dependent shearwise wavenumber associated
with the Fourier coefficient $\bar{f}_{n_x,n_y,n_z}$ at time $t$ is
given by
\[
k_y(t)=\frac{2\pi n_y}{L_y}-St\frac{2\pi n_x}{L_x}+l\frac{2\pi
N_y}{L_y},
\]
where an integer number $l$ is chosen such that to ensure $k_y(t)$
always stays within the numerical domain in spectral space 
\[
-\frac{\pi L_y}{N_y}\leq k_y(t)\leq \frac{\pi L_y}{N_y}.
\]

\bibliography{biblio}

\end{document}